\documentclass[aps,prb,twocolumn,superscriptaddress,nobibnotes,longbibliography]{revtex4-1}
\bibpunct{[}{]}{;}{n}{}{}

\usepackage[english]{babel}
\usepackage{graphics, bm, xspace}
\usepackage{graphicx}
\usepackage{amsmath}
\usepackage{amssymb}
\usepackage{times}
\usepackage[bookmarks=false,linkcolor=blue,urlcolor=blue,colorlinks,citecolor=blue]{hyperref}
\usepackage{float}
\usepackage{gensymb}

\newcommand{\ang}    {\mbox{\AA}}
\newcommand{\beq}    {\begin{equation}}
\newcommand{\enq}    {\end{equation}}
\newcommand{\ceq}[1] {(\ref{#1})}

\newcommand{\eps}    {\epsilon}
\newcommand{\aav}    {{\bf a}}
\newcommand{\dd}     {{\bf d}}
\newcommand{\kk}     {{\bf k}}
\newcommand{\hh}     {{\bf h}}
\newcommand{\qq}     {{\bf q}}

\newcommand{\KK}     {{\bf K}}
\newcommand{\pp}     {{\bf p}}
\newcommand{\ttt}    {{\bf t}}

\newcommand{\GG}     {{\bf G}}

\newcommand{\MM}     {{\bf M}}

\newcommand{\ttau}   {{\boldsymbol\tau}}

\newcommand{\df}     {\equiv}

\newcommand{\rr}     {{\bf r}}

\newcommand{\rem}    {{\rm Re}}

\newcommand{\mos}    {${\rm MoS_2}$\xspace}
\newcommand{\mose}   {${\rm MoSe_2}$\xspace}
\newcommand{\mote}   {${\rm MoTe_2}$\xspace}

\newcommand{\wse}    {${\rm WSe_2}$\xspace}
\newcommand{\wss}    {${\rm WS_2}$\xspace}
\newcommand{\wte}    {${\rm WTe_2}$\xspace}
\newcommand{\nbse}   {${\rm NbSe_2}$\xspace}

\newcommand{\ket}[1] {\lvert #1 \rangle}

\newcommand{\ssigma}     {{\boldsymbol\sigma}}
\newcommand{\stau}     {{\boldsymbol\tau}}

\begin{document}

\title{Van der Waals heterostructures with spin-orbit coupling}
\author{Enrico Rossi}
\email[Corresponding author: ]{erossi@wm.edu}
\affiliation{Department of Physics, William \& Mary, Williamsburg, Virginia (23187), USA}
\author{Christopher Triola}
\affiliation{Department of Physics and Astronomy, Uppsala University, Box 516, S-751 20 Uppsala, Sweden}


\begin{abstract}
In this article we review recent work on van der Waals (vdW) systems in which
at least one of the components has strong spin-orbit coupling.
We focus on a selection of vdW heterostructures to exemplify the type 
of interesting electronic properties that can arise in these systems.
We first present a general effective model 
to describe the low energy electronic degrees of freedom in these systems.
We apply the model to study 
the case of (vdW) systems formed by a graphene sheet and
a topological insulator. We discuss the electronic transport properties
of such systems and show how they exhibit much stronger spin-dependent
transport effects than isolated topological insulators.
We then consider vdW systems in which the layer
with strong spin-orbit coupling is a monolayer transition metal dichalcogenide (TMD)
and briefly discuss graphene-TMD systems.
In the second part of the article we discuss the case in which the vdW system includes a superconducting layer in addition to the layer with strong spin-orbit coupling.
We show in detail how these systems can be designed to realize
odd-frequency superconducting pair correlations. Finally, we discuss
twisted graphene-\nbse bilayer systems as an example in which the strength
of the proximity-induced superconducting pairing in the normal layer, and its Ising character, 
can be tuned via the relative twist angle between the two layers forming the heterostructure.
\end{abstract}
\maketitle


\section{Introduction}
Van der Waals (vdW) heterostructures~\cite{geim2007,geim2013,liu2016van,Novoselov2016} represent a growing class of systems which are formed from two-dimensional (2D) layers of material held to one another by only van der Waals forces. One of the most common vdW heterostructures, graphite, is composed entirely of layered sheets of graphene~\cite{novoselov2004,castroneto2009,dassarma2011}, a one-atom thick 2D crystal of carbon atoms arranged in a honeycomb structure.
The nature of the vdW force implies that in a vdW system the stacking configuration of the layers is not dictated by chemistry but, to a great extent, can be tuned arbitrarily in an almost continuous way.
The explosion of interest in vdW systems stems from the fact that this is now experimentally feasible. Experimentalists are able to isolate layers of different materials only one-atom, or few-atoms, thick, and to combine such layers with increasing control of the stacking configuration.
A striking example of this tunability is the recent experimental realization of vdW systems formed by two graphene layers in which the stacking angle, {\em twist angle}, can be adjusted to within a fraction of a degree~\cite{Cao2018,Cao2018b,Yankowitz2019}.
These experiments~\cite{Cao2018,Cao2018b,Yankowitz2019} have shown that, by tuning the relative twist angle between graphene layers, the system can become superconducting or insulating. These remarkable results are just one example of the ways in which vdW heterostructures can be used to realize electronic systems with exotic and desirable properties.

Of particular interest are the recent developments involving vdW systems comprised of two layers in which one of the layers has a strong spin-orbit coupling (SOC). The main interest in these systems arises from the possibility of realizing novel electronic systems by combining such a layer with a different layer which possesses little or no SOC but with other interesting properties. For example, graphene possesses only very weak SOC but has very high electron mobility at room temperature. By combining graphene with the 2D surface of a three-dimensional (3D) topological insulator (TI)~\cite{ZhangNatPhys2009,HasanRMP2010,qi2011rmp}, the SOC in the graphene layer can be enhanced by an order of magnitude~\cite{jzhang2014,khokhriakov2018tailoring}. Additionally, as we discuss in Sec.~\ref{sec:odd-f}, vdW heterostructures with SOC appear to be ideal systems for realizing unconventional odd-frequency superconducting states~\cite{triola2016prl}. In contrast to conventional superconducting states, the defining feature of odd-frequency superconductors is that they host Cooper pairs which are odd functions in the relative time, making these states intrinsically-dynamical.

The literature on vdW systems is by now very large, for this reason we restrict ourselves to a particular subset of vdW systems that are among the most relevant to the subject of this special issue and that we have studied over the past few years. We first present, in Sec.~\ref{sec:model}, the model to describe vdW systems with SOC. To exemplify the application of this general formalism, in Sec.~\ref{sec:g-ti}, we discuss the case of vdW heterostructures formed by graphene and a TI. In Sec.~\ref{sec:g-tmd}, we discuss recent progress on vdW systems which include both graphene and a layer of transition metal dichalcogenide (TMD). We then consider, in Sec.~\ref{sec:sc}, the case in which one of the layers is a superconductor, discussing the possibility of realizing different exotic superconducting states in vdW systems with SOC.  In Sec.~\ref{sec:conclusions} we present our conclusion and outlook for future developments.

While the focus of this article is primarily on vdW systems composed of graphene, topological insulators, and TMDs, we note that much progress has also been made studying  vdW systems with other components. Notably, we do not discuss the recently discovered 2D magnetic materials\cite{burch2018magnetism,johansen2019current,gong2019two}, which include: FePS$_3$\cite{lee2016ising}, Cr$_2$GeTe$_6$\cite{gong2017discovery}, CrI$_3$\cite{huang2017layer,Klein2018}, VSe$_2$\cite{bonilla2018strong}, MnSe$_x$\cite{o2018room}, and Fe$_3$GeTe$_2$\cite{fei2018two,deng2018gate}. Similarly, we do not cover the magnetic proximity effect\cite{hauser1969magnetic} which has been discussed in a variety of vdW heterostructures\cite{lazic2016effective,liang2017magnetic,cortes2019}. We also omit discussion of the intriguing heterostructures which can be made using monolayers with buckled honeycomb structure --silicene\cite{vogt2012silicene}, germanene\cite{liu2011quantum,davila2014germanene}, and stanene\cite{zhu2015epitaxial}--systems involving nanoribbons~\cite{gani2018}, and 
structures formed by different layers of TMDs~\cite{Kosmider2013a} and twisted TMD homobilayers~\cite{wang2017interlayer,li2017ultrafast,wu2019topological,yu2019giant}.


\section{Model}
\label{sec:model}
The Hamiltonian describing a generic double-layer vdW system can be written as $H = H_1 + H_2 + H_t$, where $H_\ell$ ($\ell=1,2$) is the Hamiltonian associated with layer $\ell$ and $H_t$ describes tunneling processes between the two layers. Here, we write $H_1 = \sum_{\kk\alpha\alpha'}c^\dagger_{\kk\alpha}h(\kk)_{1;\alpha\alpha'}c_{\kk\alpha'}$, $H_2 = \sum_{\kk\alpha\alpha'}d^\dagger_{\kk\alpha}h(\kk)_{2;\alpha\alpha'}d_{\kk\alpha'}$, where $c^\dagger_{\kk\alpha}$ and $d^\dagger_{\kk\alpha}$ ($c_{\kk\alpha}$ and $d_{\kk\alpha}$) create (annihilate) single electron states in layers 1 and 2, respectively, with momentum $\kk$ and all other degrees of freedom described by the composite index $\alpha$, including spin, orbital, and particle-hole degrees of freedom. 
We assume that the tunneling between the two layers depends only on the difference between the positions, $\rr_1$, $\rr_2$, of the electrons in the two layers. As a consequence, the crystal momentum is conserved during tunneling processes and we have
\begin{align}
 H_t= \sum_{\kk_1\kk_2}\sum_{\GG_1\GG_2} \sum_{\alpha_1 \alpha_2} \sum_{s_1 s_2}& 
      t_{\alpha_1 \alpha_2}(\kk_1 + \GG_1)e^{i\GG_2\cdot\ttau_{s_2}-i\GG_1\cdot\ttau_{s_1}}  \label{eq:T} \\
      &\times c^\dagger_{\kk_1\alpha_1} d_{\kk_2 + (\GG_2 - \GG_1)\alpha_2} + \text{h.c.} \nonumber
\end{align}
where $\GG_\ell$ is the reciprocal lattice vector in layer $\ell$ and we have allowed for the possibility that the lattices making-up each of the two layers may possess basis vectors $\ttau_{s_\ell}$. For example, for the case of graphene we have two basis vectors $\ttau_s$, which we can write as ${\ttau_s}=\{(0,0);(a_0,0)\}$, where $a_0$ is the carbon-carbon distance.

In a bilayer vdW system, two very different kinds of stacking are possible: commensurate stacking, and incommensurate stacking. In the first case, the vdW bilayer has a periodic structure in real space with a large primitive cell that is commensurate with the primitive cells of both layers. In the incommensurate case, no such periodicity exists in real space. For the communsurate case we obtain a well defined Moir\'e pattern. 

Given two layers ($\ell=1,2$) with primitive lattice vectors ${\aav_{\ell i}}$ ($i=1,2$), in order for a stacking to be commensurate there must exist four integers $m_1, m_2, n_1, n_2$ such that:
\beq
 m_1 \aav_{11} +  m_2 \aav_{12} = n_1 \aav_{21} + n_2 \aav_{22}.
 \label{eq:comm01}
\enq
Without loss of generality, using complex numbers to represent 2D vectors, we can write $\aav_{11}=a_{10} e^{-i\theta_1}$, $\aav_{12}=a_{10} e^{+i\theta_1}$, $\aav_{21}=a_{20} e^{-i(\theta_2-\theta)}$, $\aav_{22}=a_{20}e^{+i(\theta_2+\theta)}$, where $a_{\ell 0}$ is the lattice constant of layer $\ell$, $2\theta_\ell$ is the angle between the primitive lattice vectors of layer $\ell$, and $\theta$ is the twist angle between the two layers. Using this notation, considering that the magnitudes of the two vectors on the left and right hand side of Eq.~\ceq{eq:comm01} have to be same we obtain the following Diophantine equation constraining the integers ${m_1,m_2,n_1,n_2}$:  
\beq
 \left(\tfrac{a_{10}}{a_{20}}\right)^2\left(m_1^2 + m_2^2 + 2m_1 m_2 \cos{2\theta_1}\right) = n_1^2 + n_2^2 + 2 n_1 n_2\cos{2\theta_2}.
 \label{eq:comm02}
\enq 
Notice that Eq.~\ceq{eq:comm02} does not depend on $\theta$. For a commensurate stacking the twist angle and the integers ${m_1,m_2,n_1,n_2}$ are related
via the equation
\beq
 \theta = \ln\left[\frac{a_{10}(m_1 e^{-i\theta_1} + m_2 e^{i\theta_1})}{a_{20}(n_1 e^{-i\theta_2}+ n_2 e^{i\theta_2})}  \right].
 \label{eq:comm03}
\enq 
The number of commensurate stackings forms a set of measure zero in the whole space of possible stackings. However, the experimental evidence suggests \cite{Carr2018,Jiang2019,hunt2013,Jung2015,Yankowitz2016} that 2D vdW crystals tend to relax into stacking configurations that are, at least locally, commensurate.

The size of the primitive cell increases very rapidly as the twist angle $\theta$ decreases for $\theta<5\degree$. For this reason, to treat vdW systems with small twist angles, it is more efficient to use an effective model in momentum space~\cite{dossantos2007,morell2010,bistritzer2011} in which only the dominant interlayer tunneling processes are kept. For small twist angles an accurate description is obtained by keeping only those tunneling processes for which $|\kk_1 - \kk_2|=|\GG_1-\GG_2|$ is smallest. For the practically important case in which the 2D crystals are triangular lattices and the low energy states (i.e. the states closest to the Fermi energy) are located at the corners ($\KK$ and $\KK'$ points) of the hexagonal Brillouin zone (BZ), as for graphene, the minimum value of $|\GG_1-\GG_2|$ is equal to $2K\sin(\theta/2)$, where $K=|\KK|$, and there are three vectors $\qq_i = \GG_1-\GG_2$ ($i=1,2,3)$ for which $|\qq_i|=q=2K\sin(\theta/2)$. If we account for all tunneling processes with $|\kk_1 - \kk_2|=q$ we find that the higher-order tunneling processes generate a honeycomb structure in momentum space with nearest neighbor points connected by the vectors $\qq_i$~\cite{bistritzer2011}. 

The relative contributions of higher-order recursive tunneling processes is controlled by the parameter $\gamma\equiv t/(\eps(q))$, where $t$ is the interlayer tunneling strength between states in the two layers with momenta that differ by at most $q$, and $\eps(q)={\rm min}[\eps_1(q),\eps_2(q)]$, where $\eps_\ell(q)$ is the energy of electrons with momentum of magnitude $q$ in layer $\ell$. The value of $\gamma$, therefore, controls the size of this momentum space lattice that one needs to consider to obtain an accurate band structure for the vdW system. For $\gamma<1$ the size of the required lattice in momentum space can be quite small. For very small $\gamma$ it is sufficient to keep just the first ``ring'' of interlayer hopping processes. In this case the effective Hamiltonian matrix takes the simple form:
\begin{align}
 \hat{H}_{\mathbf{k}} & =\left(\begin{array}{cccc}
 \hat{h}_1(\mathbf{k}) & \hat{t}_{1}^{\dagger} & \hat{t}_{2}^{\dagger} & \hat{t}_{3}^{\dagger}\\
 \hat{t}_{1} & \hat{h}_2(\mathbf{q}_{1}+\mathbf{k}) & 0 & 0\\
 \hat{t}_{2} & 0 & \hat{h}_2(\mathbf{q}_{2}+\mathbf{k}) & 0\\
 \hat{t}_{3} & 0 & 0 & \hat{h}_2(\mathbf{q}_{3}+\mathbf{k})
 \end{array}\right),
 \label{eq:H02}
\end{align}
where $\hat{t}_{i}$ are matrices containing the interlayer tunneling elements.

The model described above can be generalized to the case when one (or more) of the layers forming the vdW system is superconducting. For concreteness, let us consider the case when the vdW system is formed by only two layers and only one is superconducting; it is fairly straightforward to generalize the formalism to more complex situations, such as when both layers are superconducting. We assume layer 2 is a superconductor so that $H_2\to H_{SC}$, and that this superconducting layer is a 2D crystal with triangular lattice for which the low energy states are located close to the corners of the BZ, the $\KK$ and $\KK$ points.
We also assume simple s-wave pairing so that the superconducting order parameter $\Delta_{SC}$ couples states at opposite valleys with opposite spin and momentum $\kk$, measured from $\KK$ ($\KK'$).
The Hamiltonian for the vdW system is $H= \sum_{\kk}\Psi^\dagger_{\kk SC} \hat H_{12}^{\rm SC}(\kk)\Psi_{\kk SC}$,
with
$\Psi^\dagger_{\kk SC} =(\psi^\dagger_{\kk}, \psi^T_{-\kk}$),
\beq
 \psi^\dagger_{\kk,\alpha,\beta_1,\beta_2,\beta_3} =
  (c^\dagger_{\kk,\alpha},d^\dagger_{\kk+\qq_1,\alpha'_1},d^\dagger_{\kk+\qq_2,\alpha'_2},d^\dagger_{\kk+\qq_3+\alpha'_3})
\enq
and, to lowest order in $\gamma$:
\begin{eqnarray}
\hat H_{12}^{\rm SC}(\kk)  = \left(
 \begin{array}{cc}
  \hat h_{12}(\kk)          &  \hat \Delta_{SC} \hat \Lambda \\
  \hat \Delta_{SC} \hat \Lambda^\dagger  &  -\hat h^T_{12}(-\kk)                      
\end{array}
\right),
\label{eq:HKSC}
\end{eqnarray}
where $\hat h_{12}$ is the Hamiltonian given by Eq.~\ceq{eq:H02} and $\hat \Lambda$ is the matrix with indices describing the internal structure of the superconducting pairs in layer 2. For the case in which layer 1 has spin and sublattice degrees of freedom $\alpha$ and in layer 2, at low energies, we can assume that the only internal degree of freedom is the spin, $\hat \Lambda$ is a $10\times 10$ block diagonal matrix given by~\cite{gani2019}:
\begin{equation}
\hat \Lambda  = \text{diag}\left(0,i\hat\sigma_2,i\hat\sigma_2,i\hat\sigma_2 \right)
\label{eq:lambda}
\end{equation}
where the first block is a $4\times 4$ matrix of zeros and $\hat\sigma_2$ is a $2\times 2$ Pauli matrix in spin space.
As we will discuss in Sec.~\ref{sec:ising_SC}, the form for $\hat \Lambda$ given by Eq.~\ceq{eq:lambda} is applicable to the case of a vdW structure composed of a superconducting TMD monolayer coupled to a graphenic layer, such SLG or BLG.


\section{Graphene-TI heterostructures}
\label{sec:g-ti}
In this section we follow the theoretical treatment given in Ref.~\cite{jzhang2014} to examine the electronic properties of 
graphene-TI vdW heterostructures~\cite{steinberg2015,bian2016}. 
The original motivation for studying the properties of a graphene layer coupled to the 2D surface of a 3DTI was the possibility of inducing strong SOC or novel spin textures in the graphene layer
predicted by theoretical models\cite{jin2013,jzhang2014,song2018spin}. 
Excitingly, this proximity-induced SOC has recently been observed using a combination of transport measurements and ab initio calculations\cite{khokhriakov2018tailoring}. Additionally, we also note that the possibility of inducing Dirac states with very low Fermi velocity, in which interaction effects could be greatly enhanced\cite{cao2016heavy}, has also been discussed. 

In graphene the carbon atoms are arranged in a 2D honeycomb structure formed by two triangular sublattices, $A$ and $B$, 
with lattice constant $a_g=\sqrt{3}a=2.46\ang$, with $a=1.42\ang$ the carbon-carbon atomic distance. 
The low energy states of graphene are located at the $\KK$ and $\KK'$ points of the BZ: 
$\KK=(4\pi/(3a_g),0)$, $\KK'=(-4\pi/(3a_g),0)$ (and equivalent points connected by reciprocal lattice wave vectors). 
At low energies close to the $\KK$ and $\KK'$ points in graphene, the electrons are well described 
as massless Dirac fermions with Hamiltonians 
\begin{equation}
\begin{aligned}
H^{g,K} &= \sum_{\kk}\sum_{\tau\tau'}\sum_{\sigma\sigma'}c^\dagger_{\KK+\kk,\tau\sigma}  h^{g,K}_{\textbf{k};\tau\tau',\sigma\sigma'}  c_{\KK+\kk,\tau'\sigma'}, \\
 \hat h^{g,K}_{\textbf{k}} &= \left(\hbar v_F\kk\cdot\stau-\mu_g \hat\tau_0\right)\otimes \hat\sigma_0,
\end{aligned} 
 \label{eq:HgK}
\end{equation}
and
\begin{equation}
    \begin{aligned}
H^{g,K'}&= \sum_{\kk}\sum_{\tau\tau'}\sum_{\sigma\sigma'}c^\dagger_{\KK'+\kk,\tau\sigma} h^{g,K'}_{\textbf{k};\tau\tau',\sigma\sigma'} c_{\KK'+\kk,\tau'\sigma'}, \\    
 \hat h^{g,K'}_{\textbf{k}}&=-\left(\hbar v_F\kk\cdot\stau^*+\mu_g\hat\tau_0\right)\otimes \hat\sigma_0,
    \end{aligned}
     \label{eq:HgKp} 
\end{equation}
where 
$c^\dagger_{\pp,\tau\sigma}$ ($c_{\pp,\tau\sigma}$) is the creation (annihilation) operator for an electron, in the graphene sheet, with spin $\sigma$ and two-dimensional momentum $\hbar\pp=\hbar(p_x,p_y)$, $\kk$ is a wave vector measured from $\KK$ ($\KK'$), $v_F=10^6$~m/s is the Fermi velocity, $\mu_g$ is the chemical potential, and ${\hat \tau_i}$ and ${\hat \sigma_i}$ ($i=0,1,2,3)$) are the $2\times2$ Pauli matrices in sublattice and spin space, respectively.

One class of materials for which the effect of spin-orbit coupling on the low energy fermionic states
is particularly strong is the one of three dimensional (3D) topological insulators (TIs)~\cite{ZhangNatPhys2009,HasanRMP2010,qi2011rmp}.
In these materials the combination of spin-orbit coupling and time reversal symmetry guarantees
the presence of topologically protected 2D surface states within the band gap of the bulk states.
For this reason, 3D TI materials are in many respects ideal materials for the creation of novel 
vdW heterostructures in which the effect of SOC is significant.
The 2D states at the TI's surface (TIS)
are well described as massless Dirac fermions with Hamiltonian~\cite{zhang2009ti,liucx2010}:
\begin{equation}
\begin{aligned}
 H^{\text{TIS}}&=\sum_{\mathbf{k},\sigma\sigma'}d_{\mathbf{k},\sigma}^{\dagger}h^{\text{TIS}}_{\textbf{k};\sigma\sigma'}d_{\mathbf{k},\sigma'}, \\
\hat h^{\text{TIS}}_\textbf{k}&=\hbar   
 v_{TI}\left(\boldsymbol{\sigma}\times\mathbf{k}\right)\cdot\hat{\mathbf{z}}-\mu_{TI}\hat\sigma_0,
\end{aligned}
 \label{eq:HTI}
\end{equation}
where $d_{\mathbf{k},\sigma}^{\dagger}$ ($d_{\mathbf{k},\sigma}$)
creates (annihilates) a surface massless Dirac fermion with spin $\sigma$
at wave vector $\mathbf{k}=(k_x,k_y,0)$ measured from the zone center ($\overline{\Gamma}$-point) of the surface-projected (BZ),
$\boldsymbol{\sigma}=\left(\hat\sigma_1,\hat \sigma_2,\hat \sigma_3\right)$ is the vector of Pauli matrices acting on spin space, $\hat{\mathbf{z}}$ is the
unit vector along the $z$ direction, and $\mu_{TI}$ is the chemical potential.

\begin{figure}[htb]
 \begin{center}
  \includegraphics[width=1.0\columnwidth]{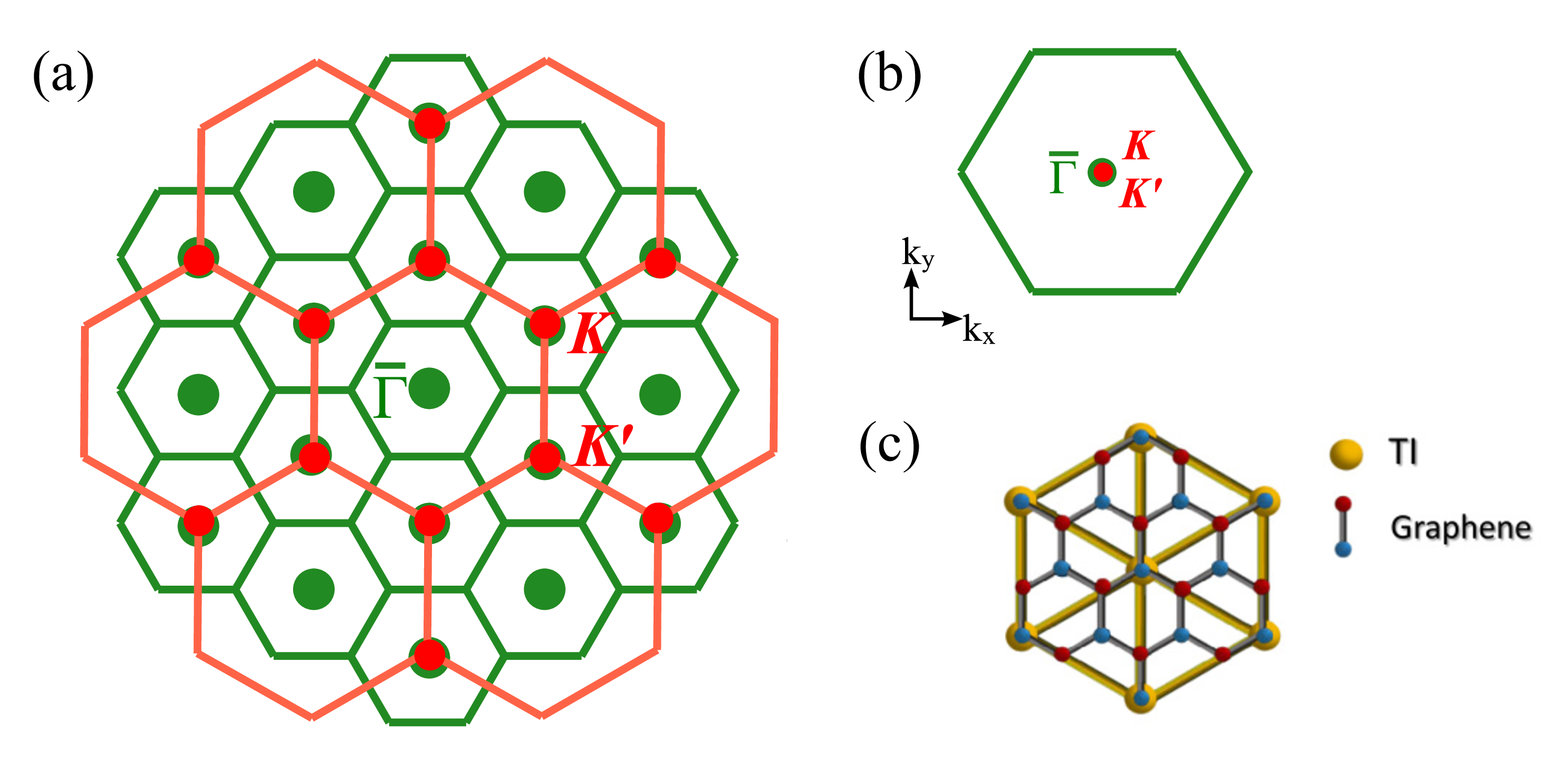}
  \caption{(a) Schematic of the first seven extended Brillouin zones of graphene, shown in red, superimposed on the extended Brillouin zones of a TI, shown in green, for the case of $\sqrt{3}\times\sqrt{3}$ commensurate stacking. (b) Depiction of the folded Brillouin zone for the same stacking as in (a). (c) The real space picture associated with the $\sqrt{3}\times\sqrt{3}$ graphene-TI commensurate stacking. Adapted with permission from Phys. Rev. Lett. 112, 096802 (2014).
         } 
  \label{fig:BZ-nesting}
 \end{center}
\end{figure}

For our purposes, a particularly interesting class of 3D TIs is the one of the tetradymites, such as $\mathrm{Bi_{2}Se_{3}}$, $\mathrm{Bi_{2}Te_{3}}$, and $\mathrm{Sb_{2}Te_{3}}$. These 3D TIs have the exceptional property that 
the lattice constant of the 111 surface, $a_{TI}$, is such that: 
$a_{TI}/(\sqrt{3}a_{g})=1+\delta$ with $\delta<1\%$ for $\mathrm{Sb_{2}Te_{3}}$ and $\delta\approx-3\%$ ($\delta\approx+3\%$) for $\mathrm{Bi_{2}Se_{3}}$ ($\mathrm{Bi_{2}Te_{3}}$)~\cite{jzhang2014}. As a consequence, in the limit $\delta\to 0$, graphene and the TI surface can be stacked in a commensurate arrangement as shown, in momentum space, in Fig.~\ref{fig:BZ-nesting}~(a). For such a stacking arrangement the corners of graphene's BZ are precisely above the $\bar\Gamma$ points of the TI surface BZ. For this stacking the primitive cell of the heterostructure corresponds to the primitive cell of the TI's surface and therefore the BZ of the resulting vdW system is equal to the BZ of the TI's surface. The $\KK$ and $\KK'$ points of the graphene BZ are folded back to the $\bar\Gamma$ point, as shown in Fig.~\ref{fig:BZ-nesting}~(b).

In the remainder of this section we limit our discussion to the cases in which the TI is $\mathrm{Bi_{2}Se_{3}}$, $\mathrm{Bi_{2}Te_{3}}$, or $\mathrm{Sb_{2}Te_{3}}$, for which $v_{TI}\approx v_g/2$, given that these materials allow the formation of a graphene-TI vdW system with commensurate stacking and therefore significant hybridization between the graphene and TI surface states.
Experimentally it turns out to be difficult to pin the Fermi energy of $\mathrm{Bi_{2}Se_{3}}$, $\mathrm{Bi_{2}Te_{3}}$, $\mathrm{Sb_{2}Te_{3}}$ in the middle of the bulk gap so that only the surface states play an active role. One way in which this problem has been overcome is by considering the corresponding ternary and quaternary compounds~\cite{ren2010, arakane2012, xia2013, segawa2012, xu2014d, durand2016, xu2016}.

Different commensurate stacking configurations, corresponding to Fig.~\ref{fig:BZ-nesting},
can be realized by rigid relative shifts of the graphene and TI lattices.
Ab-initio results~\cite{jin2013} suggest that the lowest energy stacking is the one for which the TI surface atoms are located at the center of the hexagons forming the graphene structure. However, the binding energy for the structure in which the TI surface atoms are directly below the carbon atoms of graphene (either the ones forming the A sublattice, or the ones forming the B sublattice) is only marginally higher~\cite{jin2013}.
Considering that, experimentally, vdW systems are obtained via mechanical exfoliation that allows the realization of, long-lived, metastable states, and the fact that when the TI's surface atoms are directly below the carbon atoms of graphene a stronger hybridization of the graphene and TI states is realized, it is interesting to consider this situation.

In the $\sqrt{3}\times\sqrt{3}$ commensurate stacking, in which each atom on the TI surface is directly underneath a carbon atom, the dominant interlayer tunneling term is the one between the atoms on sublattice A (or B) and the TI atoms so that, in momentum space, the tunneling Hamiltonian can be written as
$H_{t}=\sum_{\mathbf{k},\lambda,\tau,\sigma}t_{\tau}d_{\mathbf{k},\sigma}^{\dagger}c_{\lambda,\mathbf{k},\tau,\sigma}+h.c.$,
where $\lambda=K,K'$ and $t_{A}=t,\ t_{B}=0$ are the tunneling matrix elements assumed to be spin and momentum independent. The Hamiltonian matrix for such a structure takes the form
\begin{equation}
\hat{H}_{\mathbf{k}}=\left(\begin{array}{ccc}
\hat{h}_{\mathbf{k}}^{\text{g},K} & 0 & \hat{t}^{\dagger}\\
0 & \hat{h}_{\mathbf{k}}^{\text{g},K'} & \hat{t}^{\dagger}\\
\hat{t} & \hat{t} & \hat{h}_{\mathbf{k}}^{\text{TIS}}
\end{array}\right),\ \ \hat{t}=\left(\begin{array}{cccc}
t & 0 & 0 & 0\\
0 & 0 & t & 0
\end{array}\right),
\label{eq:H01}
\end{equation}
where the graphene blocks are $4\times4$ matrices in sublattice and spin space and the block describing the TI's surface states is a $2\times 2$ matrix in spin space. We note that an analogous Hamiltonian can also be constructed for the similar vdW heterostructure in which single layer graphene (SLG) is replaced by bilayer graphene (BLG) to form a BLG-TI system\cite{jzhang2014}.

\begin{figure}[htb]
 \begin{center}
  \includegraphics[width=1.0\columnwidth]{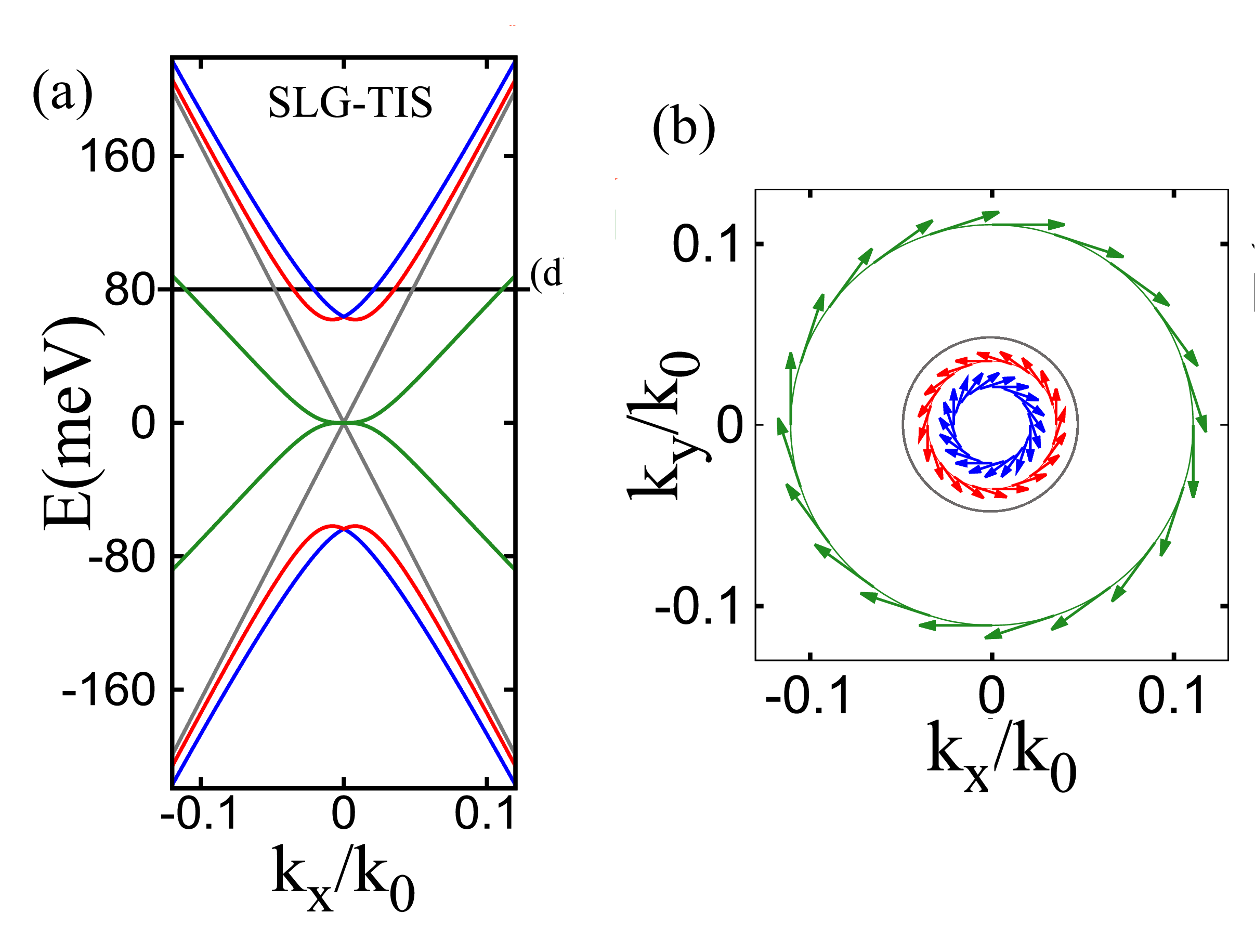}
  \caption{(a) Band structure for the $\sqrt{3}\times\sqrt{3}$ commensurate graphene-TI structure described by Eq.~\ceq{eq:H01}, assuming $\mu_g=\mu_{TI}=0$ and $t=45$~meV. (b) The in-plane spin projection of the eigenstates associated with the bands in (a), evaluated at the energy $E=80$meV. Adapted with permission from Phys. Rev. Lett. 112, 096802 (2014).
         } 
  \label{fig:bands01}
 \end{center}
\end{figure}

The simple Hamiltonian in Eq.~\ceq{eq:H01} allows us to understand the qualitative features of the bands resulting from the hybridization between the graphene states and the states of the TI surface.
Fig.~\ref{fig:bands01}~(a) shows the bands obtained by diagonalizing $\hat{H}_{\mathbf{k}}$ assuming $\mu_g=\mu_{TI}=0$ and $t=45$~meV.
From Fig.~\ref{fig:bands01}~(a), we see that the 4-fold degeneracy of the graphene states (spin and valley degrees of freedom) is partially lifted as two spin-split Rashba bands appear (shown in red and blue). However, we also see that two of the original graphene states at the $\KK$ and $\KK'$ points remain spin degenerate (shown in grey), due to the fact that in the chosen configuration one sublattice does not couple to the TI. Moreover, the TI surface bands (shown in green) become quadratic at low energy as a consequence of hybridization with the graphene states.

\begin{figure}[htb]
 \begin{center}
  \includegraphics[width=1.0\columnwidth]{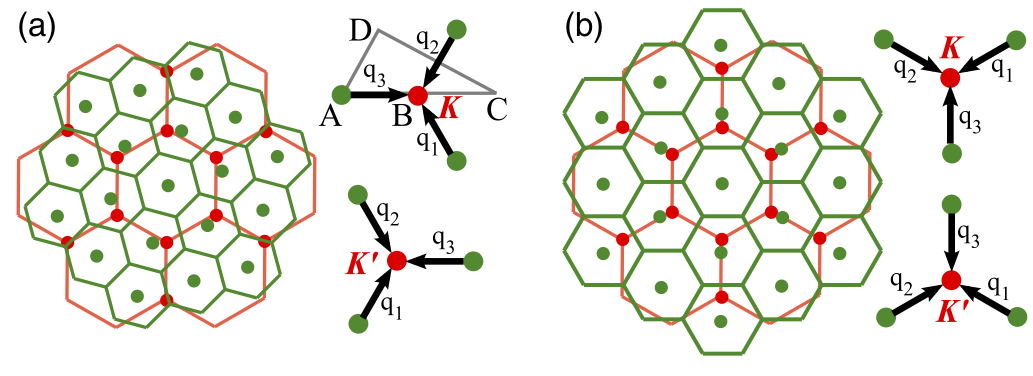}
  \caption{(a) Schematic of the first seven extended Brillouin zones of graphene (red) superimposed on the extended Brillouin zones of a TI (green) with a slight rotation away from precise $\sqrt{3}\times\sqrt{3}$ commensurate stacking. Also shown the vectors $\textbf{q}_i$ describing the displacement of the corners of the first Brillouin zone of graphene from the nearest $\bar{\Gamma}$ points in the TI surface. 
(b) Similar schematic to (a) for the case when the deviation from the $\sqrt{3}\times\sqrt{3}$ commensurate stacking is caused by a lattice mismatch. 
Also shown the vectors $\textbf{q}_i$ describing the displacement of the corners of the first Brillouin zone of graphene from the nearest $\bar{\Gamma}$ points in the TI surface for this case. Adapted with permission from Phys. Rev. Lett. 112, 096802 (2014).
         } 
  \label{fig:BZ02}
 \end{center}
\end{figure}

We now discuss the case of stacking configurations that deviate from the $\sqrt{3}\times\sqrt{3}$ configurations discussed above either because of a small twist angle $\theta$ or because of a mismatch of the graphene and TI lattice constants.
Figs.~\ref{fig:BZ02}~(a) and (b) show how the orientations of the TI and graphene BZs are affected by the presence of a twist angle and a lattice mismatch, respectively. We see that, due to the conservation of the crystal momentum, the states at the $\KK$ ($\KK'$) point of graphene now tunnel to the TI surface states with momentum $\qq_i$ ($i=1,2,3$). For the twisted case, the magnitude of this vector is $|\mathbf{q}_{j}|\equiv q=2K\sin(\theta/2)$, while for the case of a lattice mismatch we have $q=\left|\delta/(1+\delta)\right|K$.

\begin{figure}[htb]
 \begin{center}
  \includegraphics[width=1.0\columnwidth]{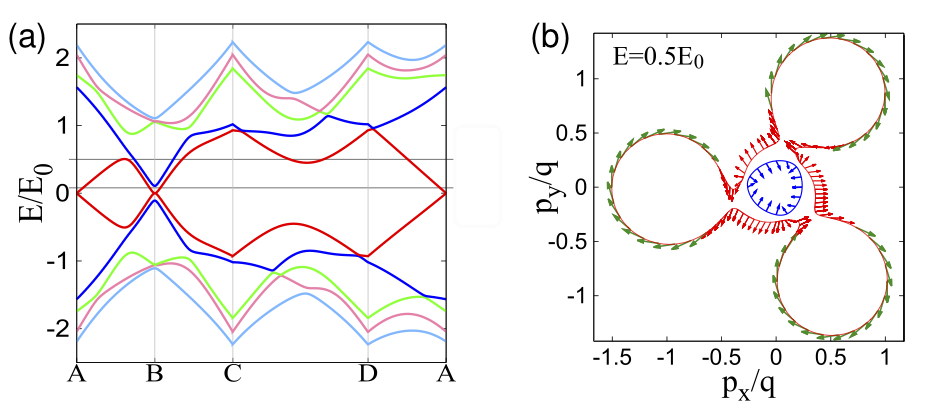}
  \caption{(a) Band structure for a twisted graphene-TI system for the case when $\gamma=0.2$, calculated by diagonalizing the Hamiltonian in Eq.~\ceq{eq:H02} along the ABCDA path indicated in Fig.~\ref{fig:BZ02}. (b) The in-plane spin projection of the eigenstates associated with the bands in (a), evaluated at the energy $E=0.5E_0$ where $E_0$ is indicated in (a). Adapted with permission from Phys. Rev. Lett. 112, 096802 (2014).
         } 
  \label{fig:bands03}
 \end{center}
\end{figure}

Assuming $\gamma<<1$, we can use the simple Hamiltonian in Eq.~\ceq{eq:H02}, with $\hat h_1=\hat h^{g,K}$, $\hat h_2=h^{\rm TIS}$ and 
\begin{align}
\hat{t}_{1} & =\left(\begin{array}{cccc}
t' & t' & 0 & 0\\
0 & 0 & t' & t'
\end{array}\right),\ \hat{t}_{2}=\left(\begin{array}{cccc}
t' & t'e^{-i\frac{2\pi}{3}} & 0 & 0\\
0 & 0 & t' & t'e^{-i\frac{2\pi}{3}}
\end{array}\right),\nonumber \\
\hat{t}_{3} & =\left(\begin{array}{cccc}
t' & t'e^{i\frac{2\pi}{3}} & 0 & 0\\
0 & 0 & t' & t'e^{i\frac{2\pi}{3}}
\end{array}\right).\nonumber 
\end{align}
with $t'=t/3$. A similar Hamiltonian is valid for the $K'$-valley \cite{jzhang2014}. By diagonalizing the resulting Hamiltonian we obtain the low energy band structure.
Figure~\ref{fig:bands03}~(a) shows the bands   along the path ABCDA shown in Fig.~\ref{fig:BZ02}~(a) for the case when $\gamma=0.2$.
Assuming $t=45$~meV, this value of $\gamma$ corresponds to a deviation from the $\sqrt{3}\times\sqrt{3}$ stacking by a twist angle $\theta=0.76\degree$.
Fig.~\ref{fig:bands03}~(b) shows the spin texture on the Fermi surface with $\eps_F=v_{TI}{q}/2$. Similar to the $\sqrt{3}\times\sqrt{3}$ commensurate case shown in Fig.~\ref{fig:bands01}, we see that the strong SOC of the TI induces a strong spin polarization of the states of the hybridized system even when the stacking configuration deviates from the ideal $\sqrt{3}\times\sqrt{3}$ one.

%


\subsection{Transport properties}

\begin{figure}[htb]
 \begin{center}
  \includegraphics[width=1.0\columnwidth]{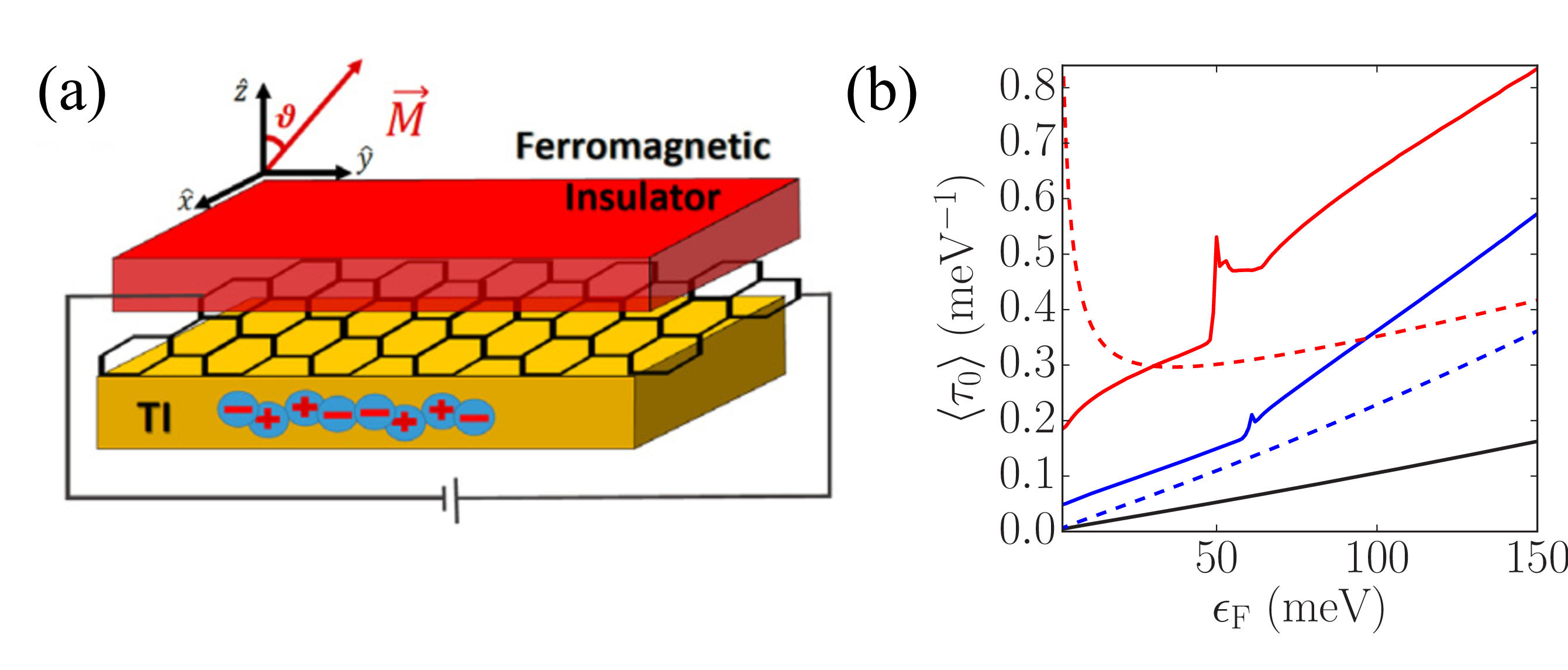}
  \caption{(a) Sketch of a TI-graphene-FM vdW system.
           (b) $\langle\tau_{0}(\epsilon_F)\rangle$
		for $\Delta=0$, $\delta \mu=0$, $n_{\rm imp}=10^{12}$ cm$^{-2}$, and $d=1$~nm. The dashed lines
		show the results for the limit $t=0$, the solid one the ones for  $t=45$~meV.
           Adapted with permission from Phys. Rev. B {\bf 96}, 235419 (2017).
         } 
  \label{fig:Gr-TI-FM}
 \end{center}
\end{figure}

The intrinsic SOC in graphene and bilayer graphene is negligible so that the spin degree of freedom does not affect
significantly the transport properties~\cite{dassarma2011,nomura2006,adam2007,rossi2009,dassarma2010,li2011,rossi2011,rossi2012,LiQiuzi2012}.
In the presence of SOC, charge and spin transport become coupled~\cite{burkov2003,burkov2010,sinova2015}. Therefore, in vdW systems with SOC we expect that, in general, transport properties will be spin-dependent and that such properties could differ considerably from isolated systems with SOC.
Exemplary in this respect are the vdW systems formed by a layer with strong SOC, such as a TMD monolayer, and either graphene or bilayer graphene. In graphene-TI vdW systems the combination of the high mobility of graphene, the strong SOC of the TI, and the increased 
screening~\cite{hwang2006b,shklovskii2007,borghi2009,abergel2012,triola2012} of impurities due to the presence of the graphene layer~\cite{lu2016}, can lead to very strong spin-dependent effects, such as a giant Edelstein effect~\cite{rodriguez2017}.
To illustrate the potential for realizing spin-dependent transport effects in vdW systems in which one layer has a strong SOC we discuss the case of a graphene-TI-Ferromagnet vdW system~\cite{rodriguez2017}, shown schematically in Fig.~\ref{fig:Gr-TI-FM}~(a). 

Throughout this section, we assume that the ferromagnet (FM) is insulating so that its presence does not significantly alter the bands of the graphene-TI part of the vdW heterostructure. The main effect of the insulating FM layer is to induce an exchange field for the electrons in the heterostructure.
This effect is captured by adding the term $H_{ex}\propto \MM\cdot\ssigma$, where $\MM$ is  the magnetization of the insulating FM,
to the Hamiltonian discussed in Sec.~\ref{sec:g-ti}.
The presence of this additional term simply causes a spin-splitting that we denote by $\Delta$.

In a graphene-TI-FM vdW system the dominant source of scattering, at low temperatures, is the presence of 
charge impurities~\cite{rossi2008}
close to the surface of the TI~\cite{culcer2010,li2012b,adam2012,skinner2012,skinner2013}. 
In the absence of screening, the bare scattering potential due to an isolated charge impurity, in momentum space, is $v(q)=2\pi e^2 e^{-q d}/(\kappa q)$, where $d$ is the average distance of the impurities from the TI's surface, and $\kappa = (\kappa_{TI} +\kappa_0)/2 $ is the average dielectric constant with $\kappa_{TI}\approx 100$~\cite{butch2010,beidenkopf2011,kim2012b,li2012b} the dielectric constant for the TIS and $\kappa_0=1$ the dielectric constant of vacuum.
Accounting for screening, the scattering potential becomes $v(q)/\epsilon(q)$ where $\epsilon(q)$ is the dielectric function. For temperatures much lower than the Fermi temperature we can assume $\epsilon(q)\approx 1+v_c(q)\nu(\epsilon_F)$, where $v_c(q)=2\pi e^2/(\kappa q)$ and $\nu(\epsilon_F)$ is the density of states at the Fermi energy.
Using this form for the scattering potential we can calculate the lifetime $\tau_{0a}(\kk)$ of a quasiparticle in band $a$ with momentum $\kk$, in the first Born approximation, using 
\beq
\frac{\hbar}{\tau_{0a}(\kk)} = 2\pi \sum_{a' \qq} n_{\rm imp} \left|  \frac{v(q)}{\epsilon(q)} \right| ^2 \lvert \langle a'  \kk+\qq | a \kk  \rangle \rvert ^2 \delta(\epsilon_{a, \kk}-\epsilon_{a', \kk+\qq}),
\label{eq:singleparttime}
\enq
where $n_{\rm imp}$  is the impurity density, $| a \kk  \rangle$ is the Bloch state with momentum $\kk$ and band index $a$, and $\epsilon_{a, \kk}$ is the energy for a quasiparticle with momentum $\kk$ in band $a$. For typical TI's samples we have $n_{\rm imp}\approx 10^{12}{\rm cm}^{-2}$ \cite{kim2012b}. It is useful to define an average $\langle\tau_0\rangle$ of $\tau_0$ over the bands at the Fermi energy as $\langle \tau_{0}(\epsilon_F)\rangle\df \sum_{\kk a}\tau_{0a}(\kk)\delta(\epsilon_F-\epsilon_{\kk a})/\sum_{\kk a}\delta(\epsilon_F-\epsilon_{\kk a})$. We allow for an offset between the charge neutrality point of the SLG (BLG) and the TI surface given by $\delta\mu$.

\begin{figure}[htb]
 \begin{center}
  \includegraphics[width=1.0\columnwidth]{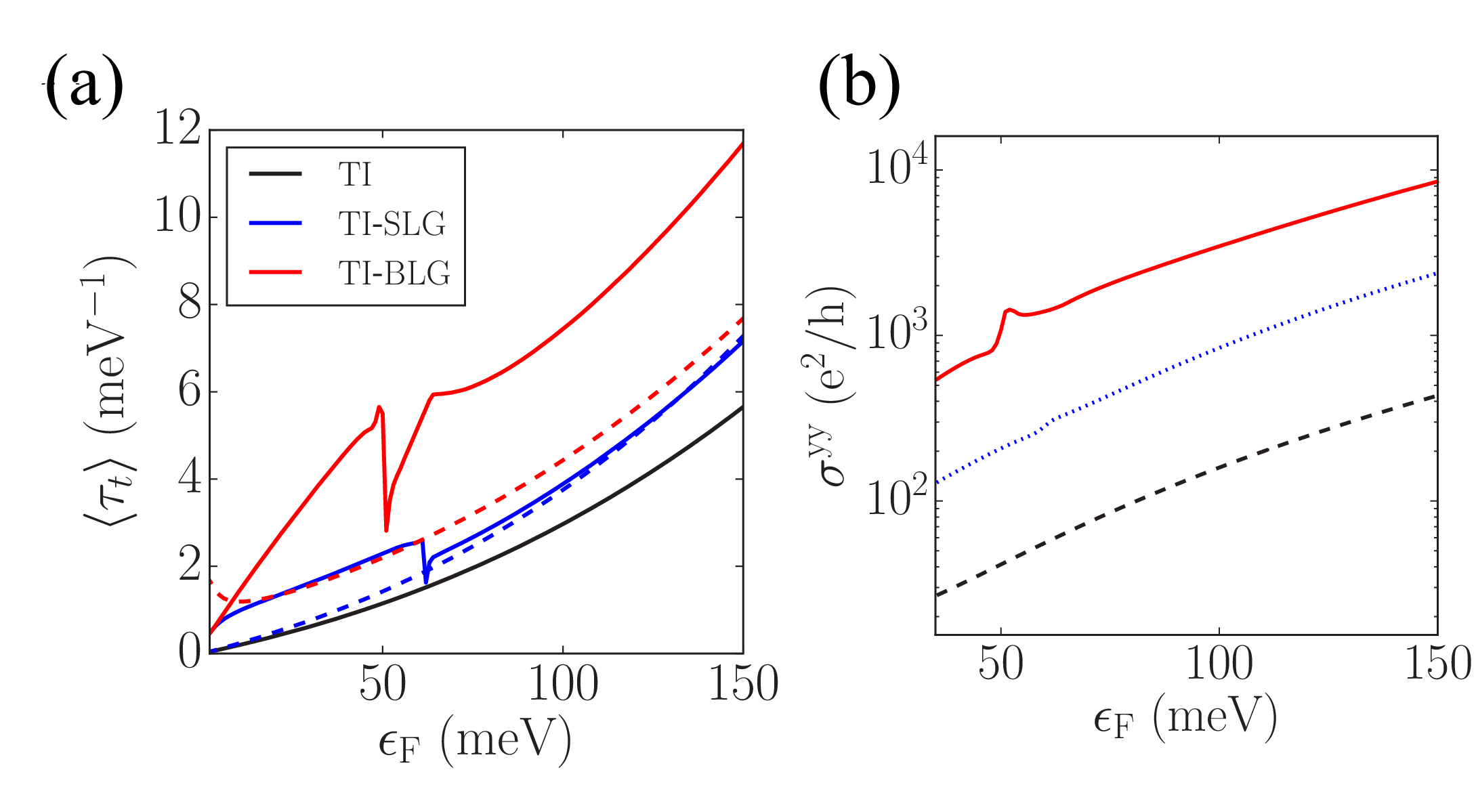}
  \caption{(a)  $\langle\tau_{t}(\epsilon_F)\rangle$
		for $\Delta=0$, $\delta \mu=0$, and $n_{\rm imp}=10^{12}$ cm$^{-2}$. The dashed lines
		show the results for the limit $t=0$, the solid one the ones for  $t=45$~meV.
            (b) $\sigma^{yy}(\epsilon_F)$, for TI (dashed line), TI-SLG (dotted line), and TI-BLG (solid line)
		for $\Delta=0$,  $\delta \mu=0$, $n_{\rm imp}=10^{12}$ cm$^{-2}$, $d=1$~nm, and
		$t=45$~meV, and 
          Adapted with permission from Phys. Rev. B {\bf 96}, 235419 (2017).
         } 
  \label{fig:Gr-TI-FM-02}
 \end{center}
\end{figure}

Figure~\ref{fig:Gr-TI-FM}~(b) shows how $\langle\tau_0(\eps_F)\rangle$ compares for: an isolated TI surface, a SLG-TI vdW system, and BLG-TI vdW system. Solid lines denote the cases when the tunneling between the SLG/BLG and the TI is finite, while dashed lines denote the cases with zero interlayer coupling.
We see that the presence of SLG or BLG significantly increases the quasiparticle lifetime, even in the limit when the interlayer tunneling between the TI and SLG (BLG) is zero. This is due to the additional screening in the presence of SLG or BLG which affects the disorder potential created by the charge impurities~\cite{rodriguez2014,lu2016}.
We expect that such an increase in the quasiparticle lifetime will lead to enhancements in some of the spin-dependent transport phenomena.
Similarly, by including the $[1-\kk\cdot(\kk+\qq)]$ under the sum on the right hand side of Eq.~\ceq{eq:singleparttime} we can obtain the transport time $\tau_{ta}(\kk)$, and then, after averaging over the bands at the Fermi energy, the corresponding average $\langle\tau_t\rangle$. 
Figure~\ref{fig:Gr-TI-FM-02}~(a) shows $\langle\tau_t\rangle$ as a function of $\eps_F$ for an isolated TI surface together with results for, both, TI-SLG, and TI-BLG vdW systems.
We see that the presence of SLG or BLG increases the transport time, a consequence of the additional screening due to the graphenic layer. Moreover, we note that this enhancement is particularly significant for the case of TI-BLG.
Using linear-response theory in the long-wavelength regime, the d.c. longitudinal conductivity is given by 
\begin{align} 
\sigma^{ii} & \approx   \frac{e^2}{2 \pi \Omega} \rem \sum_{\kk, a}  v^i_{aa}(\kk)\tilde{v}^i_{aa}(\kk) G^A_{\kk a}G^R_{\kk a}\;.
\label{eqn:conductivity}
\end{align}
where $\Omega$ is the area of the BZ, $v^i_{aa}(\kk)\df \langle a\kk|v_i|a\kk\rangle$ is the expectation value of the {\em i}-th component of the velocity operator ${\bf v}=\hbar^{-1} \partial H_{\kk}/\partial \kk$, $\tilde{v}^i_{aa}(\kk) = (\tau_a/\tau_{0a})_\kk v^i_{aa}(\kk)$ is the disorder-renormalized velocity (at the ladder approximation level), and $G^{R/A}_{\kk a}=(\epsilon_F-\epsilon_{\kk a}\pm i\hbar/2\tau_{0a}(\kk))^{-1}$ is the retarded/advanced Green's function, for electrons with momentum $\kk$ and band index $a$.
The increase of $\tau_t$ due to the presence of SLG or BLG is also reflected in an increase of the d.c. conductivity, as can be seen in Fig.~\ref{fig:Gr-TI-FM-02}~(b). We note that for finite values of the spin-splitting, $\Delta$, the results for $\tau_0$, $\tau_t$, and $\sigma^{ii}$ are very similar to the ones shown in Figs. \ref{fig:Gr-TI-FM}~(b),~\ref{fig:Gr-TI-FM-02}~\cite{rodriguez2017}. 

\begin{figure}[htb]
 \begin{center}
  \includegraphics[width=1.0\columnwidth]{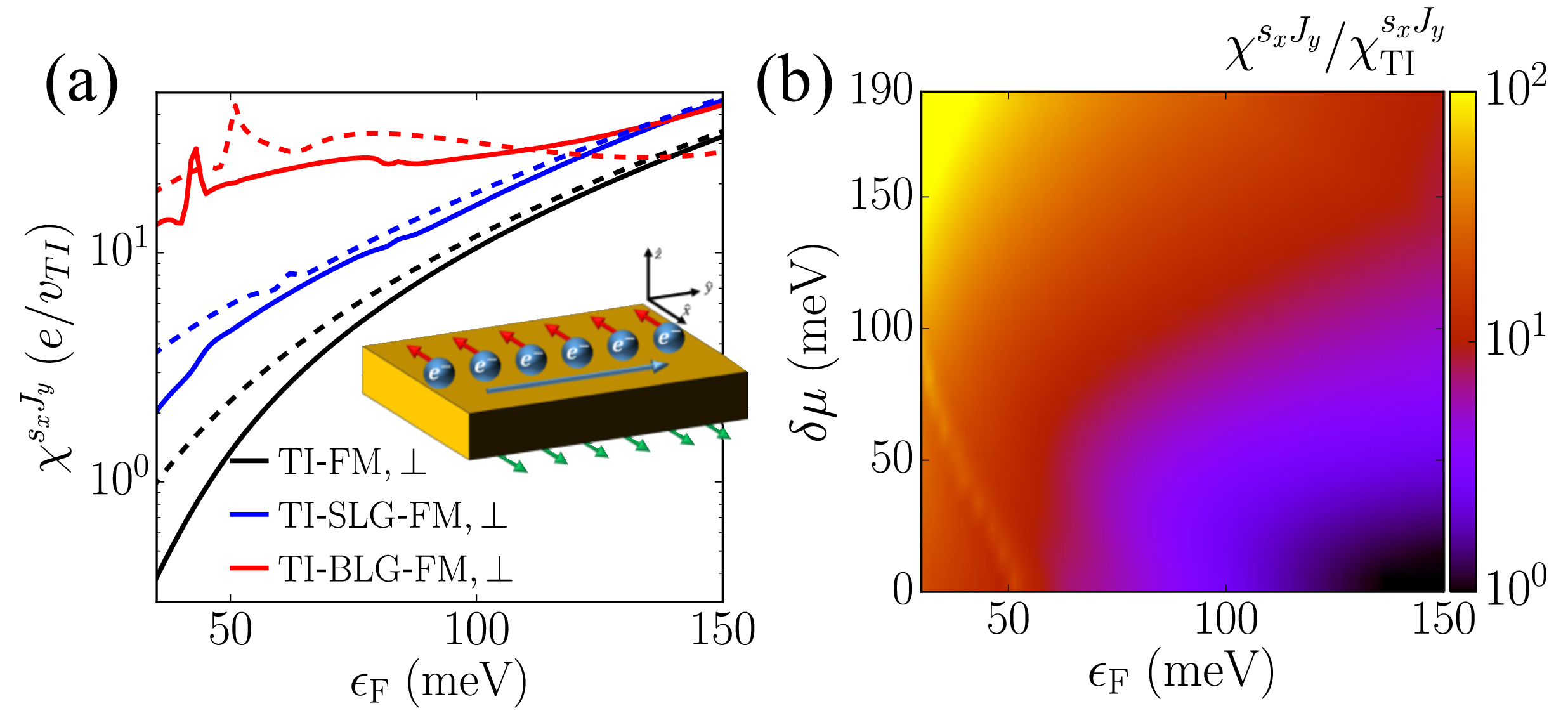}
  \caption{	(a) $\chi^{s_xJ_y}$ as a  function of $\epsilon_F$ for $\delta\mu=0$ and $\Delta=20$~meV ($\Delta=0$), solid (dashed) lines.
                Inset: sketch showing the spin density accumulation on the top and bottom surface of the TI
		induced by a current in the $y$ direction.
		(b) Enhancement of $\chi^{s_xJ_y}$ in a TI-BLG system compared to TI alone as a function of $\epsilon_F$ and $\delta\mu$
		for $\Delta=0$.
		For both the panels
		$n_{\rm imp}=10^{12}$ cm$^{-2}$, $d=1$~nm.
           Adapted with permission from Phys. Rev. B {\bf 96}, 235419 (2017).
         } 
  \label{fig:Gr-TI-FM-03}
 \end{center}
\end{figure}

Having established the basic charge transport properties we now briefly discuss spin-dependent transport. A signature effect of the coupling between charge and spin transport that can take place in systems with spin-orbit coupling is the inverse 
Edelstein effect~\cite{dyakonov1971,edelstein1990}. 
In this effect a charge current causes a spin accumulation transverse to the direction of the current. In the long wavelength, dc, limit, of the linear response regime, such an effect is encoded by the spin-current response function:
\begin{align} 
\chi^{s_x J_y} & \approx   \frac{e}{2 \pi \Omega} \rem \sum_{\kk, a}  s^x_{aa}(\kk)\tilde{v}^y_{aa}(\kk) G^A_{\kk a}G^R_{\kk a}\;,
\end{align}
where $s^i_{aa}(\kk)\df \langle a\kk|s_i|a\kk\rangle$ is the expectation value of the {\em i}-th component of the spin density operator.
Figure~\ref{fig:Gr-TI-FM-03}~(a) displays a significant enhancement of the spin-current response $\chi^{s_x J_y}$ in SLG-TI (BLG-TI) vdW system, compared to an isolated TI's surface. This enhancement is due to the presence of the graphenic layer and occurs for, both, the case of finite (solid lines) and zero $\Delta$ (dashed lines).
As the sketch in the inset shows, a current in the $y$-direction causes a spin accumulation in the transverse direction $x$. An increase of $\delta\mu$ can significantly enhance the spin-current response, as shown in Fig.~\ref{fig:Gr-TI-FM-03}~(b), whereas $\Delta$ has a small effect~\cite{rodriguez2017}.

The fact that the Edelstein effect can be much stronger in vdW systems like TI-SLG and TI-BLG than in isolated TIs~\cite{garate2010, yokoyama2010, sakai2014, fischer2013,mellnik2014b,fan2016} is due to the fact that the bands for the vdW system retain a strong SOC, even after the hybridization of the states in the TI and the graphenic layer, and the fact that the quasiparticle relaxation time can be greatly increased when the dominant source of scattering is charge impurities, thanks to the additional screening provided by the graphenic layer. 
As mentioned above, the enhancement of the lifetime gained from the additional screening makes it possible for vdW systems like TI-BLG to have quite stronger spin-charge coupled transport than isolated systems like TIs even when the tunneling between the two layers of the vdW system is negligible.

So far we have considered the case in which the interlayer tunneling is constant,
however, spin and charge transport can be coupled in vdW systems such as graphene-TI heterostructures
even in the limit when the interlayer tunneling is predominantly random~\cite{rodriguez2019}. 
In this case the evolution, in the diffusive limit,  of the graphene's charge density, $n$,  
and spin density, {\bf s}, are given by the following equations~\cite{rodriguez2019}:
\begin{align} 
\partial_t   n   =   & \bar{{D}} \nabla^2 (n+2\rho V) 
                            +\Gamma_{ns}l_{\rm TI}(\hat{z}\times 
                            \boldsymbol{\nabla})({\bf s}-2\rho{\bf h}) \nonumber \\
                            &-2\frac{\rho \Gamma^t}{\Gamma^t+\Gamma^t_{\rm TI}}\partial_t(V-V_{\rm TI})
                            \label{eq:diff_1} \\ 
\partial_t {\bf s} = & \left({D}\nabla^2-\Gamma^t\right)({\bf s}-2\rho{\bf h})
                            +\alpha\Gamma^t l_{\rm TI}(\hat{z}\times \boldsymbol{\nabla})\times ({\bf s}-2\rho{\bf h})\nonumber\\
                            &+\Gamma^tl_{\rm TI}(\hat{z}\times \boldsymbol{\nabla})
                                \left[l_{\rm TI}(\boldsymbol{\nabla}\times({\bf s}-2\rho{\bf h}) )_z +(n+2\rho V)/2\right]
	 \label{eq:diff_2} 
\end{align}
where $V$, $V_{\rm TI}$ are external driving potentials for the charge in the graphene layer and TI's surface, respectively,
$\rho$ is graphene's density of states at the Fermi energy, and {\bf h} is an external driving potential for the spin in the graphene layer.
$\bar{D}$ is the  weighted average of the diffusion constants of graphene, $D$, and TI's surface, $D_{\rm TI}$
$$
   \bar{{D}}  = \frac{\Gamma^tD_{\rm TI}+\Gamma^t_{\rm TI}D}{\Gamma^t+\Gamma^t_{\rm TI}}      
$$
where $\Gamma^t$, $\Gamma^t_{\rm TI}$, are the tunneling rates for graphene and the TI, respectively:
$\Gamma^t=\pi\rho_{\rm TI}t^2$, $\Gamma^t_{\rm TI}=\pi\rho t^2$, with $\rho_{TI}$ the density of states of the TI's surface
at the Fermi energy. The interlayer tunneling processes
are assumed to be well localized in space so that in momentum space 
the disorder average of the second moment of the interlayer 
tunneling matrix is just a constant, $t^2$.
The second term on the right hand side (r.h.s.) of Eq.~\ceq{eq:diff_1} describes the coupling between charge and spin transport with
$$
\Gamma_{ns}=2\frac{\Gamma^t\Gamma^t_{\rm TI}}{\Gamma^t+\Gamma^t_{\rm TI}}
$$ 
and $l_{\rm TI}$ the electron's mean free path on the TI's surface.
The last term on the r.h.s. of Eq.~\ceq{eq:diff_1} describes the effect of time-dependent driving potentials for the charge.
The second term on the r.h.s. of Eq.~\ceq{eq:diff_2} is due to the spin-orbit coupling term that is induced by proximity in the layer with
no spin-orbit coupling even when the interlayer tunneling is predominantly random. The coefficient of this term is 
$$
 \alpha=\frac{{\eps_F\tau^0}}{2\pi^2\rho_{\rm TI}D_{\rm TI}}
$$
where $\eps_F$ is the Fermi energy and $\tau^0$ is the quasiparticle relaxation time in the graphene layer due to intralayer disorder.

Equations~\ceq{eq:diff_1},~\ceq{eq:diff_2} are valid when the interlayer tunneling rate is much smaller than the intralayer scattering rate,
and for time scales much longer than the largest relaxation time $\tau$ ($\omega\tau\ll 1$).
They show that even when the interlayer tunneling is random, charge and spin transport in the layer with no SOC are coupled.
We can then conclude that even for vdW systems in which, due to the low quality of the interfaces, the interlayer
tunneling is random, spintronics effects, such as the Edelstein effect, can be realized~\cite{rodriguez2019}.
The fact that $\Gamma_{ns}$ depends on the intralayer and interlayer scattering rates, that in turn are directly proportional
to the density of states at the Fermi energy, implies that in systems in which for at least one of the layers the density of states depends on the doping
--such as for the surface of a TI or graphene-- it is possible to tune the coupling between spin and charge transport simply via external gate voltages.


\section{Graphene-TMD heterostructures}
\label{sec:g-tmd}
In recent years, much progress has been made on the characterization of the electronic properties of graphene-TMD heterostructures, both theoretically and experimentally.
A monolayer TMD can be either metallic, such as \nbse, or a direct gap semiconductor such as \mos, and \wse. At low temperatures \nbse becomes superconducting and so we defer discussion of graphene-\nbse vdW systems to the following section, see in particular Sec.~\ref{sec:ising_SC}.
In this section we briefly summarize the main results for graphene-TMD heterostructures in which the TMD is a semiconductor.

\begin{figure}
 \begin{center}
  \centering
  \includegraphics[width=1.0\columnwidth]{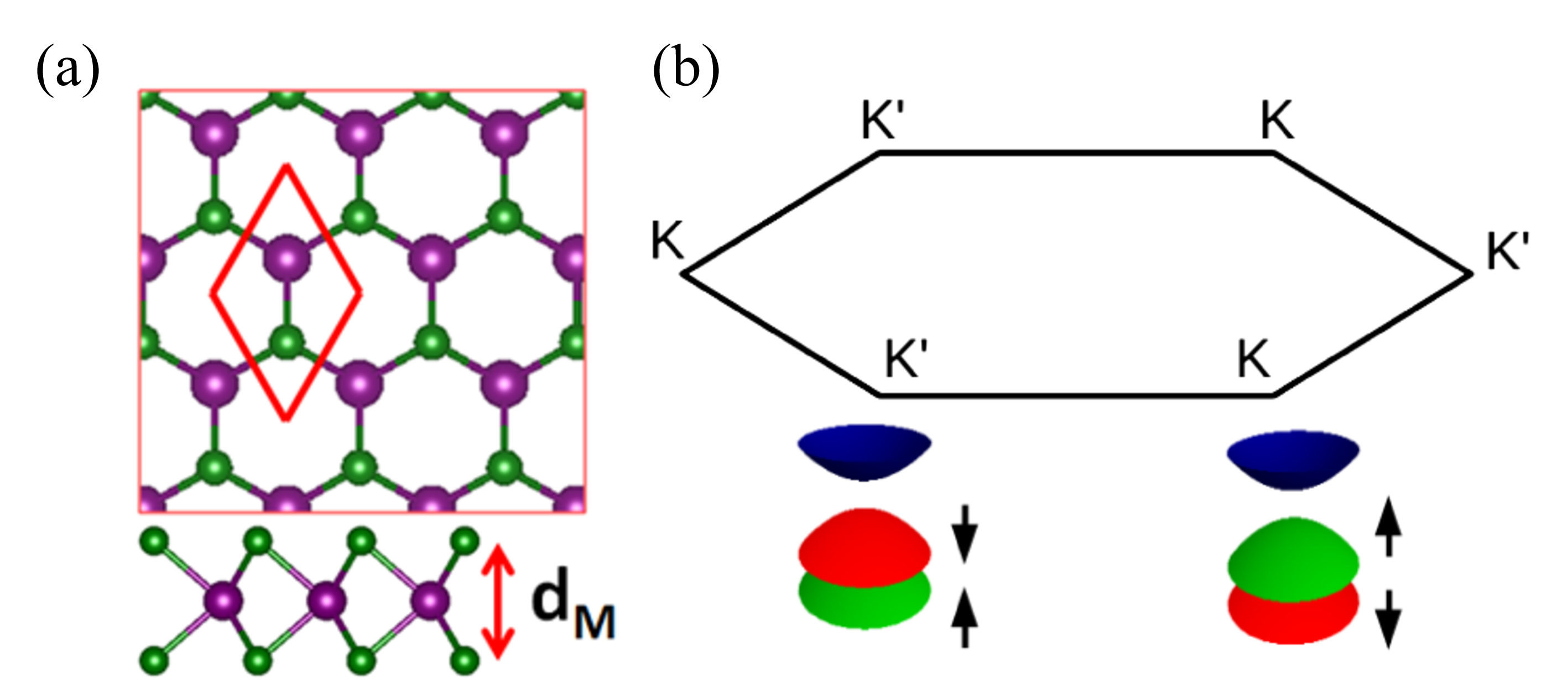}
  \caption{(a) Sketch of the typical lattice of a transition metal dichalcogenide monolayer, with purple spheres representing the metallic atoms and green spheres the chalcogen atoms. (b) Schematic of the first Brillouin zone of \mos with the low energy bands at the $K$ and $K'$ points. Spin-polarization of the valence bands denoted with arrows. Adapted with permission from Phys. Rev. Lett. 116, 257001 (2016).
         }
  \label{fig:TMD-SC}
 \end{center}
\end{figure}

The structure of a TMD monolayer is shown schematically in Fig.~\ref{fig:TMD-SC}~(a) where the purple spheres represent the metallic atoms, such as Mo in \mos, and the green spheres the chalcogen atoms, S in \mos. One of the main features of TMDs is the presence of strong SOC, which, in monolayers, induces a sizeable spin splitting of the hole bands~\cite{xiao2012,ashwin2012,strano2012,xiaodong2014,latzke2015} 
located at the corners of the BZ, as shown in Fig.~\ref{fig:TMD-SC}~(b). 
At low energies, the bands of a TMD semiconductor monolayer such as \mos, are well described by the following Hamiltonian \cite{xiao2012}:
\begin{align}
 \hat{h}^{TMD}_{\textbf{k}}=&a \gamma\left(\lambda k_x \hat\kappa_1 + k_y \hat\kappa_2 \right) + \dfrac{u}{2} \hat\kappa_3 - \mu \hat\kappa_0 \nonumber \\
 &-\dfrac{\lambda\alpha}{2}\left( \hat\kappa_3 - \hat\kappa_0 \right)\otimes \hat\sigma_3, 
 \label{eq:HTMD}
\end{align}
where $a$ is the in-plane lattice constant, $\gamma$ the in-plane hopping amplitude, $\lambda=\pm 1$ is the index denoting the valley ($\KK$ or $\KK')$, $k_x$ $k_y$ are the in-plane components of the electrons wave vector, measured from the corner of the BZ, $\hat\kappa_i$ are $2\time 2$ Pauli matrices in the orbital space~\cite{xiao2012}, $u$ is the band-gap, $\mu$ the chemical potential and $2\alpha$  is the spin-splitting of the valence bands at the $\KK$ ($\KK'$) point due to the presence of spin-orbit coupling. In the case of \mos these parameters are $a=3.193\text{ \AA}$, $\gamma\approx 1.1$~eV, $u\approx 1.65\text{ eV}$, and $2\alpha\approx 0.15\text{ eV}$~\cite{xiao2012}.

Given that the semiconducting monolayer TMDs have a fairly large gap, $\sim 2$~eV, and a large lattice mismatch with graphene, one would expect that their effect on the graphene band structure could be negligible.
Indeed, it is the case that the effects induced by proximity of the TMD to the SLG (or BLG) are quantitatively small. However, the presence of these effects can be qualitatively very significant due to the fact that in pristine SLG and BLG the spin-splitting from intrinsic SOC is very small, of the order of 10~$\mu$eV~\cite{gmitra2009}. This means that even a small enhancement of the SOC induced by the proximity of a TMD monolayer can profoundly affect the electronic properties of SLG and BLG.

Ab-initio calculations~\cite{Gmitra2015,Gmitra2016,Gmitra2017} show that semiconducting monolayer TMDs such as \mos, \mose, \mote, \wss, \wse, \wte can enhance SOC in graphene to induce spin-splittings of the order of 1~meV, i.e. orders of magnitude larger than the graphene's intrinsic SOC. The enhancement increases with the atomic number of the metal forming the TMD. For the case of graphene-\wse vdW systems the SOC induced by proximity in graphene is sufficient to create a band inversion of the spin-split bands close to the original graphene's Dirac point. Such a band inversion could lead to topological phases exhibiting the quantum spin Hall effect in SLG-TMD~\cite{Gmitra2016,alsharari2016,alsharari2018} or BLG-TMD\cite{alsharari2018a} heterostructures.
The enhancement of the SOC in graphene-TMD vdW systems has been observed, indirectly, via 
weak antilocalization~\cite{wang2015,wang2016,yang2016,yang2017,voekl2017,wakamura2018,zihlman2018} 
and spin-relaxation measurements~\cite{Avsar2014,dankert2017,Ghiasi2017a,omar2018,Benitez2018a}. It has also been suggested that by tuning the twist angle between graphene and the TMD the nature, from Zeeman-like to Rashba-like, and strength of the induced SOC can be tuned~\cite{Li2019,david2019,Gani2019b}.
Similarly to the case of graphene-TI systems, spin and charge transport are also coupled in graphene-TMD heterostructures~\cite{Offidani2017,garcia2017} as demonstrated experimentally in graphene-MoS$_2$\cite{yan2016,safeer2019}, graphene-WS$_2$\cite{ghiasi2019,benitez2019tunable}, graphene-MoTe$_2$\cite{hoque2019all}, and graphene-TaS$_2$\cite{li2019electrical} devices.

For the case of TMD-BLG vdW systems the resulting structure of the hybridized bands is richer and tunable via 
an external electric field~\cite{Gmitra2017}. For \wse-BLG ab-initio results show that the proximity of the TMD induces a gap in BLG of the order of 10~meV and SOC splittings of about 2~meV~\cite{Gmitra2017}.
Very recent experiments~\cite{Island2019} on symmetric \wse-BLG-\wse vdW systems have shown a strong enhancement of the SOC in graphene and clear signatures of the SOC-driven band inversion.


\section{Superconductor-based van der Waals systems}
\label{sec:sc}
Van der Waals heterostructures constructed with superconducting layers are extremely interesting because they allow the realization of novel superconducting states. In particular, when one of the layers has strong SOC, such heterostructures can be engineered to realize topologically non-trivial superconducting states~\cite{fu2008,sau2010,lutchyn2010,oreg2010,antipov2018} as well as states with odd-frequency pairing~\cite{bergeret2005,linder2017}. 
Given the scope of the special issue, below we focus on the cases in which the vdW heterostructure can host odd-frequency superconducting pairing.


\subsection{Odd-frequency pairing}
\label{sec:odd-f}

In general, the order parameter describing a correlated electronic state is given by a many-body wavefunction, which must be completely antisymmetric under the permutation of all quantum numbers. This antisymmetry constrains the allowed symmetries of the order parameter. In the limit of static order parameters and a single relevant band degree of freedom, this constraint implies that even-parity order parameters ($s$- or $d$-wave) must be odd in the spin index (spin-singlet) while odd-parity order parameters ($p$- or $f$-wave) must be even in spin (spin-triplet). The term odd-frequency pairing refers to the possibility that the many-body state is odd in the relative time coordinate, or, equivalently, in the relative frequency. Therefore, odd-frequency states must possess spatial and spin symmetries with the exact opposite correspondence from the static case: i.e. even-parity states must be spin-triplet and odd-parity must be spin-singlet. An odd-frequency state was first proposed by Berezinskii~\cite{berezinskii1974} as a possible superfluid state for He$^3$. Later on, Berezinskii's proposal was extended to superconducting systems~\cite{kirkpatrick_1991_prl,belitz_1992_prb,balatsky1992,coleman_1993_prl,coleman_1995_prl,belitz_1999_prb}.
However, it has been pointed out that constraints on the electron-phonon interactions inhibit odd-frequency pairing~\cite{abrahams1993interactions}, and that simple models of intrinsically odd-frequency superconducting states may be unstable~\cite{coleman1994,dolgov1994renormalization,heid1995}.

While the status of intrinsic odd-frequency states is uncertain, much progress has been made toward understanding how odd-frequency superconducting correlations can be induced using conventional superconductors in heterostructures. One theoretically well-established example can be found in superconductor-ferromagnet junctions which allow the conversion of conventional $s$-wave spin-singlet Cooper pairs to odd-frequency spin-triplet pairs, due to the breaking of spin-rotational symmetry\cite{BergeretPRL2001, bergeret2005, halterman2007odd, yokoyama2007manifestation, houzet2008ferromagnetic, EschrigNat2008, LinderPRB2008,TriolaPRB2014, crepin2015odd}. Experimental signatures of odd-frequency correlations have been observed in real systems\cite{petrashov1994conductivity,giroud1998superconducting,petrashov1999giant,aumentado2001mesoscopic,zhu2010angular,di2015signature,di2015intrinsic}. Another notable example is the interface between a conventional superconductor and a normal metal, in which odd-frequency pairing can emerge due to broken spatial translation symmetry\cite{tanaka2007theory,TanakaPRB2007}. In this case, the magnitudes of the odd-frequency correlations dominate over the even-frequency amplitudes at discrete energy levels coinciding exactly with peaks in the local density of states\cite{TanakaPRB2007}, indicating a relationship between these odd-frequency pair amplitudes and McMillan-Rowell oscillations\cite{rowell1966electron,rowell1973tunneling} as well as midgap Andreev resonances\cite{alff1997spatially,covington1997observation,wei1998directional}. Similar phenomena have also been predicted to arise in layered two-dimensional systems, like vdW heterostructures, in which one of the components is superconducting\cite{linder2010,linder2010_prb_graphene,Black-SchafferPRB2012,Black-SchafferPRB2013,parhizgar2014,triola2016prl,kuzmanovski2017multiple,rahimi2017unconventional,aliabad2018proximity}. Given the growing number of vdW systems available, the high level of tunability of their properties, and the presence of a 2D surface accessible for interrogation by experiments, vdW systems are ideal candidates for studying odd-frequency superconducting states.

A set of general criteria for the emergence of odd-frequency pairing in 2D systems were given in Ref.~\cite{triola2016prl}. To understand how their results relate to layered vdW systems we will now provide a sketch of their derivation. 

To begin, we consider a vdW system formed by a 2D crystal with Hamiltonian $H_{2D}$, and a superconductor with Hamiltonian $H_{SC}$. Without loss of generality we assume the superconductor to also be 2D. Let $H_t$ describe tunneling processes between the 2D and the SC so that the Hamiltonian $H$ for the entire bilayer system can be written as $H=H_{2D} + H_{SC} + H_t$, with
\begin{align}
  H_{2D}=&\sum_{\textbf{k},\sigma,\sigma'}c^\dagger_{\textbf{k},\sigma}\left[h_0(\textbf{k})\hat\sigma_0 +     \textbf{h}(\textbf{k})\cdot\boldsymbol\sigma\right]_{\sigma,\sigma'}c_{\textbf{k},\sigma'} \label{eq:H2DEG} \\
  H_{SC}=&\sum_{\textbf{k}\sigma\sigma'}d^\dagger_{\textbf{k}\sigma} h^{SC}_{\sigma\sigma'}(\textbf{k})d_{\textbf{k}\sigma'} +\sum_{\textbf{k}\sigma\sigma'}d^\dagger_{\textbf{k}\sigma}\Delta_{\sigma\sigma'}(\textbf{k})d^\dagger_{-\textbf{k}\sigma'} + \text{h.c.}     \label{eq:HSC} \\
  H_t=&t\sum_{\textbf{k},\sigma} d^\dagger_{\textbf{k},\sigma}c_{\textbf{k},\sigma} + \text{h.c.} \label{eq:Ht}
\end{align}
where $\hat\sigma_i$ are the $2\times 2$ Pauli matrices in spin space, $c^{\dagger}_{\textbf{k},\sigma}$ ($d^{\dagger}_{\textbf{k},\sigma}$) and $c_{\textbf{k},\sigma}$ ($d_{\textbf{k},\sigma}$) are the creation and annihilation operators, respectively, acting on the fermionic states in the 2DEG (SC) layer with momentum $\textbf{k}$ and spin $\sigma$, $h_0(\textbf{k})$ is the spin-independent part of $H_{2D}$ and $\hh(\kk)$ is the field that describes its spin-dependent part due to an exchange field and/or spin-orbit coupling. Here, $h^{SC}_{\sigma\sigma'}(\textbf{k})$ describes the quasiparticle spectrum of the normal state of the superconductor, $\Delta_{\sigma\sigma'}(\textbf{k})$ is the superconducting order parameter which, in general, has a linear combination of spin-singlet and spin-triplet terms, and $t$ is the tunneling between the 2D system and the SC, which is assumed to conserve both spin and momentum.

To examine the superconducting pairing induced in the non-superconducting 2DEG by proximity to the SC, we study the anomalous Green's function, or pair amplitude, within the 2DEG, $\hat{F}^{2D}_{\textbf{k};i\omega_n}$, which is a $2\times 2$ matrix in spin space and a function of both the crystal momentum $\textbf{k}$ and Matsubara frequency $\omega_n$. In the absence of interlayer tunneling $t$ no superconducting pairs exist in the 2DEG and therefore $\hat{F}^{2D}_{\textbf{k};i\omega_n}=0$. However, for $t\neq0$ we find that Cooper pairs can tunnel from the SC into the 2DEG, giving rise to superconducting correlations with novel symmetries which depend on the properties of the 2DEG. To understand the symmetries of these induced pairings it is sufficient to consider the leading-order terms in perturbation with respect to the tunneling strength $t$:
\begin{equation}
 \hat{F}^{2D}_{\textbf{k};i\omega_n} =t^2 \ \hat{G}^{2D}_{\textbf{k};i\omega_n} \ \hat{F}^{SC}_{\textbf{k};i\omega_n} \ \left(\hat{G}^{2D}_{-\textbf{k};-i\omega_n}\right)^T,
 \label{eq:F2DEG}
\end{equation}
where $\hat{F}^{SC}_{\textbf{k};i\omega_n}$ is the anomalous part of the Green's function for the SC, given by
\begin{equation}
 \hat{F}^{SC}_{\textbf{k};i\omega_n}=\left(s^{SC}_{\textbf{k},i\omega_n}\hat\sigma_0 + \textbf{d}_{\textbf{k},i\omega_n}\cdot\boldsymbol\sigma \right)i\hat\sigma_2
 \label{eq:FSC}
\end{equation}
where $s^{SC}_{\textbf{k},i\omega_n}$ ($\textbf{d}_{\textbf{k},i\omega_n}$) represents the spin-singlet (spin-triplet) pair amplitudes and $\hat{G}^{2D}_{\textbf{k};i\omega_n}$ is the normal Green's function for the 2DEG when $t=0$, given by
\begin{equation}
 \hat{G}^{2D}_{\textbf{k};i\omega_n}=\dfrac{(i\omega_n-h_0(\textbf{k}))\hat\sigma_0+\textbf{h}(\textbf{k})\cdot\boldsymbol\sigma}{(i\omega_n-h_0(\textbf{k}))^2 -  |\textbf{h}(\textbf{k})|^2}.
 \label{eq:G2D}
\end{equation}

Inserting Eqs.~\ceq{eq:FSC} and ~\ceq{eq:G2D} into Eq.~\ceq{eq:F2DEG} we obtain the expression for $\hat{F}^{2D}_{\textbf{k};i\omega_n}$, which can be written as
\beq
 \hat{F}^{2D}_{\textbf{k};i\omega_n} = A_{\textbf{k};i\omega_n} \left(F_{\textbf{k};i\omega_n}^{odd} + F_{\textbf{k};i\omega_n}^{even} \right),
 \label{eq:F2D02} 
\enq
where $F_{\textbf{k};i\omega_n}^{even}$ ($F_{\textbf{k};i\omega_n}^{odd}$) are strictly even (odd) functions of Matsubara frequency $\omega_n$, and
\begin{align}
 A_{\textbf{k};i\omega_n}=t^2[&[(i\omega_n+h_0(-\textbf{k}))^2-|\textbf{h}(-\textbf{k})|^2]^{-1} \nonumber \\              &\times [(i\omega_n-h_0(\textbf{k}))^2-|\textbf{h}(\textbf{k})|^2]]^{-1}
 \label{eq:A}
\end{align} 
is the spin-independent amplitude arising from the product of the denominators of the Green's functions $\hat{G}^{2D}_{\textbf{k};i\omega_n}$, $\hat{G}^{2D}_{-\textbf{k};-i\omega_n}$. For the most general form of $H_{2D}$, $A_{\textbf{k};i\omega_n}$ has both even and odd-frequency terms; however, assuming $h_0(\textbf{k})=h_0(-\textbf{k})$ and $|\textbf{h}(\textbf{k})|=|\textbf{h}(-\textbf{k})|$, which are true for most systems, $A_{\textbf{k};i\omega_n}$ becomes an even function of both frequency and momentum. In this case the relative contributions from the even- and odd-frequency pair amplitudes are given by $F_{\textbf{k};i\omega_n}^{even}$ and $F_{\textbf{k};i\omega_n}^{odd}$, respectively.

We can decompose each of these even/odd-frequency amplitudes into spin-singlet and spin-triplet components:
\begin{align}
 F_{\textbf{k};i\omega_n}^{even} =& \left(S_{\textbf{k};i\omega_n}^{even}\hat\sigma_0 + 
                                    {\bf D}_{\textbf{k};i\omega_n}^{even}\cdot\boldsymbol\sigma \right)i\hat\sigma_2,  \label{eq:Feven} \\
 F_{\textbf{k};i\omega_n}^{odd} =& i\omega_n\left(S_{\textbf{k};i\omega_n}^{odd}\hat\sigma_0 + 
                                    {\bf D}_{\textbf{k};i\omega_n}^{odd}\cdot\boldsymbol\sigma \right)i\hat\sigma_2, \label{eq:Fodd}
\end{align}
where, for the even-frequency amplitudes we have:
\begin{align}
S_{\textbf{k};i\omega_n}^{even} = &\left[ \omega_n^2 +h_0^2(\textbf{k}) 
                            -\frac{1}{4}(|\hh_+(\kk)|^2 - |\hh_-(\kk)|^2) \right]s_{\textbf{k};i\omega_n}^{SC}  \nonumber \\
                          -&\left[h_0(\kk)\hh_-(\kk)+\frac{i}{2}\hh_+(\kk)\times\hh_-(\kk)\right]\cdot \textbf{d}_{\textbf{k};i\omega_n}, \label{eq:Seven}\\
{\bf D}_{\textbf{k};i\omega}^{even}= &\left[ \omega_n^2 + h_0^2(\textbf{k}) + \frac{1}{4}(|\hh_+(\kk)|^2 - |\hh_-(\kk)|^2)\right]\textbf{d}_{\textbf{k};i\omega_n} \nonumber \\
                          -&ih_0(\kk)\hh_+(\kk)\times\textbf{d}_{\textbf{k};i\omega_n} -\frac{1}{2}\hh_+(\kk)\left(\hh_+(\kk)\cdot\textbf{d}_{\textbf{k};i\omega_n}\right) \nonumber \\
                                +&\frac{1}{2}\hh_-(\kk)\left(\hh_-(\kk)\cdot\textbf{d}_{\textbf{k};i\omega_n}\right) \nonumber \\
                               -&\left[h_0(\kk)\hh_-(\kk)-\frac{i}{2}\hh_+(\kk)\times\hh_-(\kk)\right]s_{\textbf{k};i\omega_n}^{SC}, \label{eq:Deven} 
\end{align}
and for the odd-frequency amplitudes:
\begin{align}
S_{\textbf{k};i\omega_n}^{odd} &=- \textbf{h}_{+}(\textbf{k})\cdot\textbf{d}_{\textbf{k};i\omega_n},  \label{eq:Sodd}\\
{\bf D}_{\textbf{k};i\omega}^{odd} &=- \textbf{h}_{+}(\textbf{k})s_{\textbf{k};i\omega_n}^{SC} - i\textbf{h}_{-}(\textbf{k})\times\textbf{d}_{\textbf{k};i\omega_n}, \label{eq:Dodd}
\end{align}
where $\textbf{h}_{\pm}(\textbf{k})\equiv \textbf{h}(\textbf{k})\pm \textbf{h}(-\textbf{k})$ is the even/odd parity part of the spin-dependent field in the 2DEG. Here, $\hh_+$ can be interpreted as the field arising from ferromagnetic ordering and $\hh_-$ as the field due to SOC.

Focusing first on the even-frequency spin-singlet amplitudes, we note that the first line in Eq.~\ceq{eq:Seven} shows that, as expected, if the SC layer has spin-singlet pair amplitudes, spin-singlet pairing is also induced in the 2DEG. Moreover, from the second line of Eq.~\ceq{eq:Seven} we see that, due to the presence of SOC in the 2DEG, a singlet term can also be induced by spin-triplet pairing in the SC. It is interesting to note that such contributions are only possible if $\hh_-\neq0$ and can be enhanced by adjusting the angle between $\hh_+(\kk)$ and $\hh_-(\kk)$.

Turning our attention to the induced even-frequency spin-triplet pairing, we see that the first line in  Eq.~\ceq{eq:Deven} shows an induced triplet pairing in the 2DEG directly proportional to the $\dd$ vector in the SC, as expected. The second and third lines in Eq.~\ceq{eq:Deven} show that the presence of the spin-dependent $\hh(\kk)$ field in the 2DEG layer induces a rotation of the $\dd$ vector. The last line in Eq.~\ceq{eq:Deven} shows that the presence of SOC in the 2DEG layer also converts some of the spin-singlet amplitudes in the SC to spin-triplet pairing in the 2DEG. As in Eq.~\ceq{eq:Seven}, we only find this symmetry conversion between triplet and singlet amplitudes when $\hh_-\neq 0$.

We now focus on the induced odd-frequency amplitudes. From Eq.~\ceq{eq:Sodd} we see that an odd-frequency spin-singlet amplitude is induced in the 2DEG layer when $\hh_+(\kk)\neq 0$ and a triplet component is present in the SC. In contrast to the case of induced even-frequency spin-singlet pairing, this odd-frequency amplitude emerges due to a conversion of spin-triplet amplitude in the SC to spin-singlet amplitude in the 2DEG for finite $\hh_+(\kk)$ not $\hh_-(\kk)$.  

In the case of the odd-frequency spin-triplet pairing, Eq.~\ceq{eq:Dodd} shows that two contributions are possible. One of these involves a conversion from spin-singlet pairing in the SC to triplet pairing in the 2DEG when $\hh_+(\kk)\neq 0$. We note that this term, and Eq.~\ceq{eq:Sodd} are consistent with known results for the case of ferromagnet/superconductor junctions~\cite{BergeretPRL2001, bergeret2005, halterman2007odd, yokoyama2007manifestation, houzet2008ferromagnetic, EschrigNat2008, LinderPRB2008,linder2010_magnetic,TriolaPRB2014, crepin2015odd}.
The last term in Eq.~\ceq{eq:Dodd} shows that a triplet component in the SC can also induce an odd-frequency triplet term in the 2DEG in the presence of SOC, $\hh_-(\kk)\neq0$, as long as $\hh_-$ is not parallel to $\dd_\kk$. Typically, for an isolated system, the superconducting $\dd_\kk$ vector is parallel to the direction of the SOC field and so the presence of triplet pairing and SOC is not sufficient to realize odd-frequency pairing. However, in a vdW, due to the fact that the vector fields $\hh_-(\kk)$ and $\dd_\kk$ live in different layers, the condition $\hh_-(\kk)\times \dd_\kk$ can be readily realized, as demonstrated in the concrete example below. This fact, also considering the great experimental advances in creating high quality vdW systems comprising a large variety of materials, considerably enlarges the set of systems in which odd-frequency superconducting pairing can be realized and detected.

A real system in which the condition $\hh_-(\kk)\times \dd_\kk\neq 0$ can be realized is a vdW system formed by a monolayer transition metal dichalcogenide (TMD) placed on a superconducting surface with Rashba SOC~\cite{triola2016prl}.
Considering only states close to the valence bands in the TMD layer, the Hamiltonian in Eq~\ceq{eq:HTMD} can be simplified to obtain:
\begin{equation}
\hat{h}^{TMD}_{\textbf{k},\lambda}=-\left( \dfrac{a^2\gamma^2}{u}k^2 + \dfrac{u}{2} + \mu\right) + \lambda\alpha\hat\sigma_3.
\label{eq:HTMD02}
\end{equation}
Considering that under parity $\lambda \to -\lambda$, we can see that for a 2D system described by the Hamiltonian given by Eq.~\ceq{eq:HTMD02} we have $\hh_+=0$, and $\hh_-=2\lambda\hat z$, where $\hat{z}$ is the unit vector normal to the TMD monolayer.

Using the general expressions from Eqs.~\ceq{eq:Seven}-\ceq{eq:Dodd} we can obtain the symmetry properties of the superconducting pair amplitudes induced by proximity in a vdW system formed by a hole-doped monolayer TMD and a generic superconducting layer:
\begin{align}
S_{\textbf{k},\lambda;i\omega_n}^{even} &= \left( \omega_n^2 + \xi_{\textbf{k}}^2 +\alpha^2 \right)s_{\textbf{k},\lambda;i\omega_n}^{SC}-2\lambda\alpha\xi_{\textbf{k}}\hat{z}\cdot \textbf{d}_{\textbf{k},\lambda;i\omega_n} \\
{\bf D}_{\textbf{d},\lambda;i\omega}^{even} &=\left( \omega_n^2 + \xi_{\textbf{k}}^2 -\alpha^2 \right)\textbf{d}_{\textbf{k},\lambda;i\omega_n} +2\alpha^2\left(\hat{z}\cdot\textbf{d}_{\textbf{k},\lambda;i\omega_n}\right)\hat{z} \nonumber \\
&- 2\lambda\alpha\xi_{\textbf{k}}s_{\textbf{k},\lambda;i\omega_n}^{SC} \hat{z} \\
S_{\textbf{k},\lambda;i\omega_n}^{odd} &=0 \\
{\bf D}_{\textbf{k},\lambda;i\omega}^{odd} &= -i2\lambda\alpha \hat{z}\times\textbf{d}_{\textbf{k},\lambda;i\omega_n}
\label{eq:DoddTMD}
\end{align}
As described for the general case we see that the presence of SOC, proportional to $\alpha$, mixes the singlet and triplet components for the even-frequency pair amplitudes. The SOC also generates an odd-frequency triplet pair amplitude proportional to the strength of the SOC in the TMD monolayer and the triplet component of the SC, Eq.~\ceq{eq:DoddTMD}.

We now consider an effective 2D SC with Rashba SOC~\cite{Gorkov2001}, which can be realized on the surface of Pb. The surface of this superconductor can be described using the Hamiltonian in Eq.~\ceq{eq:HSC} with  
\begin{align}
\hat h^{\rm SC}_\textbf{k}&=\epsilon_{\textbf{k}}\hat \sigma_0+\eta\hat{z}\cdot(\boldsymbol\sigma\times\textbf{k})
\label{eq:HSCN} 
\end{align}
where $\eps_k$ is the spin-independent part of the electrons' dispersion and $\eta$ is the strength of the Rashba SOC. The energy eigenvalues of $\hat h^{\rm SC}_\textbf{k}$, $E_\kk = \eps_K \pm \eta|\kk|$, identify the bands of the SC in the normal phase. For the order parameter, $\hat{\Delta}$, we assume, as is standard\cite{Gorkov2001}, that intraband pairing dominates and obtain the corresponding anomalous Green's function $F_{SC}$ in the energy eigenbasis of $\hat h^{\rm SC}_\textbf{k}$. Rotating back to the spin basis in which $\hat h^{\rm SC}_\textbf{k}$ is expressed in Eq.~\ceq{eq:HSCN} we obtain
\begin{equation}
\hat{F}^{SC}_{\textbf{k};i\omega_n}= \Delta\frac{(s_{\textbf{k};i\omega_n}^{SC}\hat\sigma_0+{\bf d}_{\textbf{k}}\cdot\boldsymbol\sigma)i\hat\sigma_2}
                               {(s_{\textbf{k};i\omega_n}^{SC})^2-|{\bf d}_{\textbf{k}}|^2} 
\label{eq:FSC_rashba}
\end{equation}
where $\Delta$ is the superconducting gap, and
\begin{align}
 s_{\textbf{k};i\omega_n}^{SC} &= \Delta^2 +\omega_n^2 + \epsilon_{\textbf{k}}^2 + \eta^2 k^2 \\
 {\bf d}_{\textbf{k}}&=2\epsilon_{\textbf{k}}\eta(-k_y,k_x,0)
\end{align}
are the singlet and triplet amplitudes, respectively. Notice that the presence of Rashba SOC gives rise to a triplet component with an in-plane $\dd$ vector, i.e. a $\dd$ vector
that is orthogonal to the $\hh_-(\kk)$ field due to SOC in the monolayer TMD.

From Eq.~\ceq{eq:DoddTMD} we see that in vdW systems composed of a TMD monolayer and an effective 2D SC with Rashba SOC an odd-frequency spin-triplet pair amplitude will be induced with strength proportional to the product of the SOC strength in the TMD and the Rashba SOC strength in the SC. In this case, the full anomalous Green's function, $F^{TMD}$, has the form
$\hat{F}^{TMD}_{\textbf{k},\lambda;i\omega_n} = A^{TMD}_{\textbf{k},\lambda;i\omega_n} \left(F_{\textbf{k},\lambda;i\omega_n}^{odd} + F_{\textbf{k},\lambda;i\omega_n}^{even} \right)$
with
$$A^{TMD}_{\textbf{k},\lambda;i\omega_n}=\frac{\Delta t^2}{[(i\omega_n-\xi_\textbf{k})^2-\alpha^2]^2[(s_{\textbf{k}+\textbf{K}_\lambda;i\omega_n}^{SC})^2-|\dd_{\textbf{k}+\textbf{K}_\lambda}|^2]},$$
where $F_{\textbf{k},\lambda;i\omega_n}^{even}$ and $F_{\textbf{k},\lambda;i\omega_n}^{odd}$ have the same form as Eqs.~\ceq{eq:Feven} and \ceq{eq:Fodd} with
\begin{align}
 S_{\textbf{k},\lambda;i\omega_n}^{even} &= \left( \omega_n^2 + \xi_{\textbf{k}}^2 +\alpha^2 \right)s_{\textbf{k}+\textbf{K}_\lambda;i\omega_n}^{SC} \\
 {\bf D}_{\textbf{k},\lambda;i\omega}^{even} &=-\left( \omega_n^2 + \xi_{\textbf{k}}^2 -\alpha^2 \right)\dd_{\textbf{k}+\textbf{K}_\lambda} -  2\lambda\alpha\xi_{\textbf{k}}s_{\textbf{k}+\textbf{K}_\lambda;i\omega_n}^{SC} \hat{z}\\
 {\bf D}_{\textbf{k},\lambda;i\omega}^{odd} &= i4\lambda\alpha\eta \epsilon_{\textbf{k}+\textbf{K}_\lambda} (\textbf{k}+\textbf{K}_\lambda)
\end{align}
where $\textbf{K}_\lambda$ is the momentum vector at the $K$ ($K'$) point for $\lambda=1$ ($\lambda=-1$). Here, ${\bf D}_{\textbf{k},\lambda;i\omega}^{odd}$ is the d-vector describing an odd-frequency spin-triplet pair amplitude. Using the above expressions, we find that this pair amplitude corresponds to a term in the anomalous Green's function of the form $F^{TMD}_{\uparrow\uparrow/\downarrow\downarrow}\sim i\omega_n\eta\alpha\epsilon_{\overline{\textbf{k}}}\lambda\left(\overline{k}_y\pm i\overline{k}_x\right)$, where $\bar{\textbf{k}}$ is the momentum measured from the center of the BZ. 

\begin{figure}
 \begin{center}
  \centering
  \includegraphics[width=0.7\columnwidth]{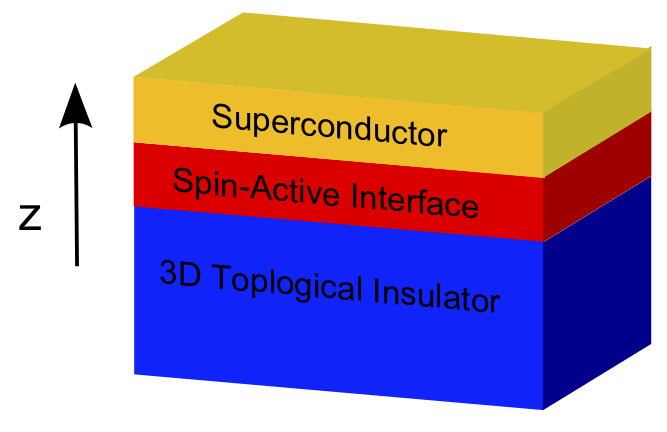}
  \caption{
           Schematic representation of a vdW system formed by a SC and the surface of a strong 3D TI. Adapted with permission from Phys. Rev. B 89, 165309 (2014).
         }
  \label{fig:TI-SC}
 \end{center}
\end{figure}

So far we have assumed that interlayer tunneling conserves spin and is entirely spin-independent. In general, tunneling between between materials with different spin eigenstates is spin-dependent. Such spin-dependence of the interlayer tunneling introduces additional mechanisms~\cite{EschrigNat2008,LinderPRB2008,LinderPRL2009,LinderPRB2010_2,TriolaPRB2014} by which odd-frequency pairing terms can be generated in a vdW system formed by a superconducting layer and a layer with strong SOC. Such mechanisms could naturally emerge in a vdW system formed, for example, by a SC and the surface of a strong 3D TI with a ``spin-active'' interface, as shown schematically in Fig.~\ref{fig:TI-SC}.

There are two basic mechanisms by which an interface can actively affect the spin state $\ket{\sigma}$: (i) it can impart a spin-dependent phase $\left|\sigma \right>_{\textbf{k}}\rightarrow e^{i\theta_{\sigma,\textbf{k}}}\left| \sigma\right>_{\textbf{k}}$ due to the precession of the spin around a magnetic moment present at the interface; (ii) the tunneling Hamiltonian can be off-diagonal in the spin basis, $\hat{t} = t_0\hat\sigma_0 + \ttt\cdot\ssigma$, where $\ttt$ is a 3-component vector, leading to spin flips. For case (i), when the interlayer tunneling induces a spin-dependent phase, we can see that a singlet state, $\left| \uparrow\right>_\textbf{k}\left| \downarrow\right>_{-\textbf{k}}-\left| \downarrow\right>_\textbf{k}\left| \uparrow\right>_{-\textbf{k}}$, is converted, after tunneling, to the state $e^{i\eta_\textbf{k}}\left(e^{i\zeta_\textbf{k}}\left| \uparrow\right>_\textbf{k}\left| \downarrow\right>_{-\textbf{k}}-e^{-i\zeta_\textbf{k}}\left| \downarrow\right>_\textbf{k}\left| \uparrow\right>_{-\textbf{k}}\right)$, where $\eta_\textbf{k}\equiv  (\theta_{\uparrow,\textbf{k}} + \theta_{\downarrow,\textbf{k}}+\theta_{\uparrow,-\textbf{k}} + \theta_{\downarrow,-\textbf{k}})/2$ and 
$\zeta_\textbf{k}\equiv (\theta_{\uparrow,\textbf{k}} - \theta_{\downarrow,\textbf{k}} - \theta_{\uparrow,-\textbf{k}} + \theta_{\downarrow,-\textbf{k}})/2$. 
Therefore, we see that a triplet component with amplitude proportional to $\sin\zeta_{\kk}$ emerges due to the spin-dependent phase introduced by a spin-active interface, even when the SC only has a singlet pairing.

Using the same leading-order perturbation theory discussed above we can deduce the symmetries of the proximity-induced pair amplitudes in a vdW system that possesses a spin-active interface. Here we briefly sketch how the approach leading to Eq.~\ceq{eq:F2DEG} must be modified in this case, see Ref.~\cite{TriolaPRB2014} for more details. To account for spin-dependent phase factors we replace the anomalous Green's function of the superconducting layer, $\hat{F}^{SC}(\textbf{k},\omega)$, with a rotated version:
\begin{equation}
 \hat{F}^{SC}_{\theta_\textbf{k}}(\textbf{k},i\omega_n)=e^{i\eta_\textbf{k}}e^{i\frac{\delta\theta_{\textbf{k}}}{2}\sigma_3}
 \hat{F}^{SC}(\textbf{k},i\omega_n)e^{i\frac{\delta\theta_{-\textbf{k}}}{2}\sigma_3}.
 \label{eq:FSCrotated}
\end{equation}
Then, we obtain the following leading-order contribution to the anomalous Green's function in the 2DEG layer:
\begin{equation}
 \hat{F}^{2D}_{\textbf{k};i\omega_n} =\hat{G}^{2D}_{\textbf{k};i\omega_n} \ \hat{t} \  \hat{F}^{SC}_{\theta_\textbf{k}}(\textbf{k},i\omega_n) \ \hat{t}^T \ \left(\hat{G}^{2D}_{-\textbf{k};-i\omega_n}\right)^T,
\label{eq:F-ind-rotated}
\end{equation}
From Eq.~\ceq{eq:F-ind-rotated} we can readily deduce the symmetries of induced pair amplitudes for a variety of vdW systems by inserting the appropriate system-specific expressions for $G^{2D}$ and $F^{SC}$. 

In Ref.~\cite{TriolaPRB2014}, the pair symmetries given by Eq.~\ceq{eq:F-ind-rotated} were examined assuming a conventional spin-singlet superconducting layer. The analysis was performed for three different 2DEG layers: (i) the surface of a 3D topological insulator (TI); (ii) a ferromagnet with in-plane magnetization, i.e. an easy plane ferromagnet (FE); and (iii) a ferromagnet with perpendicular magnetization, i.e. a z-axis ferromagnet (FZ). We summarize the results in Table~\ref{tab:odd-f}, indicating whether or not odd-frequency pairing can be realized in each of these systems for different kinds of spin-active interfaces, and its character (singlet or triplet).
\begin{center}
\begin{table}[htb]
 \begin{tabular}{|l | l | l | l|}
 \hline\hline
   Interface                             & FZ-SC        & FE-SC         & TI-SC \\
   \hline
   Not spin active                       & T          & T           & -     \\
   Spin-dep. phases                 & T \& S     & T           & T   \\
   Spin flip                             & T          & T           & -     \\
   Spin-dep. phases \& spin flip    & T \& S     & T \& S      & T   \\
  \hline\hline
\end{tabular}
\caption{Conditions for the realization of odd-frequency pairing, and its spin symmetry, singlet (S) or triplet (T), for three superconductor-based vdW heterostructures,
FZ-SC, FE-SC, and TI-SC, including the effect of a spin-active interface. A ``-'' indicates that no odd-frequency pairing is present.}
\label{tab:odd-f}
\end{table}
\end{center}

From Table~\ref{tab:odd-f} we see that when the 2DEG layer is either an FZ or FE, odd-frequency pairing is induced for any kind of interface, while odd-frequency pairing can only be induced in the TI in the presence of spin-dependent phases. Comparing the different symmetries for the different kinds of interfaces we see that for both the FZ and FE, spin-triplet pairing is always induced. However, spin-singlet pairing can be induced in the FZ due to spin-dependent phases, even without spin flips, while spin-singlet pairing can only be induced in the FE in the presence of both spin-dependent phases and spin flips. For the TI, the odd-frequency pairing must be spin-triplet. 

We note that odd-frequency pairing has also been investigated in buckled honeycomb systems possessing proximity-induced superconductivity\cite{kuzmanovski2017multiple}. In contrast to the results presented in this section, in that work the presence of the superconducting layer was accounted for by adding a spin-singlet BCS order parameter to the Kane-Mele Hamiltonian\cite{haldane1988model,kane2005qshi} and computing the on-site order parameter self-consistently. For large doping, the authors found that bulk odd-frequency intersublattice pairing emerges when the sublattice symmetry is broken by, for example, an electric field perpendicular to the plane, similar to the odd-frequency interband pairing found in multiband superconductors\cite{black2013odd,asano2015odd,komendova2015experimentally,komendova2017odd,asano2018green,triola2018odd}. At low doping, when the low-energy states are localized at the edge of the sample, the authors performed their analysis in real space for different edge terminations, finding that odd-frequency pairing arises generically, even in the absence of an external field. In the case of the zig-zag edge termination, this was due to the asymmetry between the two sublattices at the edge\cite{kuzmanovski2017multiple}. In the case of arm chair terminations, the odd-frequency pairing was due to the asymmetry between every other pair of sublattices\cite{kuzmanovski2017multiple}. In both cases the odd-frequency pairing arises due to an inhomogeneity of the order parameter which naturally occurs in such finite-size systems. This phenomenon was also studied in the 2D surfaces of 3D topological insulators in the presence of an inhomogeneous superconducting order parameter\cite{Black-SchafferPRB2012}, finding qualitatively similar results.    

To conclude this subsection we note that the pair amplitudes given by both Eqs.~\ceq{eq:F2DEG} and \ceq{eq:F-ind-rotated} represent pairing between electrons in the same 2D layer. However, when the vdW system possesses more than one normal layer, interlayer pairing can also be important, as investigated in Ref.~\cite{parhizgar_2014_prb}. In that work, a similar analysis to the one leading to Eqs.~\ceq{eq:F2DEG} was performed for a bilayer system coupled to a superconducting layer. The authors explicitly investigated the possibility of interlayer pairing. Interestingly, the authors found that, in general, because tunneling between adjacent layers dominates, an asymmetry emerges between the induced gaps on the two layers. This asymmetry leads directly to odd-frequency interlayer pairing in such vdW heterostructures\cite{parhizgar_2014_prb}, similar to phenomena studied in multiband superconductors\cite{black2013odd,asano2015odd,komendova2015experimentally,komendova2017odd,asano2018green,triola2018odd}, double quantum dots\cite{sothmann2014unconventional,burset2016all}, and double nanowires\cite{ebisu2016theory,triola2019oddnw}.


\subsection{Proximity induced Ising pairing}
\label{sec:ising_SC}

We now focus on the set of superconductor-based vdW systems in which the superconducting layer is a TMD monolayer. One of the key features of superconducting TMD monolayers is that the superconducting state is extremely robust against in-plane magnetic fields: superconductivity survives for magnetic fields much larger than the Pauli paramagnetic limit. This is due to the strong spin-splitting of the bands at the Fermi surface in metallic monolayer TMDs induced by strong SOC and a lack of inversion symmetry. These conditions favor a particular spin orientation of the Cooper pairs and the resulting superconducting pairing is termed Ising pairing~\cite{ugeda2015,xi2016}. 

\begin{figure}[htb]
 \begin{center}
  \includegraphics[width=1.0\columnwidth]{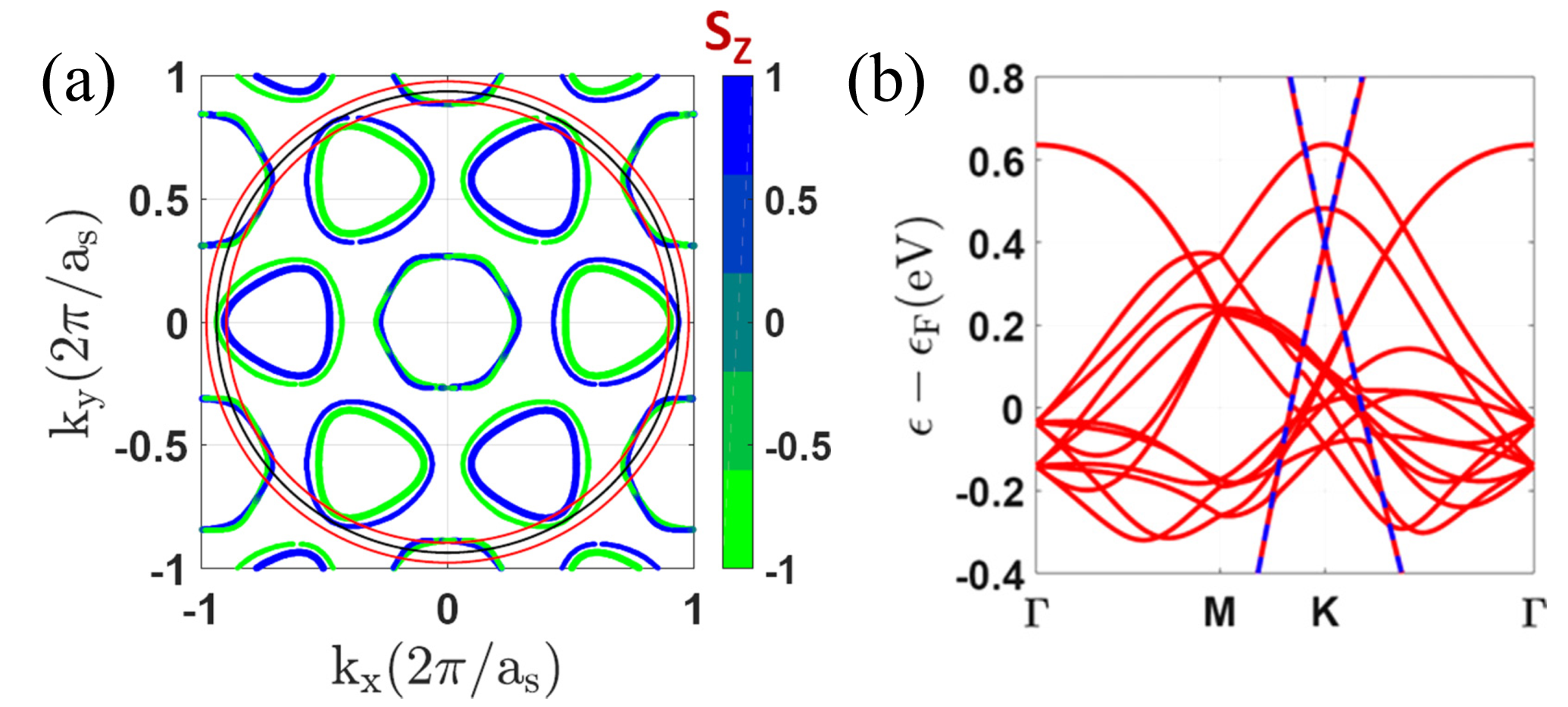}
  \caption{(a) Fermi surfaces of monolayer \nbse. The blue (green) FSs are the \nbse FSs for spin up (down) respectively,
           the black circle shows the position of the graphene Dirac point as $\theta$ is varied between 0 and 360$\degree$.
           The red circles delimit the region within which the graphene FS is confined as the twist angle is varied.
           (b) Low energy band structure, in red, of a graphene-\nbse systems obtained from ab-initio calculations including relativistic corrections for
               a commensurate stacking corresponding to $\theta=-65.2\degree$. The blue lines show the low
               energy bands of an isolated graphene layer with doping corresponding to the charge transfer occurring
               when graphene is place on \nbse. 
           Adapted with permission from Phys. Rev. B. {\bf 99}, 235404 (2019)
         } 
  \label{fig:nbse-gr-01}
 \end{center}
\end{figure}

The most commonly studied superconducting TMD is \nbse~\cite{ugeda2015,xi2016}. The lattice structure
is the same as the one shown in Fig.~\ref{fig:TMD-SC}~(a). In its monolayer form the normal state spectrum of \nbse has Fermi pockets around the $\Gamma$ point, and around the corners ($\KK$ and $\KK'$ points) of the BZ, as shown in Fig.~\ref{fig:nbse-gr-01}~(a). As the figure shows, the Fermi surfaces are spin-split due to SOC and broken inversion symmetry.
The splitting of the Fermi surface is much stronger for the $\KK$ and $\KK'$ pockets than for the $\Gamma$ pocket given that, at the $\Gamma$ point, the $\kk$ and $-\kk$ states coincide. As a consequence, in the superconducting state, the pairing at the $\KK$ and $\KK'$ pockets is much more robust against external in-plane magnetic fields than at the $\Gamma$ pocket.
Such a difference is hard to detect experimentally in isolated monolayers of \nbse given that even when the Zeeman term due to an in-plane magnetic field is large enough to completely suppress the superconducting gap at the $\Gamma$ pocket, the superconductivity arising from the states around the $\KK$ ($\KK'$) pockets survives and therefore prevents the use of transport measurements to observe the breakdown of the superconducting state at the $\Gamma$ pocket.
In vdW systems formed by one monolayer of a material such as \nbse and another, non-superconducting layer we can expect that the superconducting pairing induced by proximity in the normal layer will retain some of the properties of the pairing in \nbse and, in particular, its Ising nature.
A natural candidate system to combine with \nbse is graphene. As pointed out above, there is a large
mismatch between graphene's and TMD's lattice constants and so one would expect that no significant
hybridization between the graphene's and the \nbse's states could take place.
However, for some twist angles the Fermi pockets of \nbse are large enough to overlap with graphene's Dirac points. This is displayed in Fig.~\ref{fig:nbse-gr-01}~(a) where the 
green and blue lines show the spin-split Fermi surfaces of \nbse, and the black circle the position of graphene's Dirac points as the twist angle $\theta$ is varied between 0 and $360\degree$.
For a range of angles, $\pm 7.2\degree$~\cite{gani2019}, around $0\degree$ (and multiples of $60\degree$) the Dirac points intersect the $\KK$ (or $\KK'$) Fermi pocket of \nbse, and for a range
of angles, $\pm 3.9\degree$~\cite{gani2019}, around $21.9\degree$ (and multiples of $60\degree$)
the Dirac points intersect the $\Gamma$ Fermi pocket of \nbse. As a consequence, 
graphene can be used to probe the differences between the electronic states of the different Fermi pockets of \nbse, including properties of the superconducting pairing~\cite{dvir2018}.

Ab-initio calculations~\cite{gani2019} show that when placed on \nbse graphene becomes hole doped, Fig.~\ref{fig:nbse-gr-01}~(b), so that the Fermi energy in the graphene layer is $\sim 400$~meV below the Dirac point. These calculations also show~\cite{gani2019} that the interlayer tunneling between the two systems is of the order of 20~meV and that this value, and the amount of charge transfer do not depend significantly  on the twist angle.
Using these values we can construct a continuum model as described in Sec.~\ref{sec:model} to obtain the low-energy properties for generic values of the twist angle.

\begin{figure}[htb]
 \begin{center}
  \includegraphics[width=1.0\columnwidth]{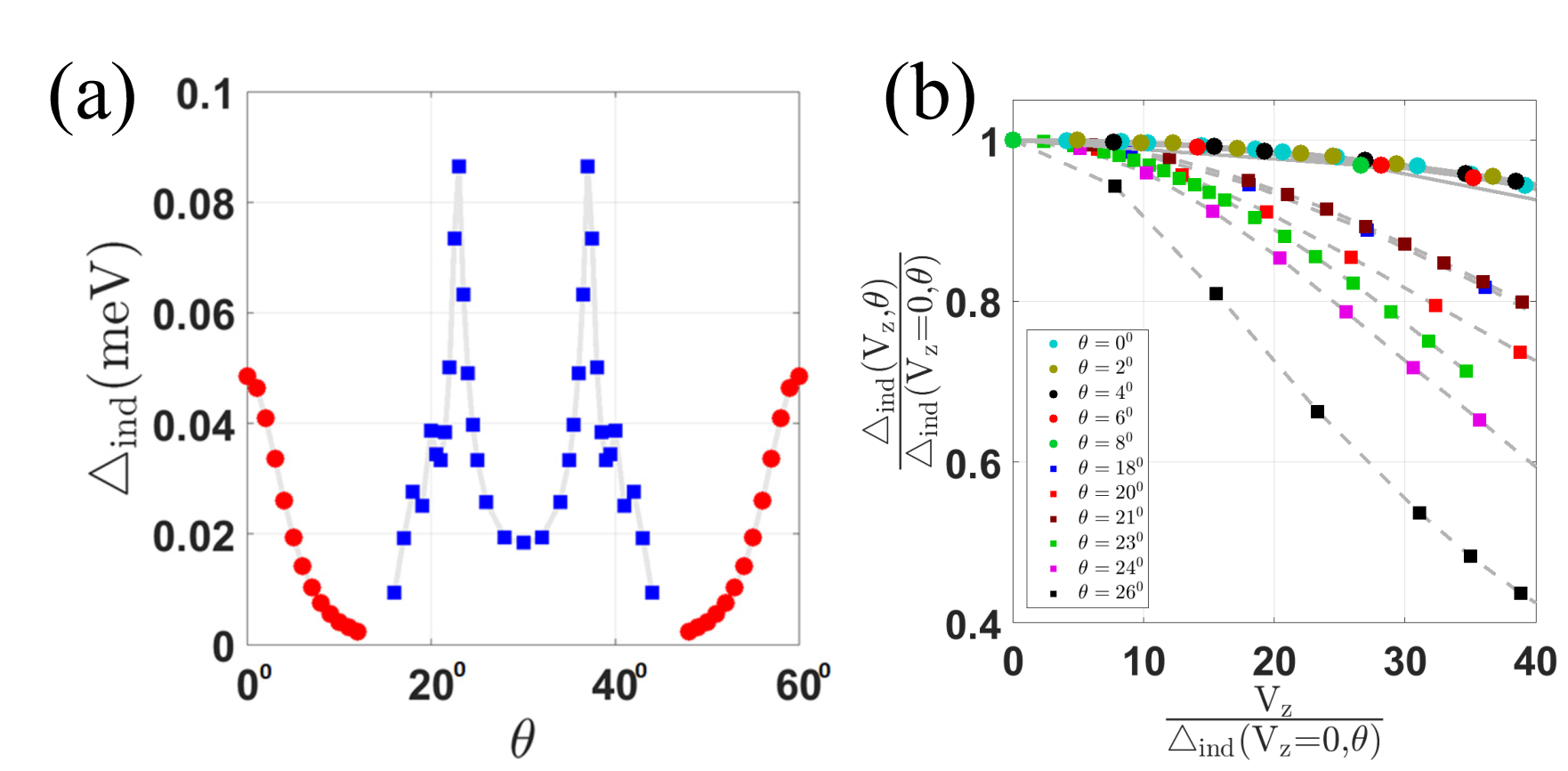}
  \caption{(a) Induced superconducting gap, $\Delta_{\rm ind}$, into the graphene layer as a function of the twist angle.
           (b) $\Delta_{\rm ind}$ as a function of Zeeman field $V_z$ The solid lines (circles) show the results for values of $\theta$ for which graphene's FS overlaps with \nbse's FS pockets around the $\KK$ ($\KK'$). The dashed lines (squares) show the results for values of $\theta$ for which graphene's FS overlaps with \nbse's FS pocket around the $\Gamma$ point.
           Adapted with permission from Phys. Rev. B. {\bf 99}, 235404 (2019)
         } 
  \label{fig:nbse-gr-02}
 \end{center}
\end{figure}

For twist angles such that the Fermi surface of graphene touches one of the Fermi pockets of \nbse,
superconducting pairing can be induced in the graphene layer.
Figure~\ref{fig:nbse-gr-02}~(a) shows the value of the induced gap, $\Delta_{\rm ind}$, in the graphene layer as a function of the twist angle. The red circles indicate the values of $\Delta_{\rm ind}$ for the cases when the graphene FS touches the $\KK$ ($\KK'$) Fermi pocket of \nbse. The blue squares denote the cases when the graphene FS touches the \nbse Fermi pocket around the $\Gamma$ point.
These results show that in superconductor-based vdW systems such as \nbse-graphene heterostructures,
the size of the gap induced by proximity can be strongly tuned by varying the twist angle.

Figure~\ref{fig:nbse-gr-02}~(b) shows how the proximity-induced superconducting gap in the
graphene layer depends on the strength of a Zeeman term, $V_z$, due to the presence of an in-plane
magnetic field, for different values of the twist angle. The solid lines with circles show the results
for values of $\theta$ such that the low energy states in graphene hybridize with the low energy states 
close to the $\KK$ ($\KK'$) point in \nbse. The dashed lines with squares
show the results for the cases where the graphene FS touches the \nbse Fermi pocket
at the $\Gamma$ point. We see that in the first case the induced superconducting gap
is much more robust against the presence of an in-plane field than in the second case.
This is a consequence of the fact that, for the first case, the graphene is effectively
probing the superconducting gap of \nbse at the $\KK$ ($\KK'$) where the spin-splitting
due to the SOC is much stronger than for the pocket around the $\Gamma$  pocket,
and therefore the Ising nature of the pairing is much more pronounced.

The results of Fig.~\ref{fig:nbse-gr-02}~(b) show that in vdW systems
like graphene-\nbse, the Ising character of the induced superconducting
pairing can be tuned via the twist angle. In addition, they show that
in these types of structures graphene can be used to probe the relative
strength of the gaps on different parts of the FS of the substrate,
and the robustness of these gaps to external
magnetic fields.


\section{Conclusions}
\label{sec:conclusions}

In this article we have reviewed recent work on heterogenous van der Waals systems in which one
of the components has strong spin-orbit coupling. The field of van der Waals systems
is now very large and so we have restricted the discussion to a few exemplary vdW systems.

We first presented the general effective model to obtain the low energy electronic spectrum
for a generic stacking configuration. The model relies on parameters that must be obtained
via ab-initio calculations, or, when possible, directly from experimental measurements.
We then discussed, in detail, the case of a van der Waals system formed by coupling a single layer of graphene to the surface of a three dimensional strong topological insulator. 
We discussed how the hybridization between the states in these two systems strongly enhances the spin-orbit coupling of the graphene layer.
We then considered the electronic transport of graphene-TI bilayers and showed how the
enhancement of the spin-orbit coupling in graphene, and the additional screening of charge impurities by the graphene layer, can lead to a considerable amplification of spin-dependent
effects, such as the Edelstein effect. We briefly discussed the 
case of heterostructures formed by graphene and semiconducting transition metal dichalcogenides.

In the second part of the work we discussed the case of heterogenous vdW systems in 
which one of the components is superconducting. Given the scope of this special issue, particular focus was placed on systems in which odd-frequency pairing can be realized.
We first presented a general analysis allowing the identification of conditions for realizing odd-frequency pairing based on a combination of proximity-induced superconductivity and spin-orbit coupling in superconductor-based vdW systems.
Based on this general treatment, we observed that vdW systems with spin-orbit coupling are ideal systems for realizing odd-frequency pair correlations. A distinct advantage over bulk superconductors with a similar degree of spin-orbit coupling is that the direction of the field describing the spin-orbit coupling and that of the $\dd_\kk$ vector describing the spin configuration of a triplet superconductor can be completely different since they belong to different layers in the vdW system.
After this general discussion, we examined a concrete example in which this condition can be realized: a vdW system formed by a monolayer transition metal dichalcogenide and a two-dimensional superconductor with Rashba spin-orbit coupling.
We then discussed the case of vdW systems in which the interface between the superconducting layer
and the normal layer causes the interlayer tunneling to be spin-dependent. We reviewed the conditions under which such a ``spin-active'' interface can lead to the formation of odd-frequency superconducting pairing.
Finally, we discussed the case in which the superconducting layer exhibits Ising superconductivity,
as in monolayer \nbse. In particular, we saw that in graphene-\nbse vdW systems the size of
the superconducting gap, and its robustness against in-plane magnetic fields, strongly
depends on the relative twist angle between the layers.

Our brief survey of the field shows that van der Waals heterostructures in which spin-orbit coupling is present constitute
a very interesting class of systems, both from a fundamental point of view, and for technological applications.
The freedom in the selection of the layers forming the heterostructure, combined with the ability to control the stacking
configuration, mean that the number of vdW systems with spin-orbit that can be realized, and that have not been studied yet is very large.
This presents the opportunity to realize vdW systems with spin-orbit coupling exhibiting novel spin dependent transport properties
or in which known spintronics effects can be optimized and deployed effectively in technological applications.
A class of vdW systems with spin-orbit coupling that would be very interesting to study, and that is still largely unexplored, 
is the one in which the presence of spin-orbit 
is coupled to strong electron-electron correlations. Such coupling could perhaps be realized in vdW heterostructures combining twisted
bilayer graphene and layers with strong spin-orbit coupling.
The study of the response to several external probes, in particular time-dependent probes, of the type of vdW systems described in this survey is still
largely unexplored.
Another interesting direction is the study of vdW systems with spin-orbit coupling in which one of the constituents is a nanostructure,
such as heterostructures formed by graphene nanoribbons and substrates with strong spin-orbit coupling. 
These structures could allow the realization of nano-devices with tunable spin-orbit coupling.

One of the challenges to understand the physics of vdW systems, and to use them in applications, is the correct
characterization of the effects of disorder. For some vdW systems the dominant source of disorder is still unknown.
In some vdW structures the twist angle is not spatially homogenous and the effect of the randomness of the twist angle on
the transport properties is not known yet.
Also still unknown is why in graphene-based structures the experimentally measured spin-relaxation time is orders of magnitude
smaller than the theoretical predictions.
In the case of superconductor-based vdW systems, a major driver in the field has been the engineering of exotic superconducting states, in particular those which possess unconventional symmetries like odd-frequency pairing. Two major barriers to understanding these exotic states are the scarcity of unambiguous experimental signatures, and the difficulty of realizing systems in which the odd-frequency pairing dominates the even-frequency. While these barriers have been overcome in particular experimental setups, as discussed in Sec. \ref{sec:odd-f}, a general solution to the problem is still lacking.

We emphasize that one of the most interesting features of van der Waals systems is that the choice of layers is, to a large extent, not constrained by chemistry. On top of the flexibility in the choice of constituent layers, recent experimental developments demonstrating that the relative twist angle between layers can be controlled within a fraction of degree allow for an incredible amount of tunability of the interlayer coupling. 
As we have seen in our discussion of just a limited sample of possible van der Waals systems with spin-orbit coupling, the ability to combine layers with different properties can be used to realize and control exotic superconducting states and engineer systems with strong spin-dependent transport effects.
By continuing to study the myriad combinations of the growing number of two-dimensional crystals, we expect many more novel and surprising electronic properties will be discovered in van der Waals structures.

\section{Acknowledgments}

It is a pleasure to thank the many collaborators with whom we have had the opportunity to work
in the past few years on the topics discussed in this article: 
D.~S.~L.~Abergel,
E.~Y.~Andrei,
D.~M.~Badiane,
A.~V.~Balatsky, 
A.~M.~Black-Schaffer, 
J.~Cayao, 
Y.~Gaucher, 
Y.~S.~Gani,
R. M.~Geilhufe, 
Y.~Kedem,
D.~Kuzmanovski, 
E.~Langmann, 
Chih-Pin Lu,
T.~L\"{o}thman, 
M.~Mashkoori, 
F.~Parhizgar
M.~Rodriguez-Vega,
G.~Schwiete,
J.~Sinova,
H.~Steinberg,
and Junhua Zhang.

The work of E.R. has been supported by  NSF CAREER Grant No. DMR- 1350663, ONR Grants No. ONR-N00014-13-1-0321, No. ONR-N00014-16-1-3158,
ARO Grants No. W911NF-16-1-0387, No. W911NF-18-1-0290, BSF Grant No 2016320, and by the computing facilities at William \& Mary which were provided by contributions from the NSF, the Commonwealth of Virginia Equipment Trust Fund, and ONR.
E.R. also thanks the Aspen Center for Physics, which is supported by National Science Foundation grant PHY-1607611, where part of this work was performed.



\bibliography{vdW_SOC}

\begin{thebibliography}{205}%
\makeatletter
\providecommand \@ifxundefined [1]{%
 \@ifx{#1\undefined}
}%
\providecommand \@ifnum [1]{%
 \ifnum #1\expandafter \@firstoftwo
 \else \expandafter \@secondoftwo
 \fi
}%
\providecommand \@ifx [1]{%
 \ifx #1\expandafter \@firstoftwo
 \else \expandafter \@secondoftwo
 \fi
}%
\providecommand \natexlab [1]{#1}%
\providecommand \enquote  [1]{``#1''}%
\providecommand \bibnamefont  [1]{#1}%
\providecommand \bibfnamefont [1]{#1}%
\providecommand \citenamefont [1]{#1}%
\providecommand \href@noop [0]{\@secondoftwo}%
\providecommand \href [0]{\begingroup \@sanitize@url \@href}%
\providecommand \@href[1]{\@@startlink{#1}\@@href}%
\providecommand \@@href[1]{\endgroup#1\@@endlink}%
\providecommand \@sanitize@url [0]{\catcode `\\12\catcode `\$12\catcode
  `\&12\catcode `\#12\catcode `\^12\catcode `\_12\catcode `\%12\relax}%
\providecommand \@@startlink[1]{}%
\providecommand \@@endlink[0]{}%
\providecommand \url  [0]{\begingroup\@sanitize@url \@url }%
\providecommand \@url [1]{\endgroup\@href {#1}{\urlprefix }}%
\providecommand \urlprefix  [0]{URL }%
\providecommand \Eprint [0]{\href }%
\providecommand \doibase [0]{http://dx.doi.org/}%
\providecommand \selectlanguage [0]{\@gobble}%
\providecommand \bibinfo  [0]{\@secondoftwo}%
\providecommand \bibfield  [0]{\@secondoftwo}%
\providecommand \translation [1]{[#1]}%
\providecommand \BibitemOpen [0]{}%
\providecommand \bibitemStop [0]{}%
\providecommand \bibitemNoStop [0]{.\EOS\space}%
\providecommand \EOS [0]{\spacefactor3000\relax}%
\providecommand \BibitemShut  [1]{\csname bibitem#1\endcsname}%
\let\auto@bib@innerbib\@empty
\bibitem [{\citenamefont {Geim}\ and\ \citenamefont
  {Novoselov}(2007)}]{geim2007}%
  \BibitemOpen
  \bibfield  {author} {\bibinfo {author} {\bibfnamefont {A.~K.}\ \bibnamefont
  {Geim}}\ and\ \bibinfo {author} {\bibfnamefont {K.~S.}\ \bibnamefont
  {Novoselov}},\ }\bibfield  {title} {\enquote {\bibinfo {title} {The rise of
  graphene},}\ }\href@noop {} {\bibfield  {journal} {\bibinfo  {journal}
  {Nature Materials}\ }\textbf {\bibinfo {volume} {6}},\ \bibinfo {pages} {183}
  (\bibinfo {year} {2007})}\BibitemShut {NoStop}%
\bibitem [{\citenamefont {Geim}\ and\ \citenamefont
  {Grigorieva}(2013)}]{geim2013}%
  \BibitemOpen
  \bibfield  {author} {\bibinfo {author} {\bibfnamefont {A.~K.}\ \bibnamefont
  {Geim}}\ and\ \bibinfo {author} {\bibfnamefont {I.~V.}\ \bibnamefont
  {Grigorieva}},\ }\bibfield  {title} {\enquote {\bibinfo {title} {Van der
  {Waals} heterostructures},}\ }\href@noop {} {\bibfield  {journal} {\bibinfo
  {journal} {Nature}\ }\textbf {\bibinfo {volume} {499}},\ \bibinfo {pages}
  {419--425} (\bibinfo {year} {2013})}\BibitemShut {NoStop}%
\bibitem [{\citenamefont {Liu}\ \emph {et~al.}(2016)\citenamefont {Liu},
  \citenamefont {Weiss}, \citenamefont {Duan}, \citenamefont {Cheng},
  \citenamefont {Huang},\ and\ \citenamefont {Duan}}]{liu2016van}%
  \BibitemOpen
  \bibfield  {author} {\bibinfo {author} {\bibfnamefont {Yuan}\ \bibnamefont
  {Liu}}, \bibinfo {author} {\bibfnamefont {Nathan~O}\ \bibnamefont {Weiss}},
  \bibinfo {author} {\bibfnamefont {Xidong}\ \bibnamefont {Duan}}, \bibinfo
  {author} {\bibfnamefont {Hung-Chieh}\ \bibnamefont {Cheng}}, \bibinfo
  {author} {\bibfnamefont {Yu}~\bibnamefont {Huang}}, \ and\ \bibinfo {author}
  {\bibfnamefont {Xiangfeng}\ \bibnamefont {Duan}},\ }\bibfield  {title}
  {\enquote {\bibinfo {title} {Van der {Waals} heterostructures and devices},}\
  }\href@noop {} {\bibfield  {journal} {\bibinfo  {journal} {Nat. Rev. Mat.}\
  }\textbf {\bibinfo {volume} {1}},\ \bibinfo {pages} {16042} (\bibinfo {year}
  {2016})}\BibitemShut {NoStop}%
\bibitem [{\citenamefont {Novoselov}\ \emph {et~al.}(2016)\citenamefont
  {Novoselov}, \citenamefont {Mishchenko}, \citenamefont {Carvalho},\ and\
  \citenamefont {{Castro Neto}}}]{Novoselov2016}%
  \BibitemOpen
  \bibfield  {author} {\bibinfo {author} {\bibfnamefont {K~S}\ \bibnamefont
  {Novoselov}}, \bibinfo {author} {\bibfnamefont {A}~\bibnamefont
  {Mishchenko}}, \bibinfo {author} {\bibfnamefont {A}~\bibnamefont {Carvalho}},
  \ and\ \bibinfo {author} {\bibfnamefont {A~H}\ \bibnamefont {{Castro
  Neto}}},\ }\bibfield  {title} {\enquote {\bibinfo {title} {{2D materials and
  van der {Waals} heterostructures}},}\ }\href
  {http://science.sciencemag.org/content/353/6298/aac9439} {\bibfield
  {journal} {\bibinfo  {journal} {Science}\ }\textbf {\bibinfo {volume}
  {353}},\ \bibinfo {pages} {aac9439} (\bibinfo {year} {2016})}\BibitemShut
  {NoStop}%
\bibitem [{\citenamefont {Novoselov}\ \emph {et~al.}(2004)\citenamefont
  {Novoselov}, \citenamefont {Geim}, \citenamefont {Morozov}, \citenamefont
  {Jiang}, \citenamefont {Zhang}, \citenamefont {Dubonos}, \citenamefont
  {Grigorieva},\ and\ \citenamefont {Firsov}}]{novoselov2004}%
  \BibitemOpen
  \bibfield  {author} {\bibinfo {author} {\bibfnamefont {K.~S.}\ \bibnamefont
  {Novoselov}}, \bibinfo {author} {\bibfnamefont {A.~K.}\ \bibnamefont {Geim}},
  \bibinfo {author} {\bibfnamefont {S.~V.}\ \bibnamefont {Morozov}}, \bibinfo
  {author} {\bibfnamefont {D.}~\bibnamefont {Jiang}}, \bibinfo {author}
  {\bibfnamefont {Y.}~\bibnamefont {Zhang}}, \bibinfo {author} {\bibfnamefont
  {S.~V.}\ \bibnamefont {Dubonos}}, \bibinfo {author} {\bibfnamefont {I.~V.}\
  \bibnamefont {Grigorieva}}, \ and\ \bibinfo {author} {\bibfnamefont {A.~A.}\
  \bibnamefont {Firsov}},\ }\bibfield  {title} {\enquote {\bibinfo {title}
  {Electric field effect in atomically thin carbon films},}\ }\href@noop {}
  {\bibfield  {journal} {\bibinfo  {journal} {Science}\ }\textbf {\bibinfo
  {volume} {306}},\ \bibinfo {pages} {666--669} (\bibinfo {year}
  {2004})}\BibitemShut {NoStop}%
\bibitem [{\citenamefont {Neto}\ \emph {et~al.}(2009)\citenamefont {Neto},
  \citenamefont {Guinea}, \citenamefont {Peres}, \citenamefont {Novoselov},\
  and\ \citenamefont {Geim}}]{castroneto2009}%
  \BibitemOpen
  \bibfield  {author} {\bibinfo {author} {\bibfnamefont {A.~H.~Castro}\
  \bibnamefont {Neto}}, \bibinfo {author} {\bibfnamefont {F.}~\bibnamefont
  {Guinea}}, \bibinfo {author} {\bibfnamefont {N.~M.~R.}\ \bibnamefont
  {Peres}}, \bibinfo {author} {\bibfnamefont {K.~S.}\ \bibnamefont
  {Novoselov}}, \ and\ \bibinfo {author} {\bibfnamefont {A.~K.}\ \bibnamefont
  {Geim}},\ }\bibfield  {title} {\enquote {\bibinfo {title} {The electronic
  properties of graphene},}\ }\href@noop {} {\bibfield  {journal} {\bibinfo
  {journal} {Rev. Mod. Phys.}\ }\textbf {\bibinfo {volume} {81}},\ \bibinfo
  {pages} {109} (\bibinfo {year} {2009})}\BibitemShut {NoStop}%
\bibitem [{\citenamefont {{Das Sarma}}\ \emph {et~al.}(2011)\citenamefont {{Das
  Sarma}}, \citenamefont {Adam}, \citenamefont {Hwang},\ and\ \citenamefont
  {Rossi}}]{dassarma2011}%
  \BibitemOpen
  \bibfield  {author} {\bibinfo {author} {\bibfnamefont {S}~\bibnamefont {{Das
  Sarma}}}, \bibinfo {author} {\bibfnamefont {Shaffique}\ \bibnamefont {Adam}},
  \bibinfo {author} {\bibfnamefont {E~H}\ \bibnamefont {Hwang}}, \ and\
  \bibinfo {author} {\bibfnamefont {Enrico}\ \bibnamefont {Rossi}},\ }\bibfield
   {title} {\enquote {\bibinfo {title} {{Electronic transport in
  two-dimensional graphene}},}\ }\href {\doibase 10.1103/RevModPhys.83.407}
  {\bibfield  {journal} {\bibinfo  {journal} {Reviews of Modern Physics}\
  }\textbf {\bibinfo {volume} {83}},\ \bibinfo {pages} {407} (\bibinfo {year}
  {2011})}\BibitemShut {NoStop}%
\bibitem [{\citenamefont {Cao}\ \emph {et~al.}(2018{\natexlab{a}})\citenamefont
  {Cao}, \citenamefont {Fatemi}, \citenamefont {Fang}, \citenamefont
  {Watanabe}, \citenamefont {Taniguchi}, \citenamefont {Kaxiras},\ and\
  \citenamefont {Jarillo-Herrero}}]{Cao2018}%
  \BibitemOpen
  \bibfield  {author} {\bibinfo {author} {\bibfnamefont {Yuan}\ \bibnamefont
  {Cao}}, \bibinfo {author} {\bibfnamefont {Valla}\ \bibnamefont {Fatemi}},
  \bibinfo {author} {\bibfnamefont {Shiang}\ \bibnamefont {Fang}}, \bibinfo
  {author} {\bibfnamefont {Kenji}\ \bibnamefont {Watanabe}}, \bibinfo {author}
  {\bibfnamefont {Takashi}\ \bibnamefont {Taniguchi}}, \bibinfo {author}
  {\bibfnamefont {Efthimios}\ \bibnamefont {Kaxiras}}, \ and\ \bibinfo {author}
  {\bibfnamefont {Pablo}\ \bibnamefont {Jarillo-Herrero}},\ }\bibfield  {title}
  {\enquote {\bibinfo {title} {Unconventional superconductivity in magic-angle
  graphene superlattices},}\ }\href {\doibase 10.1038/nature26160} {\bibfield
  {journal} {\bibinfo  {journal} {Nature}\ }\textbf {\bibinfo {volume} {556}},\
  \bibinfo {pages} {43} (\bibinfo {year} {2018}{\natexlab{a}})}\BibitemShut
  {NoStop}%
\bibitem [{\citenamefont {Cao}\ \emph {et~al.}(2018{\natexlab{b}})\citenamefont
  {Cao}, \citenamefont {Fatemi}, \citenamefont {Demir}, \citenamefont {Fang},
  \citenamefont {Tomarken}, \citenamefont {Luo}, \citenamefont
  {Sanchez-Yamagishi}, \citenamefont {Watanabe}, \citenamefont {Taniguchi},
  \citenamefont {Kaxiras}, \citenamefont {Ashoori},\ and\ \citenamefont
  {Jarillo-Herrero}}]{Cao2018b}%
  \BibitemOpen
  \bibfield  {author} {\bibinfo {author} {\bibfnamefont {Yuan}\ \bibnamefont
  {Cao}}, \bibinfo {author} {\bibfnamefont {Valla}\ \bibnamefont {Fatemi}},
  \bibinfo {author} {\bibfnamefont {Ahmet}\ \bibnamefont {Demir}}, \bibinfo
  {author} {\bibfnamefont {Shiang}\ \bibnamefont {Fang}}, \bibinfo {author}
  {\bibfnamefont {Spencer~L.}\ \bibnamefont {Tomarken}}, \bibinfo {author}
  {\bibfnamefont {Jason~Y.}\ \bibnamefont {Luo}}, \bibinfo {author}
  {\bibfnamefont {Javier~D.}\ \bibnamefont {Sanchez-Yamagishi}}, \bibinfo
  {author} {\bibfnamefont {Kenji}\ \bibnamefont {Watanabe}}, \bibinfo {author}
  {\bibfnamefont {Takashi}\ \bibnamefont {Taniguchi}}, \bibinfo {author}
  {\bibfnamefont {Efthimios}\ \bibnamefont {Kaxiras}}, \bibinfo {author}
  {\bibfnamefont {Ray~C.}\ \bibnamefont {Ashoori}}, \ and\ \bibinfo {author}
  {\bibfnamefont {Pablo}\ \bibnamefont {Jarillo-Herrero}},\ }\bibfield  {title}
  {\enquote {\bibinfo {title} {Correlated insulator behaviour at half-filling
  in magic-angle graphene superlattices},}\ }\href {\doibase
  10.1038/nature26154} {\bibfield  {journal} {\bibinfo  {journal} {Nature}\
  }\textbf {\bibinfo {volume} {556}},\ \bibinfo {pages} {80} (\bibinfo {year}
  {2018}{\natexlab{b}})}\BibitemShut {NoStop}%
\bibitem [{\citenamefont {Yankowitz}\ \emph {et~al.}(2019)\citenamefont
  {Yankowitz}, \citenamefont {Chen}, \citenamefont {Polshyn}, \citenamefont
  {Zhang}, \citenamefont {Watanabe}, \citenamefont {Taniguchi}, \citenamefont
  {Graf}, \citenamefont {Young},\ and\ \citenamefont {Dean}}]{Yankowitz2019}%
  \BibitemOpen
  \bibfield  {author} {\bibinfo {author} {\bibfnamefont {Matthew}\ \bibnamefont
  {Yankowitz}}, \bibinfo {author} {\bibfnamefont {Shaowen}\ \bibnamefont
  {Chen}}, \bibinfo {author} {\bibfnamefont {Hryhoriy}\ \bibnamefont
  {Polshyn}}, \bibinfo {author} {\bibfnamefont {Yuxuan}\ \bibnamefont {Zhang}},
  \bibinfo {author} {\bibfnamefont {K}~\bibnamefont {Watanabe}}, \bibinfo
  {author} {\bibfnamefont {T}~\bibnamefont {Taniguchi}}, \bibinfo {author}
  {\bibfnamefont {David}\ \bibnamefont {Graf}}, \bibinfo {author}
  {\bibfnamefont {Andrea~F}\ \bibnamefont {Young}}, \ and\ \bibinfo {author}
  {\bibfnamefont {Cory~R}\ \bibnamefont {Dean}},\ }\bibfield  {title} {\enquote
  {\bibinfo {title} {{Tuning superconductivity in twisted bilayer graphene}},}\
  }\href {\doibase 10.1126/science.aav1910} {\bibfield  {journal} {\bibinfo
  {journal} {Science}\ }\textbf {\bibinfo {volume} {363}},\ \bibinfo {pages}
  {1059} (\bibinfo {year} {2019})}\BibitemShut {NoStop}%
\bibitem [{\citenamefont {Zhang}\ \emph
  {et~al.}(2009{\natexlab{a}})\citenamefont {Zhang}, \citenamefont {Liu},
  \citenamefont {Qi}, \citenamefont {Dai}, \citenamefont {Fang},\ and\
  \citenamefont {Zhang}}]{ZhangNatPhys2009}%
  \BibitemOpen
  \bibfield  {author} {\bibinfo {author} {\bibfnamefont {H.}~\bibnamefont
  {Zhang}}, \bibinfo {author} {\bibfnamefont {C.X.}\ \bibnamefont {Liu}},
  \bibinfo {author} {\bibfnamefont {X.L.}\ \bibnamefont {Qi}}, \bibinfo
  {author} {\bibfnamefont {X.}~\bibnamefont {Dai}}, \bibinfo {author}
  {\bibfnamefont {Z.}~\bibnamefont {Fang}}, \ and\ \bibinfo {author}
  {\bibfnamefont {S.C.}\ \bibnamefont {Zhang}},\ }\bibfield  {title} {\enquote
  {\bibinfo {title} {Topological insulators in bi2se3, bi2te3 and sb2te3 with a
  single dirac cone on the surface},}\ }\href@noop {} {\bibfield  {journal}
  {\bibinfo  {journal} {Nature Physics}\ }\textbf {\bibinfo {volume} {5}}
  (\bibinfo {year} {2009}{\natexlab{a}})}\BibitemShut {NoStop}%
\bibitem [{\citenamefont {Hasan}\ and\ \citenamefont
  {Kane}(2010)}]{HasanRMP2010}%
  \BibitemOpen
  \bibfield  {author} {\bibinfo {author} {\bibfnamefont {M.Z.}\ \bibnamefont
  {Hasan}}\ and\ \bibinfo {author} {\bibfnamefont {C.L.}\ \bibnamefont
  {Kane}},\ }\bibfield  {title} {\enquote {\bibinfo {title} {Colloquium:
  Topological insulators},}\ }\href@noop {} {\bibfield  {journal} {\bibinfo
  {journal} {Rev. Mod. Phys.}\ }\textbf {\bibinfo {volume} {82}},\ \bibinfo
  {pages} {3045} (\bibinfo {year} {2010})}\BibitemShut {NoStop}%
\bibitem [{\citenamefont {Qi}\ and\ \citenamefont {Zhang}(2011)}]{qi2011rmp}%
  \BibitemOpen
  \bibfield  {author} {\bibinfo {author} {\bibfnamefont {Xiao-Liang}\
  \bibnamefont {Qi}}\ and\ \bibinfo {author} {\bibfnamefont {Shou-Cheng}\
  \bibnamefont {Zhang}},\ }\bibfield  {title} {\enquote {\bibinfo {title}
  {Topological insulators and superconductors},}\ }\href@noop {} {\bibfield
  {journal} {\bibinfo  {journal} {Reviews of Modern Physics}\ }\textbf
  {\bibinfo {volume} {83}},\ \bibinfo {pages} {1057} (\bibinfo {year}
  {2011})}\BibitemShut {NoStop}%
\bibitem [{\citenamefont {Zhang}\ \emph {et~al.}(2014)\citenamefont {Zhang},
  \citenamefont {Triola},\ and\ \citenamefont {Rossi}}]{jzhang2014}%
  \BibitemOpen
  \bibfield  {author} {\bibinfo {author} {\bibfnamefont {Junhua}\ \bibnamefont
  {Zhang}}, \bibinfo {author} {\bibfnamefont {C.}~\bibnamefont {Triola}}, \
  and\ \bibinfo {author} {\bibfnamefont {E.}~\bibnamefont {Rossi}},\ }\bibfield
   {title} {\enquote {\bibinfo {title} {Ppproximity effect in graphene
  topological{-}insulator heterostructures},}\ }\href
  {https://link.aps.org/doi/10.1103/PhysRevLett.112.096802} {\bibfield
  {journal} {\bibinfo  {journal} {Physical Review Letters}\ }\textbf {\bibinfo
  {volume} {112}},\ \bibinfo {pages} {096802} (\bibinfo {year}
  {2014})}\BibitemShut {NoStop}%
\bibitem [{\citenamefont {Khokhriakov}\ \emph {et~al.}(2018)\citenamefont
  {Khokhriakov}, \citenamefont {Cummings}, \citenamefont {Song}, \citenamefont
  {Vila}, \citenamefont {Karpiak}, \citenamefont {Dankert}, \citenamefont
  {Roche},\ and\ \citenamefont {Dash}}]{khokhriakov2018tailoring}%
  \BibitemOpen
  \bibfield  {author} {\bibinfo {author} {\bibfnamefont {Dmitrii}\ \bibnamefont
  {Khokhriakov}}, \bibinfo {author} {\bibfnamefont {Aron~W}\ \bibnamefont
  {Cummings}}, \bibinfo {author} {\bibfnamefont {Kenan}\ \bibnamefont {Song}},
  \bibinfo {author} {\bibfnamefont {Marc}\ \bibnamefont {Vila}}, \bibinfo
  {author} {\bibfnamefont {Bogdan}\ \bibnamefont {Karpiak}}, \bibinfo {author}
  {\bibfnamefont {Andr{\'e}}\ \bibnamefont {Dankert}}, \bibinfo {author}
  {\bibfnamefont {Stephan}\ \bibnamefont {Roche}}, \ and\ \bibinfo {author}
  {\bibfnamefont {Saroj~P}\ \bibnamefont {Dash}},\ }\bibfield  {title}
  {\enquote {\bibinfo {title} {Tailoring emergent spin phenomena in dirac
  material heterostructures},}\ }\href@noop {} {\bibfield  {journal} {\bibinfo
  {journal} {Sci. Adv.}\ }\textbf {\bibinfo {volume} {4}},\ \bibinfo {pages}
  {eaat9349} (\bibinfo {year} {2018})}\BibitemShut {NoStop}%
\bibitem [{\citenamefont {Triola}\ \emph {et~al.}(2016)\citenamefont {Triola},
  \citenamefont {Badiane}, \citenamefont {Balatsky},\ and\ \citenamefont
  {Rossi}}]{triola2016prl}%
  \BibitemOpen
  \bibfield  {author} {\bibinfo {author} {\bibfnamefont {Christopher}\
  \bibnamefont {Triola}}, \bibinfo {author} {\bibfnamefont {Driss~M.}\
  \bibnamefont {Badiane}}, \bibinfo {author} {\bibfnamefont {Alexander~V.}\
  \bibnamefont {Balatsky}}, \ and\ \bibinfo {author} {\bibfnamefont
  {E.}~\bibnamefont {Rossi}},\ }\bibfield  {title} {\enquote {\bibinfo {title}
  {General conditions for proximity-induced odd-frequency superconductivity in
  two-dimensional electronic systems},}\ }\href
  {https://link.aps.org/doi/10.1103/PhysRevLett.116.257001} {\bibfield
  {journal} {\bibinfo  {journal} {Phys. Rev. Lett.}\ }\textbf {\bibinfo
  {volume} {116}},\ \bibinfo {pages} {257001} (\bibinfo {year}
  {2016})}\BibitemShut {NoStop}%
\bibitem [{\citenamefont {Burch}\ \emph {et~al.}(2018)\citenamefont {Burch},
  \citenamefont {Mandrus},\ and\ \citenamefont {Park}}]{burch2018magnetism}%
  \BibitemOpen
  \bibfield  {author} {\bibinfo {author} {\bibfnamefont {Kenneth~S}\
  \bibnamefont {Burch}}, \bibinfo {author} {\bibfnamefont {David}\ \bibnamefont
  {Mandrus}}, \ and\ \bibinfo {author} {\bibfnamefont {Je-Geun}\ \bibnamefont
  {Park}},\ }\bibfield  {title} {\enquote {\bibinfo {title} {Magnetism in
  two-dimensional van der {Waals} materials},}\ }\href@noop {} {\bibfield
  {journal} {\bibinfo  {journal} {Nature}\ }\textbf {\bibinfo {volume} {563}},\
  \bibinfo {pages} {47} (\bibinfo {year} {2018})}\BibitemShut {NoStop}%
\bibitem [{\citenamefont {Johansen}\ \emph {et~al.}(2019)\citenamefont
  {Johansen}, \citenamefont {Risingg{\aa}rd}, \citenamefont {Sudb{\o}},
  \citenamefont {Linder},\ and\ \citenamefont {Brataas}}]{johansen2019current}%
  \BibitemOpen
  \bibfield  {author} {\bibinfo {author} {\bibfnamefont {{\O}yvind}\
  \bibnamefont {Johansen}}, \bibinfo {author} {\bibfnamefont {Vetle}\
  \bibnamefont {Risingg{\aa}rd}}, \bibinfo {author} {\bibfnamefont {Asle}\
  \bibnamefont {Sudb{\o}}}, \bibinfo {author} {\bibfnamefont {Jacob}\
  \bibnamefont {Linder}}, \ and\ \bibinfo {author} {\bibfnamefont {Arne}\
  \bibnamefont {Brataas}},\ }\bibfield  {title} {\enquote {\bibinfo {title}
  {Current control of magnetism in two-dimensional fe 3 gete 2},}\ }\href@noop
  {} {\bibfield  {journal} {\bibinfo  {journal} {Phys. Rev. Lett.}\ }\textbf
  {\bibinfo {volume} {122}},\ \bibinfo {pages} {217203} (\bibinfo {year}
  {2019})}\BibitemShut {NoStop}%
\bibitem [{\citenamefont {Gong}\ and\ \citenamefont
  {Zhang}(2019)}]{gong2019two}%
  \BibitemOpen
  \bibfield  {author} {\bibinfo {author} {\bibfnamefont {Cheng}\ \bibnamefont
  {Gong}}\ and\ \bibinfo {author} {\bibfnamefont {Xiang}\ \bibnamefont
  {Zhang}},\ }\bibfield  {title} {\enquote {\bibinfo {title} {Two-dimensional
  magnetic crystals and emergent heterostructure devices},}\ }\href@noop {}
  {\bibfield  {journal} {\bibinfo  {journal} {Science}\ }\textbf {\bibinfo
  {volume} {363}},\ \bibinfo {pages} {eaav4450} (\bibinfo {year}
  {2019})}\BibitemShut {NoStop}%
\bibitem [{\citenamefont {Lee}\ \emph {et~al.}(2016)\citenamefont {Lee},
  \citenamefont {Lee}, \citenamefont {Ryoo}, \citenamefont {Kang},
  \citenamefont {Kim}, \citenamefont {Kim}, \citenamefont {Park}, \citenamefont
  {Park},\ and\ \citenamefont {Cheong}}]{lee2016ising}%
  \BibitemOpen
  \bibfield  {author} {\bibinfo {author} {\bibfnamefont {Jae-Ung}\ \bibnamefont
  {Lee}}, \bibinfo {author} {\bibfnamefont {Sungmin}\ \bibnamefont {Lee}},
  \bibinfo {author} {\bibfnamefont {Ji~Hoon}\ \bibnamefont {Ryoo}}, \bibinfo
  {author} {\bibfnamefont {Soonmin}\ \bibnamefont {Kang}}, \bibinfo {author}
  {\bibfnamefont {Tae~Yun}\ \bibnamefont {Kim}}, \bibinfo {author}
  {\bibfnamefont {Pilkwang}\ \bibnamefont {Kim}}, \bibinfo {author}
  {\bibfnamefont {Cheol-Hwan}\ \bibnamefont {Park}}, \bibinfo {author}
  {\bibfnamefont {Je-Geun}\ \bibnamefont {Park}}, \ and\ \bibinfo {author}
  {\bibfnamefont {Hyeonsik}\ \bibnamefont {Cheong}},\ }\bibfield  {title}
  {\enquote {\bibinfo {title} {Ising-type magnetic ordering in atomically thin
  feps3},}\ }\href@noop {} {\bibfield  {journal} {\bibinfo  {journal} {Nano
  Lett.}\ }\textbf {\bibinfo {volume} {16}},\ \bibinfo {pages} {7433--7438}
  (\bibinfo {year} {2016})}\BibitemShut {NoStop}%
\bibitem [{\citenamefont {Gong}\ \emph {et~al.}(2017)\citenamefont {Gong},
  \citenamefont {Li}, \citenamefont {Li}, \citenamefont {Ji}, \citenamefont
  {Stern}, \citenamefont {Xia}, \citenamefont {Cao}, \citenamefont {Bao},
  \citenamefont {Wang}, \citenamefont {Wang} \emph
  {et~al.}}]{gong2017discovery}%
  \BibitemOpen
  \bibfield  {author} {\bibinfo {author} {\bibfnamefont {Cheng}\ \bibnamefont
  {Gong}}, \bibinfo {author} {\bibfnamefont {Lin}\ \bibnamefont {Li}}, \bibinfo
  {author} {\bibfnamefont {Zhenglu}\ \bibnamefont {Li}}, \bibinfo {author}
  {\bibfnamefont {Huiwen}\ \bibnamefont {Ji}}, \bibinfo {author} {\bibfnamefont
  {Alex}\ \bibnamefont {Stern}}, \bibinfo {author} {\bibfnamefont {Yang}\
  \bibnamefont {Xia}}, \bibinfo {author} {\bibfnamefont {Ting}\ \bibnamefont
  {Cao}}, \bibinfo {author} {\bibfnamefont {Wei}\ \bibnamefont {Bao}}, \bibinfo
  {author} {\bibfnamefont {Chenzhe}\ \bibnamefont {Wang}}, \bibinfo {author}
  {\bibfnamefont {Yuan}\ \bibnamefont {Wang}},  \emph {et~al.},\ }\bibfield
  {title} {\enquote {\bibinfo {title} {Discovery of intrinsic ferromagnetism in
  two-dimensional van der {Waals} crystals},}\ }\href@noop {} {\bibfield
  {journal} {\bibinfo  {journal} {Nature}\ }\textbf {\bibinfo {volume} {546}},\
  \bibinfo {pages} {265} (\bibinfo {year} {2017})}\BibitemShut {NoStop}%
\bibitem [{\citenamefont {Huang}\ \emph {et~al.}(2017)\citenamefont {Huang},
  \citenamefont {Clark}, \citenamefont {Navarro-Moratalla}, \citenamefont
  {Klein}, \citenamefont {Cheng}, \citenamefont {Seyler}, \citenamefont
  {Zhong}, \citenamefont {Schmidgall}, \citenamefont {McGuire}, \citenamefont
  {Cobden} \emph {et~al.}}]{huang2017layer}%
  \BibitemOpen
  \bibfield  {author} {\bibinfo {author} {\bibfnamefont {Bevin}\ \bibnamefont
  {Huang}}, \bibinfo {author} {\bibfnamefont {Genevieve}\ \bibnamefont
  {Clark}}, \bibinfo {author} {\bibfnamefont {Efr{\'e}n}\ \bibnamefont
  {Navarro-Moratalla}}, \bibinfo {author} {\bibfnamefont {Dahlia~R}\
  \bibnamefont {Klein}}, \bibinfo {author} {\bibfnamefont {Ran}\ \bibnamefont
  {Cheng}}, \bibinfo {author} {\bibfnamefont {Kyle~L}\ \bibnamefont {Seyler}},
  \bibinfo {author} {\bibfnamefont {Ding}\ \bibnamefont {Zhong}}, \bibinfo
  {author} {\bibfnamefont {Emma}\ \bibnamefont {Schmidgall}}, \bibinfo {author}
  {\bibfnamefont {Michael~A}\ \bibnamefont {McGuire}}, \bibinfo {author}
  {\bibfnamefont {David~H}\ \bibnamefont {Cobden}},  \emph {et~al.},\
  }\bibfield  {title} {\enquote {\bibinfo {title} {Layer-dependent
  ferromagnetism in a van der {Waals} crystal down to the monolayer limit},}\
  }\href@noop {} {\bibfield  {journal} {\bibinfo  {journal} {Nature}\ }\textbf
  {\bibinfo {volume} {546}},\ \bibinfo {pages} {270} (\bibinfo {year}
  {2017})}\BibitemShut {NoStop}%
\bibitem [{\citenamefont {Klein}\ \emph {et~al.}(2018)\citenamefont {Klein},
  \citenamefont {MacNeill}, \citenamefont {Lado}, \citenamefont {Soriano},
  \citenamefont {Navarro-Moratalla}, \citenamefont {Watanabe}, \citenamefont
  {Taniguchi}, \citenamefont {Manni}, \citenamefont {Canfield}, \citenamefont
  {Fern{\'{a}}ndez-Rossier},\ and\ \citenamefont
  {Jarillo-Herrero}}]{Klein2018}%
  \BibitemOpen
  \bibfield  {author} {\bibinfo {author} {\bibfnamefont {D.~R.}\ \bibnamefont
  {Klein}}, \bibinfo {author} {\bibfnamefont {D.}~\bibnamefont {MacNeill}},
  \bibinfo {author} {\bibfnamefont {J.~L.}\ \bibnamefont {Lado}}, \bibinfo
  {author} {\bibfnamefont {D.}~\bibnamefont {Soriano}}, \bibinfo {author}
  {\bibfnamefont {E.}~\bibnamefont {Navarro-Moratalla}}, \bibinfo {author}
  {\bibfnamefont {K.}~\bibnamefont {Watanabe}}, \bibinfo {author}
  {\bibfnamefont {T.}~\bibnamefont {Taniguchi}}, \bibinfo {author}
  {\bibfnamefont {S.}~\bibnamefont {Manni}}, \bibinfo {author} {\bibfnamefont
  {P.}~\bibnamefont {Canfield}}, \bibinfo {author} {\bibfnamefont
  {J.}~\bibnamefont {Fern{\'{a}}ndez-Rossier}}, \ and\ \bibinfo {author}
  {\bibfnamefont {P.}~\bibnamefont {Jarillo-Herrero}},\ }\bibfield  {title}
  {\enquote {\bibinfo {title} {{Probing magnetism in 2D van der {Waals}
  crystalline insulators via electron tunneling}},}\ }\href {\doibase
  10.1126/science.aar3617} {\bibfield  {journal} {\bibinfo  {journal}
  {Science}\ }\textbf {\bibinfo {volume} {360}},\ \bibinfo {pages} {1218--1222}
  (\bibinfo {year} {2018})}\BibitemShut {NoStop}%
\bibitem [{\citenamefont {Bonilla}\ \emph {et~al.}(2018)\citenamefont
  {Bonilla}, \citenamefont {Kolekar}, \citenamefont {Ma}, \citenamefont {Diaz},
  \citenamefont {Kalappattil}, \citenamefont {Das}, \citenamefont {Eggers},
  \citenamefont {Gutierrez}, \citenamefont {Phan},\ and\ \citenamefont
  {Batzill}}]{bonilla2018strong}%
  \BibitemOpen
  \bibfield  {author} {\bibinfo {author} {\bibfnamefont {Manuel}\ \bibnamefont
  {Bonilla}}, \bibinfo {author} {\bibfnamefont {Sadhu}\ \bibnamefont
  {Kolekar}}, \bibinfo {author} {\bibfnamefont {Yujing}\ \bibnamefont {Ma}},
  \bibinfo {author} {\bibfnamefont {Horacio~Coy}\ \bibnamefont {Diaz}},
  \bibinfo {author} {\bibfnamefont {Vijaysankar}\ \bibnamefont {Kalappattil}},
  \bibinfo {author} {\bibfnamefont {Raja}\ \bibnamefont {Das}}, \bibinfo
  {author} {\bibfnamefont {Tatiana}\ \bibnamefont {Eggers}}, \bibinfo {author}
  {\bibfnamefont {Humberto~R}\ \bibnamefont {Gutierrez}}, \bibinfo {author}
  {\bibfnamefont {Manh-Huong}\ \bibnamefont {Phan}}, \ and\ \bibinfo {author}
  {\bibfnamefont {Matthias}\ \bibnamefont {Batzill}},\ }\bibfield  {title}
  {\enquote {\bibinfo {title} {Strong room-temperature ferromagnetism in vse 2
  monolayers on van der {Waals} substrates},}\ }\href@noop {} {\bibfield
  {journal} {\bibinfo  {journal} {Nature Nano.}\ }\textbf {\bibinfo {volume}
  {13}},\ \bibinfo {pages} {289} (\bibinfo {year} {2018})}\BibitemShut
  {NoStop}%
\bibitem [{\citenamefont {O’Hara}\ \emph {et~al.}(2018)\citenamefont
  {O’Hara}, \citenamefont {Zhu}, \citenamefont {Trout}, \citenamefont
  {Ahmed}, \citenamefont {Luo}, \citenamefont {Lee}, \citenamefont {Brenner},
  \citenamefont {Rajan}, \citenamefont {Gupta}, \citenamefont {McComb} \emph
  {et~al.}}]{o2018room}%
  \BibitemOpen
  \bibfield  {author} {\bibinfo {author} {\bibfnamefont {Dante~J}\ \bibnamefont
  {O’Hara}}, \bibinfo {author} {\bibfnamefont {Tiancong}\ \bibnamefont
  {Zhu}}, \bibinfo {author} {\bibfnamefont {Amanda~H}\ \bibnamefont {Trout}},
  \bibinfo {author} {\bibfnamefont {Adam~S}\ \bibnamefont {Ahmed}}, \bibinfo
  {author} {\bibfnamefont {Yunqiu~Kelly}\ \bibnamefont {Luo}}, \bibinfo
  {author} {\bibfnamefont {Choong~Hee}\ \bibnamefont {Lee}}, \bibinfo {author}
  {\bibfnamefont {Mark~R}\ \bibnamefont {Brenner}}, \bibinfo {author}
  {\bibfnamefont {Siddharth}\ \bibnamefont {Rajan}}, \bibinfo {author}
  {\bibfnamefont {Jay~A}\ \bibnamefont {Gupta}}, \bibinfo {author}
  {\bibfnamefont {David~W}\ \bibnamefont {McComb}},  \emph {et~al.},\
  }\bibfield  {title} {\enquote {\bibinfo {title} {Room temperature intrinsic
  ferromagnetism in epitaxial manganese selenide films in the monolayer
  limit},}\ }\href@noop {} {\bibfield  {journal} {\bibinfo  {journal} {Nano
  Lett.}\ }\textbf {\bibinfo {volume} {18}},\ \bibinfo {pages} {3125--3131}
  (\bibinfo {year} {2018})}\BibitemShut {NoStop}%
\bibitem [{\citenamefont {Fei}\ \emph {et~al.}(2018)\citenamefont {Fei},
  \citenamefont {Huang}, \citenamefont {Malinowski}, \citenamefont {Wang},
  \citenamefont {Song}, \citenamefont {Sanchez}, \citenamefont {Yao},
  \citenamefont {Xiao}, \citenamefont {Zhu}, \citenamefont {May} \emph
  {et~al.}}]{fei2018two}%
  \BibitemOpen
  \bibfield  {author} {\bibinfo {author} {\bibfnamefont {Zaiyao}\ \bibnamefont
  {Fei}}, \bibinfo {author} {\bibfnamefont {Bevin}\ \bibnamefont {Huang}},
  \bibinfo {author} {\bibfnamefont {Paul}\ \bibnamefont {Malinowski}}, \bibinfo
  {author} {\bibfnamefont {Wenbo}\ \bibnamefont {Wang}}, \bibinfo {author}
  {\bibfnamefont {Tiancheng}\ \bibnamefont {Song}}, \bibinfo {author}
  {\bibfnamefont {Joshua}\ \bibnamefont {Sanchez}}, \bibinfo {author}
  {\bibfnamefont {Wang}\ \bibnamefont {Yao}}, \bibinfo {author} {\bibfnamefont
  {Di}~\bibnamefont {Xiao}}, \bibinfo {author} {\bibfnamefont {Xiaoyang}\
  \bibnamefont {Zhu}}, \bibinfo {author} {\bibfnamefont {Andrew~F}\
  \bibnamefont {May}},  \emph {et~al.},\ }\bibfield  {title} {\enquote
  {\bibinfo {title} {Two-dimensional itinerant ferromagnetism in atomically
  thin fe 3 gete 2},}\ }\href@noop {} {\bibfield  {journal} {\bibinfo
  {journal} {Nat. Mater.}\ }\textbf {\bibinfo {volume} {17}},\ \bibinfo {pages}
  {778} (\bibinfo {year} {2018})}\BibitemShut {NoStop}%
\bibitem [{\citenamefont {Deng}\ \emph {et~al.}(2018)\citenamefont {Deng},
  \citenamefont {Yu}, \citenamefont {Song}, \citenamefont {Zhang},
  \citenamefont {Wang}, \citenamefont {Sun}, \citenamefont {Yi}, \citenamefont
  {Wu}, \citenamefont {Wu}, \citenamefont {Zhu} \emph {et~al.}}]{deng2018gate}%
  \BibitemOpen
  \bibfield  {author} {\bibinfo {author} {\bibfnamefont {Yujun}\ \bibnamefont
  {Deng}}, \bibinfo {author} {\bibfnamefont {Yijun}\ \bibnamefont {Yu}},
  \bibinfo {author} {\bibfnamefont {Yichen}\ \bibnamefont {Song}}, \bibinfo
  {author} {\bibfnamefont {Jingzhao}\ \bibnamefont {Zhang}}, \bibinfo {author}
  {\bibfnamefont {Nai~Zhou}\ \bibnamefont {Wang}}, \bibinfo {author}
  {\bibfnamefont {Zeyuan}\ \bibnamefont {Sun}}, \bibinfo {author}
  {\bibfnamefont {Yangfan}\ \bibnamefont {Yi}}, \bibinfo {author}
  {\bibfnamefont {Yi~Zheng}\ \bibnamefont {Wu}}, \bibinfo {author}
  {\bibfnamefont {Shiwei}\ \bibnamefont {Wu}}, \bibinfo {author} {\bibfnamefont
  {Junyi}\ \bibnamefont {Zhu}},  \emph {et~al.},\ }\bibfield  {title} {\enquote
  {\bibinfo {title} {Gate-tunable room-temperature ferromagnetism in
  two-dimensional fe 3 gete 2},}\ }\href@noop {} {\bibfield  {journal}
  {\bibinfo  {journal} {Nature}\ }\textbf {\bibinfo {volume} {563}},\ \bibinfo
  {pages} {94} (\bibinfo {year} {2018})}\BibitemShut {NoStop}%
\bibitem [{\citenamefont {Hauser}(1969)}]{hauser1969magnetic}%
  \BibitemOpen
  \bibfield  {author} {\bibinfo {author} {\bibfnamefont {JJ}~\bibnamefont
  {Hauser}},\ }\bibfield  {title} {\enquote {\bibinfo {title} {Magnetic
  proximity effect},}\ }\href@noop {} {\bibfield  {journal} {\bibinfo
  {journal} {Phys. Rev.}\ }\textbf {\bibinfo {volume} {187}},\ \bibinfo {pages}
  {580} (\bibinfo {year} {1969})}\BibitemShut {NoStop}%
\bibitem [{\citenamefont {Lazi{\'c}}\ \emph {et~al.}(2016)\citenamefont
  {Lazi{\'c}}, \citenamefont {Belashchenko},\ and\ \citenamefont
  {{\v{Z}}uti{\'c}}}]{lazic2016effective}%
  \BibitemOpen
  \bibfield  {author} {\bibinfo {author} {\bibfnamefont {Predrag}\ \bibnamefont
  {Lazi{\'c}}}, \bibinfo {author} {\bibfnamefont {KD}~\bibnamefont
  {Belashchenko}}, \ and\ \bibinfo {author} {\bibfnamefont {Igor}\ \bibnamefont
  {{\v{Z}}uti{\'c}}},\ }\bibfield  {title} {\enquote {\bibinfo {title}
  {Effective gating and tunable magnetic proximity effects in two-dimensional
  heterostructures},}\ }\href@noop {} {\bibfield  {journal} {\bibinfo
  {journal} {Phys. Rev. B}\ }\textbf {\bibinfo {volume} {93}},\ \bibinfo
  {pages} {241401} (\bibinfo {year} {2016})}\BibitemShut {NoStop}%
\bibitem [{\citenamefont {Liang}\ \emph {et~al.}(2017)\citenamefont {Liang},
  \citenamefont {Deng}, \citenamefont {Huang}, \citenamefont {Tang},
  \citenamefont {Wang}, \citenamefont {Zhu}, \citenamefont {Qin}, \citenamefont
  {Zhang}, \citenamefont {Peng},\ and\ \citenamefont {Bi}}]{liang2017magnetic}%
  \BibitemOpen
  \bibfield  {author} {\bibinfo {author} {\bibfnamefont {Xiao}\ \bibnamefont
  {Liang}}, \bibinfo {author} {\bibfnamefont {Longjiang}\ \bibnamefont {Deng}},
  \bibinfo {author} {\bibfnamefont {Fei}\ \bibnamefont {Huang}}, \bibinfo
  {author} {\bibfnamefont {Tingting}\ \bibnamefont {Tang}}, \bibinfo {author}
  {\bibfnamefont {Chuangtang}\ \bibnamefont {Wang}}, \bibinfo {author}
  {\bibfnamefont {Yupeng}\ \bibnamefont {Zhu}}, \bibinfo {author}
  {\bibfnamefont {Jun}\ \bibnamefont {Qin}}, \bibinfo {author} {\bibfnamefont
  {Yan}\ \bibnamefont {Zhang}}, \bibinfo {author} {\bibfnamefont
  {Bo}~\bibnamefont {Peng}}, \ and\ \bibinfo {author} {\bibfnamefont {Lei}\
  \bibnamefont {Bi}},\ }\bibfield  {title} {\enquote {\bibinfo {title} {The
  magnetic proximity effect and electrical field tunable valley degeneracy in
  mos 2/eus van der {Waals} heterojunctions},}\ }\href@noop {} {\bibfield
  {journal} {\bibinfo  {journal} {Nanoscale}\ }\textbf {\bibinfo {volume}
  {9}},\ \bibinfo {pages} {9502--9509} (\bibinfo {year} {2017})}\BibitemShut
  {NoStop}%
\bibitem [{\citenamefont {Cort\'es}\ \emph {et~al.}(2019)\citenamefont
  {Cort\'es}, \citenamefont {\'Avalos-Ovando}, \citenamefont {Rosales},
  \citenamefont {Orellana},\ and\ \citenamefont {Ulloa}}]{cortes2019}%
  \BibitemOpen
  \bibfield  {author} {\bibinfo {author} {\bibfnamefont {Natalia}\ \bibnamefont
  {Cort\'es}}, \bibinfo {author} {\bibfnamefont {O.}~\bibnamefont
  {\'Avalos-Ovando}}, \bibinfo {author} {\bibfnamefont {L.}~\bibnamefont
  {Rosales}}, \bibinfo {author} {\bibfnamefont {P.~A.}\ \bibnamefont
  {Orellana}}, \ and\ \bibinfo {author} {\bibfnamefont {S.~E.}\ \bibnamefont
  {Ulloa}},\ }\bibfield  {title} {\enquote {\bibinfo {title} {Tunable
  spin-polarized edge currents in proximitized transition metal
  dichalcogenides},}\ }\href {\doibase 10.1103/PhysRevLett.122.086401}
  {\bibfield  {journal} {\bibinfo  {journal} {Phys. Rev. Lett.}\ }\textbf
  {\bibinfo {volume} {122}},\ \bibinfo {pages} {086401} (\bibinfo {year}
  {2019})}\BibitemShut {NoStop}%
\bibitem [{\citenamefont {Vogt}\ \emph {et~al.}(2012)\citenamefont {Vogt},
  \citenamefont {De~Padova}, \citenamefont {Quaresima}, \citenamefont {Avila},
  \citenamefont {Frantzeskakis}, \citenamefont {Asensio}, \citenamefont
  {Resta}, \citenamefont {Ealet},\ and\ \citenamefont
  {Le~Lay}}]{vogt2012silicene}%
  \BibitemOpen
  \bibfield  {author} {\bibinfo {author} {\bibfnamefont {Patrick}\ \bibnamefont
  {Vogt}}, \bibinfo {author} {\bibfnamefont {Paola}\ \bibnamefont {De~Padova}},
  \bibinfo {author} {\bibfnamefont {Claudio}\ \bibnamefont {Quaresima}},
  \bibinfo {author} {\bibfnamefont {Jose}\ \bibnamefont {Avila}}, \bibinfo
  {author} {\bibfnamefont {Emmanouil}\ \bibnamefont {Frantzeskakis}}, \bibinfo
  {author} {\bibfnamefont {Maria~Carmen}\ \bibnamefont {Asensio}}, \bibinfo
  {author} {\bibfnamefont {Andrea}\ \bibnamefont {Resta}}, \bibinfo {author}
  {\bibfnamefont {B{\'e}n{\'e}dicte}\ \bibnamefont {Ealet}}, \ and\ \bibinfo
  {author} {\bibfnamefont {Guy}\ \bibnamefont {Le~Lay}},\ }\bibfield  {title}
  {\enquote {\bibinfo {title} {Silicene: compelling experimental evidence for
  graphenelike two-dimensional silicon},}\ }\href@noop {} {\bibfield  {journal}
  {\bibinfo  {journal} {Phys. Rev. Lett.}\ }\textbf {\bibinfo {volume} {108}},\
  \bibinfo {pages} {155501} (\bibinfo {year} {2012})}\BibitemShut {NoStop}%
\bibitem [{\citenamefont {Liu}\ \emph {et~al.}(2011)\citenamefont {Liu},
  \citenamefont {Feng},\ and\ \citenamefont {Yao}}]{liu2011quantum}%
  \BibitemOpen
  \bibfield  {author} {\bibinfo {author} {\bibfnamefont {Cheng-Cheng}\
  \bibnamefont {Liu}}, \bibinfo {author} {\bibfnamefont {Wanxiang}\
  \bibnamefont {Feng}}, \ and\ \bibinfo {author} {\bibfnamefont {Yugui}\
  \bibnamefont {Yao}},\ }\bibfield  {title} {\enquote {\bibinfo {title}
  {Quantum spin hall effect in silicene and two-dimensional germanium},}\
  }\href@noop {} {\bibfield  {journal} {\bibinfo  {journal} {Phys. Rev. Lett.}\
  }\textbf {\bibinfo {volume} {107}},\ \bibinfo {pages} {076802} (\bibinfo
  {year} {2011})}\BibitemShut {NoStop}%
\bibitem [{\citenamefont {D{\'a}vila}\ \emph {et~al.}(2014)\citenamefont
  {D{\'a}vila}, \citenamefont {Xian}, \citenamefont {Cahangirov}, \citenamefont
  {Rubio},\ and\ \citenamefont {Le~Lay}}]{davila2014germanene}%
  \BibitemOpen
  \bibfield  {author} {\bibinfo {author} {\bibfnamefont {ME}~\bibnamefont
  {D{\'a}vila}}, \bibinfo {author} {\bibfnamefont {Lede}\ \bibnamefont {Xian}},
  \bibinfo {author} {\bibfnamefont {Seymur}\ \bibnamefont {Cahangirov}},
  \bibinfo {author} {\bibfnamefont {Angel}\ \bibnamefont {Rubio}}, \ and\
  \bibinfo {author} {\bibfnamefont {Guy}\ \bibnamefont {Le~Lay}},\ }\bibfield
  {title} {\enquote {\bibinfo {title} {Germanene: a novel two-dimensional
  germanium allotrope akin to graphene and silicene},}\ }\href@noop {}
  {\bibfield  {journal} {\bibinfo  {journal} {New J. Phys.}\ }\textbf {\bibinfo
  {volume} {16}},\ \bibinfo {pages} {095002} (\bibinfo {year}
  {2014})}\BibitemShut {NoStop}%
\bibitem [{\citenamefont {Zhu}\ \emph {et~al.}(2015)\citenamefont {Zhu},
  \citenamefont {Chen}, \citenamefont {Xu}, \citenamefont {Gao}, \citenamefont
  {Guan}, \citenamefont {Liu}, \citenamefont {Qian}, \citenamefont {Zhang},\
  and\ \citenamefont {Jia}}]{zhu2015epitaxial}%
  \BibitemOpen
  \bibfield  {author} {\bibinfo {author} {\bibfnamefont {Feng-feng}\
  \bibnamefont {Zhu}}, \bibinfo {author} {\bibfnamefont {Wei-jiong}\
  \bibnamefont {Chen}}, \bibinfo {author} {\bibfnamefont {Yong}\ \bibnamefont
  {Xu}}, \bibinfo {author} {\bibfnamefont {Chun-lei}\ \bibnamefont {Gao}},
  \bibinfo {author} {\bibfnamefont {Dan-dan}\ \bibnamefont {Guan}}, \bibinfo
  {author} {\bibfnamefont {Can-hua}\ \bibnamefont {Liu}}, \bibinfo {author}
  {\bibfnamefont {Dong}\ \bibnamefont {Qian}}, \bibinfo {author} {\bibfnamefont
  {Shou-Cheng}\ \bibnamefont {Zhang}}, \ and\ \bibinfo {author} {\bibfnamefont
  {Jin-feng}\ \bibnamefont {Jia}},\ }\bibfield  {title} {\enquote {\bibinfo
  {title} {Epitaxial growth of two-dimensional stanene},}\ }\href@noop {}
  {\bibfield  {journal} {\bibinfo  {journal} {Nat. Mater.}\ }\textbf {\bibinfo
  {volume} {14}},\ \bibinfo {pages} {1020} (\bibinfo {year}
  {2015})}\BibitemShut {NoStop}%
\bibitem [{\citenamefont {Gani}\ \emph {et~al.}(2018)\citenamefont {Gani},
  \citenamefont {Abergel},\ and\ \citenamefont {Rossi}}]{gani2018}%
  \BibitemOpen
  \bibfield  {author} {\bibinfo {author} {\bibfnamefont {Yohanes~S.}\
  \bibnamefont {Gani}}, \bibinfo {author} {\bibfnamefont {D.~S.~L.}\
  \bibnamefont {Abergel}}, \ and\ \bibinfo {author} {\bibfnamefont {Enrico}\
  \bibnamefont {Rossi}},\ }\bibfield  {title} {\enquote {\bibinfo {title}
  {Electronic structure of graphene nanoribbons on hexagonal boron nitride},}\
  }\href {\doibase 10.1103/PhysRevB.98.205415} {\bibfield  {journal} {\bibinfo
  {journal} {Phys. Rev. B}\ }\textbf {\bibinfo {volume} {98}},\ \bibinfo
  {pages} {205415} (\bibinfo {year} {2018})}\BibitemShut {NoStop}%
\bibitem [{\citenamefont {Ko\ifmmode~\acute{s}\else \'{s}\fi{}mider}\ and\
  \citenamefont {Fern\'andez-Rossier}(2013)}]{Kosmider2013a}%
  \BibitemOpen
  \bibfield  {author} {\bibinfo {author} {\bibfnamefont {K.}~\bibnamefont
  {Ko\ifmmode~\acute{s}\else \'{s}\fi{}mider}}\ and\ \bibinfo {author}
  {\bibfnamefont {J.}~\bibnamefont {Fern\'andez-Rossier}},\ }\bibfield  {title}
  {\enquote {\bibinfo {title} {Electronic properties of the
  mos${}_{2}$-ws${}_{2}$ heterojunction},}\ }\href {\doibase
  10.1103/PhysRevB.87.075451} {\bibfield  {journal} {\bibinfo  {journal} {Phys.
  Rev. B}\ }\textbf {\bibinfo {volume} {87}},\ \bibinfo {pages} {075451}
  (\bibinfo {year} {2013})}\BibitemShut {NoStop}%
\bibitem [{\citenamefont {Wang}\ \emph {et~al.}(2017)\citenamefont {Wang},
  \citenamefont {Wang}, \citenamefont {Yao}, \citenamefont {Liu},\ and\
  \citenamefont {Yu}}]{wang2017interlayer}%
  \BibitemOpen
  \bibfield  {author} {\bibinfo {author} {\bibfnamefont {Yong}\ \bibnamefont
  {Wang}}, \bibinfo {author} {\bibfnamefont {Zhan}\ \bibnamefont {Wang}},
  \bibinfo {author} {\bibfnamefont {Wang}\ \bibnamefont {Yao}}, \bibinfo
  {author} {\bibfnamefont {Gui-Bin}\ \bibnamefont {Liu}}, \ and\ \bibinfo
  {author} {\bibfnamefont {Hongyi}\ \bibnamefont {Yu}},\ }\bibfield  {title}
  {\enquote {\bibinfo {title} {Interlayer coupling in commensurate and
  incommensurate bilayer structures of transition-metal dichalcogenides},}\
  }\href@noop {} {\bibfield  {journal} {\bibinfo  {journal} {Phys. Rev. B}\
  }\textbf {\bibinfo {volume} {95}},\ \bibinfo {pages} {115429} (\bibinfo
  {year} {2017})}\BibitemShut {NoStop}%
\bibitem [{\citenamefont {Li}\ \emph {et~al.}(2017)\citenamefont {Li},
  \citenamefont {Cui}, \citenamefont {Ceballos}, \citenamefont {Lane},
  \citenamefont {Qi},\ and\ \citenamefont {Zhao}}]{li2017ultrafast}%
  \BibitemOpen
  \bibfield  {author} {\bibinfo {author} {\bibfnamefont {Yuanyuan}\
  \bibnamefont {Li}}, \bibinfo {author} {\bibfnamefont {Qiannan}\ \bibnamefont
  {Cui}}, \bibinfo {author} {\bibfnamefont {Frank}\ \bibnamefont {Ceballos}},
  \bibinfo {author} {\bibfnamefont {Samuel~D}\ \bibnamefont {Lane}}, \bibinfo
  {author} {\bibfnamefont {Zeming}\ \bibnamefont {Qi}}, \ and\ \bibinfo
  {author} {\bibfnamefont {Hui}\ \bibnamefont {Zhao}},\ }\bibfield  {title}
  {\enquote {\bibinfo {title} {Ultrafast interlayer electron transfer in
  incommensurate transition metal dichalcogenide homobilayers},}\ }\href@noop
  {} {\bibfield  {journal} {\bibinfo  {journal} {Nano Lett.}\ }\textbf
  {\bibinfo {volume} {17}},\ \bibinfo {pages} {6661--6666} (\bibinfo {year}
  {2017})}\BibitemShut {NoStop}%
\bibitem [{\citenamefont {Wu}\ \emph {et~al.}(2019)\citenamefont {Wu},
  \citenamefont {Lovorn}, \citenamefont {Tutuc}, \citenamefont {Martin},\ and\
  \citenamefont {MacDonald}}]{wu2019topological}%
  \BibitemOpen
  \bibfield  {author} {\bibinfo {author} {\bibfnamefont {Fengcheng}\
  \bibnamefont {Wu}}, \bibinfo {author} {\bibfnamefont {Timothy}\ \bibnamefont
  {Lovorn}}, \bibinfo {author} {\bibfnamefont {Emanuel}\ \bibnamefont {Tutuc}},
  \bibinfo {author} {\bibfnamefont {Ivar}\ \bibnamefont {Martin}}, \ and\
  \bibinfo {author} {\bibfnamefont {AH}~\bibnamefont {MacDonald}},\ }\bibfield
  {title} {\enquote {\bibinfo {title} {Topological insulators in twisted
  transition metal dichalcogenide homobilayers},}\ }\href@noop {} {\bibfield
  {journal} {\bibinfo  {journal} {Phys. Rev. Lett.}\ }\textbf {\bibinfo
  {volume} {122}},\ \bibinfo {pages} {086402} (\bibinfo {year}
  {2019})}\BibitemShut {NoStop}%
\bibitem [{\citenamefont {Yu}\ \emph {et~al.}(2019)\citenamefont {Yu},
  \citenamefont {Chen},\ and\ \citenamefont {Yao}}]{yu2019giant}%
  \BibitemOpen
  \bibfield  {author} {\bibinfo {author} {\bibfnamefont {Hongyi}\ \bibnamefont
  {Yu}}, \bibinfo {author} {\bibfnamefont {Mingxing}\ \bibnamefont {Chen}}, \
  and\ \bibinfo {author} {\bibfnamefont {Wang}\ \bibnamefont {Yao}},\
  }\bibfield  {title} {\enquote {\bibinfo {title} {Giant magnetic field from
  moir$\backslash$'e induced berry phase in homobilayer semiconductors},}\
  }\href@noop {} {\bibfield  {journal} {\bibinfo  {journal} {arXiv preprint
  1906.05499}\ } (\bibinfo {year} {2019})}\BibitemShut {NoStop}%
\bibitem [{\citenamefont {Carr}\ \emph {et~al.}(2018)\citenamefont {Carr},
  \citenamefont {Fang}, \citenamefont {Jarillo-Herrero},\ and\ \citenamefont
  {Kaxiras}}]{Carr2018}%
  \BibitemOpen
  \bibfield  {author} {\bibinfo {author} {\bibfnamefont {Stephen}\ \bibnamefont
  {Carr}}, \bibinfo {author} {\bibfnamefont {Shiang}\ \bibnamefont {Fang}},
  \bibinfo {author} {\bibfnamefont {Pablo}\ \bibnamefont {Jarillo-Herrero}}, \
  and\ \bibinfo {author} {\bibfnamefont {Efthimios}\ \bibnamefont {Kaxiras}},\
  }\bibfield  {title} {\enquote {\bibinfo {title} {{Pressure dependence of the
  magic twist angle in graphene superlattices}},}\ }\href {\doibase
  10.1103/PhysRevB.98.085144} {\bibfield  {journal} {\bibinfo  {journal}
  {Physical Review B}\ }\textbf {\bibinfo {volume} {98}},\ \bibinfo {pages}
  {1--5} (\bibinfo {year} {2018})}\BibitemShut {NoStop}%
\bibitem [{\citenamefont {Jiang}\ \emph {et~al.}(2019)\citenamefont {Jiang},
  \citenamefont {Mao}, \citenamefont {Lai}, \citenamefont {Watanabe},
  \citenamefont {Taniguchi}, \citenamefont {Haule},\ and\ \citenamefont
  {Andrei}}]{Jiang2019}%
  \BibitemOpen
  \bibfield  {author} {\bibinfo {author} {\bibfnamefont {Yuhang}\ \bibnamefont
  {Jiang}}, \bibinfo {author} {\bibfnamefont {Jinhai}\ \bibnamefont {Mao}},
  \bibinfo {author} {\bibfnamefont {Xinyuan}\ \bibnamefont {Lai}}, \bibinfo
  {author} {\bibfnamefont {Kenji}\ \bibnamefont {Watanabe}}, \bibinfo {author}
  {\bibfnamefont {Takashi}\ \bibnamefont {Taniguchi}}, \bibinfo {author}
  {\bibfnamefont {Kristjan}\ \bibnamefont {Haule}}, \ and\ \bibinfo {author}
  {\bibfnamefont {Eva~Y.}\ \bibnamefont {Andrei}},\ }\bibfield  {title}
  {\enquote {\bibinfo {title} {{Evidence of charge-ordering and broken
  rotational symmetry in magic angle twisted bilayer graphene}},}\ }\href
  {\doibase 10.1038/s41586-019-1460-4} {\bibfield  {journal} {\bibinfo
  {journal} {Nature}\ }\textbf {\bibinfo {volume} {4}} (\bibinfo {year}
  {2019}),\ 10.1038/s41586-019-1460-4}\BibitemShut {NoStop}%
\bibitem [{\citenamefont {Hunt}\ \emph {et~al.}(2013)\citenamefont {Hunt},
  \citenamefont {Sanchez-Yamagishi}, \citenamefont {Young}, \citenamefont
  {Yankowitz}, \citenamefont {LeRoy}, \citenamefont {Watanabe}, \citenamefont
  {Taniguchi}, \citenamefont {Moon}, \citenamefont {Koshino}, \citenamefont
  {Jarillo-Herrero},\ and\ \citenamefont {Ashoori}}]{hunt2013}%
  \BibitemOpen
  \bibfield  {author} {\bibinfo {author} {\bibfnamefont {B.}~\bibnamefont
  {Hunt}}, \bibinfo {author} {\bibfnamefont {J.~D.}\ \bibnamefont
  {Sanchez-Yamagishi}}, \bibinfo {author} {\bibfnamefont {A.~F.}\ \bibnamefont
  {Young}}, \bibinfo {author} {\bibfnamefont {M.}~\bibnamefont {Yankowitz}},
  \bibinfo {author} {\bibfnamefont {B.~J.}\ \bibnamefont {LeRoy}}, \bibinfo
  {author} {\bibfnamefont {K.}~\bibnamefont {Watanabe}}, \bibinfo {author}
  {\bibfnamefont {T.}~\bibnamefont {Taniguchi}}, \bibinfo {author}
  {\bibfnamefont {P.}~\bibnamefont {Moon}}, \bibinfo {author} {\bibfnamefont
  {M.}~\bibnamefont {Koshino}}, \bibinfo {author} {\bibfnamefont
  {P.}~\bibnamefont {Jarillo-Herrero}}, \ and\ \bibinfo {author} {\bibfnamefont
  {R.~C.}\ \bibnamefont {Ashoori}},\ }\bibfield  {title} {\enquote {\bibinfo
  {title} {Massive dirac fermions and hofstadter butterfly in a van der {Waals}
  heterostructure},}\ }\href@noop {} {\bibfield  {journal} {\bibinfo  {journal}
  {Science}\ }\textbf {\bibinfo {volume} {340}},\ \bibinfo {pages} {1427}
  (\bibinfo {year} {2013})}\BibitemShut {NoStop}%
\bibitem [{\citenamefont {Jung}\ \emph {et~al.}(2015)\citenamefont {Jung},
  \citenamefont {Dasilva}, \citenamefont {Macdonald},\ and\ \citenamefont
  {Adam}}]{Jung2015}%
  \BibitemOpen
  \bibfield  {author} {\bibinfo {author} {\bibfnamefont {Jeil}\ \bibnamefont
  {Jung}}, \bibinfo {author} {\bibfnamefont {Ashley~M.}\ \bibnamefont
  {Dasilva}}, \bibinfo {author} {\bibfnamefont {Allan~H.}\ \bibnamefont
  {Macdonald}}, \ and\ \bibinfo {author} {\bibfnamefont {Shaffique}\
  \bibnamefont {Adam}},\ }\bibfield  {title} {\enquote {\bibinfo {title}
  {{Origin of band gaps in graphene on hexagonal boron nitride}},}\ }\href
  {\doibase 10.1038/ncomms7308} {\bibfield  {journal} {\bibinfo  {journal}
  {Nature Communications}\ }\textbf {\bibinfo {volume} {6}},\ \bibinfo {pages}
  {1--11} (\bibinfo {year} {2015})}\BibitemShut {NoStop}%
\bibitem [{\citenamefont {Yankowitz}\ \emph {et~al.}(2016)\citenamefont
  {Yankowitz}, \citenamefont {Watanabe}, \citenamefont {Taniguchi},
  \citenamefont {San-Jose},\ and\ \citenamefont {LeRoy}}]{Yankowitz2016}%
  \BibitemOpen
  \bibfield  {author} {\bibinfo {author} {\bibfnamefont {Matthew}\ \bibnamefont
  {Yankowitz}}, \bibinfo {author} {\bibfnamefont {K.}~\bibnamefont {Watanabe}},
  \bibinfo {author} {\bibfnamefont {T.}~\bibnamefont {Taniguchi}}, \bibinfo
  {author} {\bibfnamefont {Pablo}\ \bibnamefont {San-Jose}}, \ and\ \bibinfo
  {author} {\bibfnamefont {Brian~J.}\ \bibnamefont {LeRoy}},\ }\bibfield
  {title} {\enquote {\bibinfo {title} {{Pressure-induced commensurate stacking
  of graphene on boron nitride}},}\ }\href {\doibase 10.1038/ncomms13168}
  {\bibfield  {journal} {\bibinfo  {journal} {Nature Communications}\ }\textbf
  {\bibinfo {volume} {7}},\ \bibinfo {pages} {13168} (\bibinfo {year}
  {2016})}\BibitemShut {NoStop}%
\bibitem [{\citenamefont {Lopes~dos Santos}\ \emph {et~al.}(2007)\citenamefont
  {Lopes~dos Santos}, \citenamefont {Peres},\ and\ \citenamefont
  {Castro~Neto}}]{dossantos2007}%
  \BibitemOpen
  \bibfield  {author} {\bibinfo {author} {\bibfnamefont {J.M.B.}\ \bibnamefont
  {Lopes~dos Santos}}, \bibinfo {author} {\bibfnamefont {N.M.R.}\ \bibnamefont
  {Peres}}, \ and\ \bibinfo {author} {\bibfnamefont {A.H.}\ \bibnamefont
  {Castro~Neto}},\ }\bibfield  {title} {\enquote {\bibinfo {title} {Graphene
  bilayer with a twist: {Electronic} structure},}\ }\href@noop {} {\bibfield
  {journal} {\bibinfo  {journal} {Phys. Rev. Lett.}\ }\textbf {\bibinfo
  {volume} {99}},\ \bibinfo {pages} {256802} (\bibinfo {year}
  {2007})}\BibitemShut {NoStop}%
\bibitem [{\citenamefont {Morell}\ \emph {et~al.}(2010)\citenamefont {Morell},
  \citenamefont {Correa}, \citenamefont {Vargas}, \citenamefont {Pacheco},\
  and\ \citenamefont {Barticevic}}]{morell2010}%
  \BibitemOpen
  \bibfield  {author} {\bibinfo {author} {\bibfnamefont {E.~S.}\ \bibnamefont
  {Morell}}, \bibinfo {author} {\bibfnamefont {J.~D.}\ \bibnamefont {Correa}},
  \bibinfo {author} {\bibfnamefont {P.}~\bibnamefont {Vargas}}, \bibinfo
  {author} {\bibfnamefont {M.}~\bibnamefont {Pacheco}}, \ and\ \bibinfo
  {author} {\bibfnamefont {Z.}~\bibnamefont {Barticevic}},\ }\bibfield  {title}
  {\enquote {\bibinfo {title} {Flat bands in slightly twisted bilayer graphene:
  {Tight-binding} calculations},}\ }\href@noop {} {\bibfield  {journal}
  {\bibinfo  {journal} {Phys. Rev. B}\ }\textbf {\bibinfo {volume} {82}},\
  \bibinfo {pages} {121407} (\bibinfo {year} {2010})}\BibitemShut {NoStop}%
\bibitem [{\citenamefont {Bistritzer}\ and\ \citenamefont
  {MacDonald}(2011)}]{bistritzer2011}%
  \BibitemOpen
  \bibfield  {author} {\bibinfo {author} {\bibfnamefont {R.}~\bibnamefont
  {Bistritzer}}\ and\ \bibinfo {author} {\bibfnamefont {A.~H.}\ \bibnamefont
  {MacDonald}},\ }\bibfield  {title} {\enquote {\bibinfo {title} {Moire bands
  in twisted double-layer graphene},}\ }\href@noop {} {\bibfield  {journal}
  {\bibinfo  {journal} {Proc. National Acad. Sciences United States Am.}\
  }\textbf {\bibinfo {volume} {108}},\ \bibinfo {pages} {12233--12237}
  (\bibinfo {year} {2011})}\BibitemShut {NoStop}%
\bibitem [{\citenamefont {Gani}\ \emph {et~al.}(2019)\citenamefont {Gani},
  \citenamefont {Steinberg},\ and\ \citenamefont {Rossi}}]{gani2019}%
  \BibitemOpen
  \bibfield  {author} {\bibinfo {author} {\bibfnamefont {Yohanes~S.}\
  \bibnamefont {Gani}}, \bibinfo {author} {\bibfnamefont {Hadar}\ \bibnamefont
  {Steinberg}}, \ and\ \bibinfo {author} {\bibfnamefont {Enrico}\ \bibnamefont
  {Rossi}},\ }\bibfield  {title} {\enquote {\bibinfo {title}
  {Ssuperconductivity in twisted graphene {NbSe$_2$} heterostructures},}\
  }\href {\doibase 10.1103/PhysRevB.99.235404} {\bibfield  {journal} {\bibinfo
  {journal} {Phys. Rev. B}\ }\textbf {\bibinfo {volume} {99}},\ \bibinfo
  {pages} {235404} (\bibinfo {year} {2019})}\BibitemShut {NoStop}%
\bibitem [{\citenamefont {Steinberg}\ \emph {et~al.}(2015)\citenamefont
  {Steinberg}, \citenamefont {Orona}, \citenamefont {Fatemi}, \citenamefont
  {Sanchez-Yamagishi}, \citenamefont {Watanabe}, \citenamefont {Taniguchi},\
  and\ \citenamefont {Jarillo-Herrero}}]{steinberg2015}%
  \BibitemOpen
  \bibfield  {author} {\bibinfo {author} {\bibfnamefont {H.}~\bibnamefont
  {Steinberg}}, \bibinfo {author} {\bibfnamefont {L.~A.}\ \bibnamefont
  {Orona}}, \bibinfo {author} {\bibfnamefont {V.}~\bibnamefont {Fatemi}},
  \bibinfo {author} {\bibfnamefont {J.~D.}\ \bibnamefont {Sanchez-Yamagishi}},
  \bibinfo {author} {\bibfnamefont {K.}~\bibnamefont {Watanabe}}, \bibinfo
  {author} {\bibfnamefont {T.}~\bibnamefont {Taniguchi}}, \ and\ \bibinfo
  {author} {\bibfnamefont {P.}~\bibnamefont {Jarillo-Herrero}},\ }\bibfield
  {title} {\enquote {\bibinfo {title} {Tunneling in graphene--topological
  insulator hybrid devices},}\ }\href {\doibase 10.1103/PhysRevB.92.241409}
  {\bibfield  {journal} {\bibinfo  {journal} {Phys. Rev. B}\ }\textbf {\bibinfo
  {volume} {92}},\ \bibinfo {pages} {241409} (\bibinfo {year}
  {2015})}\BibitemShut {NoStop}%
\bibitem [{\citenamefont {Bian}\ \emph {et~al.}(2016)\citenamefont {Bian},
  \citenamefont {Chung}, \citenamefont {Chen}, \citenamefont {Liu},
  \citenamefont {Chang}, \citenamefont {Wu}, \citenamefont {Belopolski},
  \citenamefont {Zheng}, \citenamefont {Xu}, \citenamefont {Sanchez},
  \citenamefont {Alidoust}, \citenamefont {Pierce}, \citenamefont {Quilliams},
  \citenamefont {Barletta}, \citenamefont {Lorcy}, \citenamefont {Avila},
  \citenamefont {Chang}, \citenamefont {Lin}, \citenamefont {Jeng},
  \citenamefont {Asensio}, \citenamefont {Chen},\ and\ \citenamefont
  {Hasan}}]{bian2016}%
  \BibitemOpen
  \bibfield  {author} {\bibinfo {author} {\bibfnamefont {Guang}\ \bibnamefont
  {Bian}}, \bibinfo {author} {\bibfnamefont {Ting-Fung}\ \bibnamefont {Chung}},
  \bibinfo {author} {\bibfnamefont {Chaoyu}\ \bibnamefont {Chen}}, \bibinfo
  {author} {\bibfnamefont {Chang}\ \bibnamefont {Liu}}, \bibinfo {author}
  {\bibfnamefont {Tay-Rong}\ \bibnamefont {Chang}}, \bibinfo {author}
  {\bibfnamefont {Tailung}\ \bibnamefont {Wu}}, \bibinfo {author}
  {\bibfnamefont {Ilya}\ \bibnamefont {Belopolski}}, \bibinfo {author}
  {\bibfnamefont {Hao}\ \bibnamefont {Zheng}}, \bibinfo {author} {\bibfnamefont
  {Su-Yang}\ \bibnamefont {Xu}}, \bibinfo {author} {\bibfnamefont {Daniel~S}\
  \bibnamefont {Sanchez}}, \bibinfo {author} {\bibfnamefont {Nasser}\
  \bibnamefont {Alidoust}}, \bibinfo {author} {\bibfnamefont {Jonathan}\
  \bibnamefont {Pierce}}, \bibinfo {author} {\bibfnamefont {Bryson}\
  \bibnamefont {Quilliams}}, \bibinfo {author} {\bibfnamefont {Philip~P}\
  \bibnamefont {Barletta}}, \bibinfo {author} {\bibfnamefont {Stephane}\
  \bibnamefont {Lorcy}}, \bibinfo {author} {\bibfnamefont {Jos{\'{e}}}\
  \bibnamefont {Avila}}, \bibinfo {author} {\bibfnamefont {Guoqing}\
  \bibnamefont {Chang}}, \bibinfo {author} {\bibfnamefont {Hsin}\ \bibnamefont
  {Lin}}, \bibinfo {author} {\bibfnamefont {Horng-Tay}\ \bibnamefont {Jeng}},
  \bibinfo {author} {\bibfnamefont {Maria-Carmen}\ \bibnamefont {Asensio}},
  \bibinfo {author} {\bibfnamefont {Yong~P}\ \bibnamefont {Chen}}, \ and\
  \bibinfo {author} {\bibfnamefont {M~Zahid}\ \bibnamefont {Hasan}},\
  }\bibfield  {title} {\enquote {\bibinfo {title} {{Experimental observation of
  two massless Dirac-fermion gases in graphene-topological insulator
  heterostructure}},}\ }\href {\doibase 10.1088/2053-1583/3/2/021009}
  {\bibfield  {journal} {\bibinfo  {journal} {2D Materials}\ }\textbf {\bibinfo
  {volume} {3}},\ \bibinfo {pages} {021009} (\bibinfo {year}
  {2016})}\BibitemShut {NoStop}%
\bibitem [{\citenamefont {Jin}\ and\ \citenamefont {Jhi}(2013)}]{jin2013}%
  \BibitemOpen
  \bibfield  {author} {\bibinfo {author} {\bibfnamefont {Kyung-Hwan}\
  \bibnamefont {Jin}}\ and\ \bibinfo {author} {\bibfnamefont {Seung-Hoon}\
  \bibnamefont {Jhi}},\ }\bibfield  {title} {\enquote {\bibinfo {title}
  {Proximity-induced giant spin-orbit interaction in epitaxial graphene on a
  topological insulator},}\ }\href@noop {} {\bibfield  {journal} {\bibinfo
  {journal} {Phys. Rev. B}\ }\textbf {\bibinfo {volume} {87}},\ \bibinfo
  {pages} {075442} (\bibinfo {year} {2013})}\BibitemShut {NoStop}%
\bibitem [{\citenamefont {Song}\ \emph {et~al.}(2018)\citenamefont {Song},
  \citenamefont {Soriano}, \citenamefont {Cummings}, \citenamefont {Robles},
  \citenamefont {Ordej{\'o}n},\ and\ \citenamefont {Roche}}]{song2018spin}%
  \BibitemOpen
  \bibfield  {author} {\bibinfo {author} {\bibfnamefont {Kenan}\ \bibnamefont
  {Song}}, \bibinfo {author} {\bibfnamefont {David}\ \bibnamefont {Soriano}},
  \bibinfo {author} {\bibfnamefont {Aron~W}\ \bibnamefont {Cummings}}, \bibinfo
  {author} {\bibfnamefont {Roberto}\ \bibnamefont {Robles}}, \bibinfo {author}
  {\bibfnamefont {Pablo}\ \bibnamefont {Ordej{\'o}n}}, \ and\ \bibinfo {author}
  {\bibfnamefont {Stephan}\ \bibnamefont {Roche}},\ }\bibfield  {title}
  {\enquote {\bibinfo {title} {Spin proximity effects in graphene/topological
  insulator heterostructures},}\ }\href@noop {} {\bibfield  {journal} {\bibinfo
   {journal} {Nano Lett.}\ }\textbf {\bibinfo {volume} {18}},\ \bibinfo {pages}
  {2033--2039} (\bibinfo {year} {2018})}\BibitemShut {NoStop}%
\bibitem [{\citenamefont {Cao}\ \emph {et~al.}(2016)\citenamefont {Cao},
  \citenamefont {Zhang}, \citenamefont {Tang}, \citenamefont {Yang},
  \citenamefont {Sofo}, \citenamefont {Duan},\ and\ \citenamefont
  {Liu}}]{cao2016heavy}%
  \BibitemOpen
  \bibfield  {author} {\bibinfo {author} {\bibfnamefont {Wendong}\ \bibnamefont
  {Cao}}, \bibinfo {author} {\bibfnamefont {Rui-Xing}\ \bibnamefont {Zhang}},
  \bibinfo {author} {\bibfnamefont {Peizhe}\ \bibnamefont {Tang}}, \bibinfo
  {author} {\bibfnamefont {Gang}\ \bibnamefont {Yang}}, \bibinfo {author}
  {\bibfnamefont {Jorge}\ \bibnamefont {Sofo}}, \bibinfo {author}
  {\bibfnamefont {Wenhui}\ \bibnamefont {Duan}}, \ and\ \bibinfo {author}
  {\bibfnamefont {Chao-Xing}\ \bibnamefont {Liu}},\ }\bibfield  {title}
  {\enquote {\bibinfo {title} {Heavy dirac fermions in a graphene/topological
  insulator hetero-junction},}\ }\href@noop {} {\bibfield  {journal} {\bibinfo
  {journal} {2D Materials}\ }\textbf {\bibinfo {volume} {3}},\ \bibinfo {pages}
  {034006} (\bibinfo {year} {2016})}\BibitemShut {NoStop}%
\bibitem [{\citenamefont {Zhang}\ \emph
  {et~al.}(2009{\natexlab{b}})\citenamefont {Zhang}, \citenamefont {Liu},
  \citenamefont {Qi}, \citenamefont {Dai}, \citenamefont {Fang},\ and\
  \citenamefont {Zhang}}]{zhang2009ti}%
  \BibitemOpen
  \bibfield  {author} {\bibinfo {author} {\bibfnamefont {H.~J.}\ \bibnamefont
  {Zhang}}, \bibinfo {author} {\bibfnamefont {C.~X.}\ \bibnamefont {Liu}},
  \bibinfo {author} {\bibfnamefont {X.~L.}\ \bibnamefont {Qi}}, \bibinfo
  {author} {\bibfnamefont {X.}~\bibnamefont {Dai}}, \bibinfo {author}
  {\bibfnamefont {Z.}~\bibnamefont {Fang}}, \ and\ \bibinfo {author}
  {\bibfnamefont {S.~C.}\ \bibnamefont {Zhang}},\ }\bibfield  {title} {\enquote
  {\bibinfo {title} {Topological insulators in {Bi2Se3,} {Bi2Te3} and {Sb2Te3}
  with a single {Dirac} cone on the surface},}\ }\href@noop {} {\bibfield
  {journal} {\bibinfo  {journal} {Nature Phys.}\ }\textbf {\bibinfo {volume}
  {5}},\ \bibinfo {pages} {438} (\bibinfo {year}
  {2009}{\natexlab{b}})}\BibitemShut {NoStop}%
\bibitem [{\citenamefont {Liu}\ \emph {et~al.}(2010)\citenamefont {Liu},
  \citenamefont {Qi}, \citenamefont {Zhang}, \citenamefont {Dai}, \citenamefont
  {Fang},\ and\ \citenamefont {Zhang}}]{liucx2010}%
  \BibitemOpen
  \bibfield  {author} {\bibinfo {author} {\bibfnamefont {C-X.}\ \bibnamefont
  {Liu}}, \bibinfo {author} {\bibfnamefont {X-L.}\ \bibnamefont {Qi}}, \bibinfo
  {author} {\bibfnamefont {H.}~\bibnamefont {Zhang}}, \bibinfo {author}
  {\bibfnamefont {X.}~\bibnamefont {Dai}}, \bibinfo {author} {\bibfnamefont
  {Z.}~\bibnamefont {Fang}}, \ and\ \bibinfo {author} {\bibfnamefont {S-C.}\
  \bibnamefont {Zhang}},\ }\bibfield  {title} {\enquote {\bibinfo {title}
  {Model hamiltonian for topological insulators},}\ }\href@noop {} {\bibfield
  {journal} {\bibinfo  {journal} {Phys. Rev. B}\ }\textbf {\bibinfo {volume}
  {82}},\ \bibinfo {pages} {045122} (\bibinfo {year} {2010})}\BibitemShut
  {NoStop}%
\bibitem [{\citenamefont {Ren}\ \emph {et~al.}(2010)\citenamefont {Ren},
  \citenamefont {Taskin}, \citenamefont {Sasaki}, \citenamefont {Segawa},\ and\
  \citenamefont {Ando}}]{ren2010}%
  \BibitemOpen
  \bibfield  {author} {\bibinfo {author} {\bibfnamefont {Z.}~\bibnamefont
  {Ren}}, \bibinfo {author} {\bibfnamefont {A.~A.}\ \bibnamefont {Taskin}},
  \bibinfo {author} {\bibfnamefont {S.}~\bibnamefont {Sasaki}}, \bibinfo
  {author} {\bibfnamefont {K.}~\bibnamefont {Segawa}}, \ and\ \bibinfo {author}
  {\bibfnamefont {Y.}~\bibnamefont {Ando}},\ }\bibfield  {title} {\enquote
  {\bibinfo {title} {Large bulk resistivity and surface quantum oscillations in
  the topological insulator bi2te2se},}\ }\href@noop {} {\bibfield  {journal}
  {\bibinfo  {journal} {Phys. Rev. B}\ }\textbf {\bibinfo {volume} {82}},\
  \bibinfo {pages} {241306} (\bibinfo {year} {2010})}\BibitemShut {NoStop}%
\bibitem [{\citenamefont {Arakane}\ \emph {et~al.}(2012)\citenamefont
  {Arakane}, \citenamefont {Sato}, \citenamefont {Souma}, \citenamefont
  {Kosaka}, \citenamefont {Nakayama}, \citenamefont {Komatsu}, \citenamefont
  {Takahashi}, \citenamefont {Ren}, \citenamefont {Segawa},\ and\ \citenamefont
  {Ando}}]{arakane2012}%
  \BibitemOpen
  \bibfield  {author} {\bibinfo {author} {\bibfnamefont {T.}~\bibnamefont
  {Arakane}}, \bibinfo {author} {\bibfnamefont {T.}~\bibnamefont {Sato}},
  \bibinfo {author} {\bibfnamefont {S.}~\bibnamefont {Souma}}, \bibinfo
  {author} {\bibfnamefont {K.}~\bibnamefont {Kosaka}}, \bibinfo {author}
  {\bibfnamefont {K.}~\bibnamefont {Nakayama}}, \bibinfo {author}
  {\bibfnamefont {M.}~\bibnamefont {Komatsu}}, \bibinfo {author} {\bibfnamefont
  {T.}~\bibnamefont {Takahashi}}, \bibinfo {author} {\bibfnamefont
  {Z.}~\bibnamefont {Ren}}, \bibinfo {author} {\bibfnamefont {K.}~\bibnamefont
  {Segawa}}, \ and\ \bibinfo {author} {\bibfnamefont {Y.}~\bibnamefont
  {Ando}},\ }\bibfield  {title} {\enquote {\bibinfo {title} {Tunable {Dirac}
  cone in the topological insulator bi2-xsbxte3-ysey},}\ }\href@noop {}
  {\bibfield  {journal} {\bibinfo  {journal} {Nature Comm.}\ }\textbf {\bibinfo
  {volume} {3}},\ \bibinfo {pages} {636} (\bibinfo {year} {2012})}\BibitemShut
  {NoStop}%
\bibitem [{\citenamefont {Xia}\ \emph {et~al.}(2013)\citenamefont {Xia},
  \citenamefont {Ren}, \citenamefont {Sulaev}, \citenamefont {Liu},
  \citenamefont {Shen},\ and\ \citenamefont {Wang}}]{xia2013}%
  \BibitemOpen
  \bibfield  {author} {\bibinfo {author} {\bibfnamefont {B.}~\bibnamefont
  {Xia}}, \bibinfo {author} {\bibfnamefont {P.}~\bibnamefont {Ren}}, \bibinfo
  {author} {\bibfnamefont {A.}~\bibnamefont {Sulaev}}, \bibinfo {author}
  {\bibfnamefont {P.}~\bibnamefont {Liu}}, \bibinfo {author} {\bibfnamefont
  {S.~Q.}\ \bibnamefont {Shen}}, \ and\ \bibinfo {author} {\bibfnamefont
  {L.}~\bibnamefont {Wang}},\ }\bibfield  {title} {\enquote {\bibinfo {title}
  {Indications of surface-dominated transport in single crystalline nanoflake
  devices of topological insulator bi1.5sb0.5te1.8se1.2},}\ }\href@noop {}
  {\bibfield  {journal} {\bibinfo  {journal} {Phys. Rev. B}\ }\textbf {\bibinfo
  {volume} {87}},\ \bibinfo {pages} {085442} (\bibinfo {year}
  {2013})}\BibitemShut {NoStop}%
\bibitem [{\citenamefont {Segawa}\ \emph {et~al.}(2012)\citenamefont {Segawa},
  \citenamefont {Ren}, \citenamefont {Sasaki}, \citenamefont {Tsuda},
  \citenamefont {Kuwabata},\ and\ \citenamefont {Ando}}]{segawa2012}%
  \BibitemOpen
  \bibfield  {author} {\bibinfo {author} {\bibfnamefont {K.}~\bibnamefont
  {Segawa}}, \bibinfo {author} {\bibfnamefont {Z.}~\bibnamefont {Ren}},
  \bibinfo {author} {\bibfnamefont {S.}~\bibnamefont {Sasaki}}, \bibinfo
  {author} {\bibfnamefont {T.}~\bibnamefont {Tsuda}}, \bibinfo {author}
  {\bibfnamefont {S.}~\bibnamefont {Kuwabata}}, \ and\ \bibinfo {author}
  {\bibfnamefont {Y.}~\bibnamefont {Ando}},\ }\bibfield  {title} {\enquote
  {\bibinfo {title} {Ambipolar transport in bulk crystals of a topological
  insulator by gating with ionic liquid},}\ }\href@noop {} {\bibfield
  {journal} {\bibinfo  {journal} {Phys. Rev. B}\ }\textbf {\bibinfo {volume}
  {86}},\ \bibinfo {pages} {075306} (\bibinfo {year} {2012})}\BibitemShut
  {NoStop}%
\bibitem [{\citenamefont {Xu}\ \emph {et~al.}(2014{\natexlab{a}})\citenamefont
  {Xu}, \citenamefont {Miotkowski}, \citenamefont {Liu}, \citenamefont {Tian},
  \citenamefont {Nam}, \citenamefont {Alidoust}, \citenamefont {Hu},
  \citenamefont {Shih}, \citenamefont {Hasan},\ and\ \citenamefont
  {Chen}}]{xu2014d}%
  \BibitemOpen
  \bibfield  {author} {\bibinfo {author} {\bibfnamefont {Yang}\ \bibnamefont
  {Xu}}, \bibinfo {author} {\bibfnamefont {Ireneusz}\ \bibnamefont
  {Miotkowski}}, \bibinfo {author} {\bibfnamefont {Chang}\ \bibnamefont {Liu}},
  \bibinfo {author} {\bibfnamefont {Jifa}\ \bibnamefont {Tian}}, \bibinfo
  {author} {\bibfnamefont {Hyoungdo}\ \bibnamefont {Nam}}, \bibinfo {author}
  {\bibfnamefont {Nasser}\ \bibnamefont {Alidoust}}, \bibinfo {author}
  {\bibfnamefont {Jiuning}\ \bibnamefont {Hu}}, \bibinfo {author}
  {\bibfnamefont {Chih-Kang}\ \bibnamefont {Shih}}, \bibinfo {author}
  {\bibfnamefont {M.~Zahid}\ \bibnamefont {Hasan}}, \ and\ \bibinfo {author}
  {\bibfnamefont {Yong~P.}\ \bibnamefont {Chen}},\ }\bibfield  {title}
  {\enquote {\bibinfo {title} {{Observation of topological surface state
  quantum Hall effect in an intrinsic three-dimensional topological
  insulator}},}\ }\href {\doibase 10.1038/nphys3140} {\bibfield  {journal}
  {\bibinfo  {journal} {Nature Physics}\ }\textbf {\bibinfo {volume} {10}},\
  \bibinfo {pages} {956} (\bibinfo {year} {2014}{\natexlab{a}})}\BibitemShut
  {NoStop}%
\bibitem [{\citenamefont {Durand}\ \emph {et~al.}(2016)\citenamefont {Durand},
  \citenamefont {Zhang}, \citenamefont {Hus}, \citenamefont {Ma}, \citenamefont
  {McGuire}, \citenamefont {Xu}, \citenamefont {Cao}, \citenamefont
  {Miotkowski}, \citenamefont {Chen},\ and\ \citenamefont {Li}}]{durand2016}%
  \BibitemOpen
  \bibfield  {author} {\bibinfo {author} {\bibfnamefont {Corentin}\
  \bibnamefont {Durand}}, \bibinfo {author} {\bibfnamefont {X.-G.}\
  \bibnamefont {Zhang}}, \bibinfo {author} {\bibfnamefont {Saban~M.}\
  \bibnamefont {Hus}}, \bibinfo {author} {\bibfnamefont {Chuanxu}\ \bibnamefont
  {Ma}}, \bibinfo {author} {\bibfnamefont {Michael~A.}\ \bibnamefont
  {McGuire}}, \bibinfo {author} {\bibfnamefont {Yang}\ \bibnamefont {Xu}},
  \bibinfo {author} {\bibfnamefont {Helin}\ \bibnamefont {Cao}}, \bibinfo
  {author} {\bibfnamefont {Ireneusz}\ \bibnamefont {Miotkowski}}, \bibinfo
  {author} {\bibfnamefont {Yong~P.}\ \bibnamefont {Chen}}, \ and\ \bibinfo
  {author} {\bibfnamefont {An-Ping}\ \bibnamefont {Li}},\ }\bibfield  {title}
  {\enquote {\bibinfo {title} {Differentiation of surface and bulk
  conductivities in topological insulators via four-probe spectroscopy},}\
  }\href@noop {} {\bibfield  {journal} {\bibinfo  {journal} {Nano Letters}\
  }\textbf {\bibinfo {volume} {16}},\ \bibinfo {pages} {2213--2220} (\bibinfo
  {year} {2016})}\BibitemShut {NoStop}%
\bibitem [{\citenamefont {Xu}\ \emph {et~al.}(2016)\citenamefont {Xu},
  \citenamefont {Miotkowski},\ and\ \citenamefont {Chen}}]{xu2016}%
  \BibitemOpen
  \bibfield  {author} {\bibinfo {author} {\bibfnamefont {Yang}\ \bibnamefont
  {Xu}}, \bibinfo {author} {\bibfnamefont {Ireneusz}\ \bibnamefont
  {Miotkowski}}, \ and\ \bibinfo {author} {\bibfnamefont {Yong~P.}\
  \bibnamefont {Chen}},\ }\bibfield  {title} {\enquote {\bibinfo {title}
  {Quantum transport of two-species dirac fermions in dual-gated
  three-dimensional topological insulators},}\ }\href
  {http://dx.doi.org/10.1038/ncomms11434} {\bibfield  {journal} {\bibinfo
  {journal} {Nat Commun}\ }\textbf {\bibinfo {volume} {7}} (\bibinfo {year}
  {2016})}\BibitemShut {NoStop}%
\bibitem [{\citenamefont {Nomura}\ and\ \citenamefont
  {MacDonald}(2006)}]{nomura2006}%
  \BibitemOpen
  \bibfield  {author} {\bibinfo {author} {\bibfnamefont {K.}~\bibnamefont
  {Nomura}}\ and\ \bibinfo {author} {\bibfnamefont {A.~H.}\ \bibnamefont
  {MacDonald}},\ }\bibfield  {title} {\enquote {\bibinfo {title} {Quantum hall
  ferromagnetism in graphene},}\ }\href@noop {} {\bibfield  {journal} {\bibinfo
   {journal} {Phys. Rev. Lett.}\ }\textbf {\bibinfo {volume} {96}},\ \bibinfo
  {pages} {256602} (\bibinfo {year} {2006})}\BibitemShut {NoStop}%
\bibitem [{\citenamefont {Adam}\ \emph {et~al.}(2007)\citenamefont {Adam},
  \citenamefont {Hwang}, \citenamefont {Galitski},\ and\ \citenamefont {{\mbox
  Das Sarma}}}]{adam2007}%
  \BibitemOpen
  \bibfield  {author} {\bibinfo {author} {\bibfnamefont {S.}~\bibnamefont
  {Adam}}, \bibinfo {author} {\bibfnamefont {E.~H.}\ \bibnamefont {Hwang}},
  \bibinfo {author} {\bibfnamefont {V.~M.}\ \bibnamefont {Galitski}}, \ and\
  \bibinfo {author} {\bibfnamefont {S.}~\bibnamefont {{\mbox Das Sarma}}},\
  }\bibfield  {title} {\enquote {\bibinfo {title} {A self-consistent theory for
  graphene transport},}\ }\href@noop {} {\bibfield  {journal} {\bibinfo
  {journal} {Proc. Natl. Acad. Sci. USA}\ }\textbf {\bibinfo {volume} {104}},\
  \bibinfo {pages} {18392} (\bibinfo {year} {2007})}\BibitemShut {NoStop}%
\bibitem [{\citenamefont {Rossi}\ \emph {et~al.}(2009)\citenamefont {Rossi},
  \citenamefont {Adam},\ and\ \citenamefont {Das~Sarma}}]{rossi2009}%
  \BibitemOpen
  \bibfield  {author} {\bibinfo {author} {\bibfnamefont {Enrico}\ \bibnamefont
  {Rossi}}, \bibinfo {author} {\bibfnamefont {S.}~\bibnamefont {Adam}}, \ and\
  \bibinfo {author} {\bibfnamefont {S.}~\bibnamefont {Das~Sarma}},\ }\bibfield
  {title} {\enquote {\bibinfo {title} {Effective medium theory for disordered
  two-dimensional graphene},}\ }\href {\doibase 10.1103/PhysRevB.79.245423}
  {\bibfield  {journal} {\bibinfo  {journal} {Phys. Rev. B}\ }\textbf {\bibinfo
  {volume} {79}},\ \bibinfo {pages} {245423} (\bibinfo {year}
  {2009})}\BibitemShut {NoStop}%
\bibitem [{\citenamefont {{Das Sarma}}\ \emph {et~al.}(2010)\citenamefont {{Das
  Sarma}}, \citenamefont {Hwang},\ and\ \citenamefont {Rossi}}]{dassarma2010}%
  \BibitemOpen
  \bibfield  {author} {\bibinfo {author} {\bibfnamefont {S.}~\bibnamefont {{Das
  Sarma}}}, \bibinfo {author} {\bibfnamefont {E.~H.}\ \bibnamefont {Hwang}}, \
  and\ \bibinfo {author} {\bibfnamefont {E.}~\bibnamefont {Rossi}},\ }\bibfield
   {title} {\enquote {\bibinfo {title} {{Theory of carrier transport in bilayer
  graphene}},}\ }\href {\doibase 10.1103/PhysRevB.81.161407} {\bibfield
  {journal} {\bibinfo  {journal} {Physical Review B}\ }\textbf {\bibinfo
  {volume} {81}},\ \bibinfo {pages} {161407} (\bibinfo {year}
  {2010})}\BibitemShut {NoStop}%
\bibitem [{\citenamefont {Li}\ \emph {et~al.}(2011)\citenamefont {Li},
  \citenamefont {Hwang}, \citenamefont {Rossi},\ and\ \citenamefont {{Das
  Sarma}}}]{li2011}%
  \BibitemOpen
  \bibfield  {author} {\bibinfo {author} {\bibfnamefont {Qiuzi}\ \bibnamefont
  {Li}}, \bibinfo {author} {\bibfnamefont {E.~H.}\ \bibnamefont {Hwang}},
  \bibinfo {author} {\bibfnamefont {E.}~\bibnamefont {Rossi}}, \ and\ \bibinfo
  {author} {\bibfnamefont {S.}~\bibnamefont {{Das Sarma}}},\ }\bibfield
  {title} {\enquote {\bibinfo {title} {{Theory of 2D Transport in Graphene for
  Correlated Disorder}},}\ }\href {\doibase 10.1103/PhysRevLett.107.156601}
  {\bibfield  {journal} {\bibinfo  {journal} {Physical Review Letters}\
  }\textbf {\bibinfo {volume} {107}},\ \bibinfo {pages} {156601} (\bibinfo
  {year} {2011})}\BibitemShut {NoStop}%
\bibitem [{\citenamefont {Rossi}\ and\ \citenamefont {{Das
  Sarma}}(2011)}]{rossi2011}%
  \BibitemOpen
  \bibfield  {author} {\bibinfo {author} {\bibfnamefont {E.}~\bibnamefont
  {Rossi}}\ and\ \bibinfo {author} {\bibfnamefont {S.}~\bibnamefont {{Das
  Sarma}}},\ }\bibfield  {title} {\enquote {\bibinfo {title} {{Inhomogenous
  Electronic Structure, Transport Gap, and Percolation Threshold in Disordered
  Bilayer Graphene}},}\ }\href {\doibase 10.1103/PhysRevLett.107.155502}
  {\bibfield  {journal} {\bibinfo  {journal} {Physical Review Letters}\
  }\textbf {\bibinfo {volume} {107}},\ \bibinfo {pages} {155502} (\bibinfo
  {year} {2011})}\BibitemShut {NoStop}%
\bibitem [{\citenamefont {Rossi}\ \emph {et~al.}(2012)\citenamefont {Rossi},
  \citenamefont {Bardarson}, \citenamefont {Fuhrer},\ and\ \citenamefont {{Das
  Sarma}}}]{rossi2012}%
  \BibitemOpen
  \bibfield  {author} {\bibinfo {author} {\bibfnamefont {E}~\bibnamefont
  {Rossi}}, \bibinfo {author} {\bibfnamefont {J~H}\ \bibnamefont {Bardarson}},
  \bibinfo {author} {\bibfnamefont {M~S}\ \bibnamefont {Fuhrer}}, \ and\
  \bibinfo {author} {\bibfnamefont {S}~\bibnamefont {{Das Sarma}}},\ }\bibfield
   {title} {\enquote {\bibinfo {title} {{Universal Conductance Fluctuations in
  Dirac Materials in the Presence of Long-range Disorder}},}\ }\href
  {https://link.aps.org/doi/10.1103/PhysRevLett.109.096801} {\bibfield
  {journal} {\bibinfo  {journal} {Physical Review Letters}\ }\textbf {\bibinfo
  {volume} {109}},\ \bibinfo {pages} {096801} (\bibinfo {year}
  {2012})}\BibitemShut {NoStop}%
\bibitem [{\citenamefont {Li}\ \emph {et~al.}(2012{\natexlab{a}})\citenamefont
  {Li}, \citenamefont {Hwang},\ and\ \citenamefont {Rossi}}]{LiQiuzi2012}%
  \BibitemOpen
  \bibfield  {author} {\bibinfo {author} {\bibfnamefont {Qiuzi}\ \bibnamefont
  {Li}}, \bibinfo {author} {\bibfnamefont {E.H.}\ \bibnamefont {Hwang}}, \ and\
  \bibinfo {author} {\bibfnamefont {E.}~\bibnamefont {Rossi}},\ }\bibfield
  {title} {\enquote {\bibinfo {title} {{Effect of charged impurity correlations
  on transport in monolayer and bilayer graphene}},}\ }\href
  {http://dx.doi.org/10.1016/j.ssc.2012.04.053} {\bibfield  {journal} {\bibinfo
   {journal} {Solid State Communications}\ }\textbf {\bibinfo {volume} {152}},\
  \bibinfo {pages} {1390} (\bibinfo {year} {2012}{\natexlab{a}})}\BibitemShut
  {NoStop}%
\bibitem [{\citenamefont {Burkov}\ \emph {et~al.}(2004)\citenamefont {Burkov},
  \citenamefont {N\'u\~nez},\ and\ \citenamefont {MacDonald}}]{burkov2003}%
  \BibitemOpen
  \bibfield  {author} {\bibinfo {author} {\bibfnamefont {A.~A.}\ \bibnamefont
  {Burkov}}, \bibinfo {author} {\bibfnamefont {Alvaro~S.}\ \bibnamefont
  {N\'u\~nez}}, \ and\ \bibinfo {author} {\bibfnamefont {A.~H.}\ \bibnamefont
  {MacDonald}},\ }\bibfield  {title} {\enquote {\bibinfo {title} {Theory of
  spin-charge-coupled transport in a two-dimensional electron gas with rashba
  spin-orbit interactions},}\ }\href@noop {} {\bibfield  {journal} {\bibinfo
  {journal} {Phys. Rev. B}\ }\textbf {\bibinfo {volume} {70}},\ \bibinfo
  {pages} {155308} (\bibinfo {year} {2004})}\BibitemShut {NoStop}%
\bibitem [{\citenamefont {Burkov}\ and\ \citenamefont
  {Hawthorn}(2010)}]{burkov2010}%
  \BibitemOpen
  \bibfield  {author} {\bibinfo {author} {\bibfnamefont {A.~A.}\ \bibnamefont
  {Burkov}}\ and\ \bibinfo {author} {\bibfnamefont {D.~G.}\ \bibnamefont
  {Hawthorn}},\ }\bibfield  {title} {\enquote {\bibinfo {title} {Spin and
  {Charge} {Transport} on the {Surface} of a {Topological} insulator},}\
  }\href@noop {} {\bibfield  {journal} {\bibinfo  {journal} {Phys. Rev. Lett.}\
  }\textbf {\bibinfo {volume} {105}},\ \bibinfo {pages} {066802} (\bibinfo
  {year} {2010})}\BibitemShut {NoStop}%
\bibitem [{\citenamefont {Sinova}\ \emph {et~al.}(2015)\citenamefont {Sinova},
  \citenamefont {Valenzuela}, \citenamefont {Wunderlich}, \citenamefont
  {Back},\ and\ \citenamefont {Jungwirth}}]{sinova2015}%
  \BibitemOpen
  \bibfield  {author} {\bibinfo {author} {\bibfnamefont {J.}~\bibnamefont
  {Sinova}}, \bibinfo {author} {\bibfnamefont {S.~O.}\ \bibnamefont
  {Valenzuela}}, \bibinfo {author} {\bibfnamefont {J.}~\bibnamefont
  {Wunderlich}}, \bibinfo {author} {\bibfnamefont {C.~H.}\ \bibnamefont
  {Back}}, \ and\ \bibinfo {author} {\bibfnamefont {T.}~\bibnamefont
  {Jungwirth}},\ }\bibfield  {title} {\enquote {\bibinfo {title} {Spin {Hall}
  effects},}\ }\href@noop {} {\bibfield  {journal} {\bibinfo  {journal} {Rev.
  Modern Phys.}\ }\textbf {\bibinfo {volume} {87}},\ \bibinfo {pages}
  {1213--1259} (\bibinfo {year} {2015})}\BibitemShut {NoStop}%
\bibitem [{\citenamefont {Hwang}\ and\ \citenamefont {{Das
  Sarma}}(2007)}]{hwang2006b}%
  \BibitemOpen
  \bibfield  {author} {\bibinfo {author} {\bibfnamefont {E.~H.}\ \bibnamefont
  {Hwang}}\ and\ \bibinfo {author} {\bibfnamefont {S.}~\bibnamefont {{Das
  Sarma}}},\ }\bibfield  {title} {\enquote {\bibinfo {title} {Dielectric
  function, screening and plasmons in 2d graphene},}\ }\href@noop {} {\bibfield
   {journal} {\bibinfo  {journal} {Phys. Rev. B}\ }\textbf {\bibinfo {volume}
  {75}},\ \bibinfo {pages} {205418} (\bibinfo {year} {2007})}\BibitemShut
  {NoStop}%
\bibitem [{\citenamefont {Shklovskii}(2007)}]{shklovskii2007}%
  \BibitemOpen
  \bibfield  {author} {\bibinfo {author} {\bibfnamefont {B.~I.}\ \bibnamefont
  {Shklovskii}},\ }\bibfield  {title} {\enquote {\bibinfo {title} {Simple model
  of coulomb disorder and screening in graphene},}\ }\href@noop {} {\bibfield
  {journal} {\bibinfo  {journal} {Phys. Rev. B}\ }\textbf {\bibinfo {volume}
  {76}},\ \bibinfo {pages} {233411} (\bibinfo {year} {2007})}\BibitemShut
  {NoStop}%
\bibitem [{\citenamefont {Borghi}\ \emph {et~al.}(2009)\citenamefont {Borghi},
  \citenamefont {Polini}, \citenamefont {Asgari},\ and\ \citenamefont
  {MacDonald}}]{borghi2009}%
  \BibitemOpen
  \bibfield  {author} {\bibinfo {author} {\bibfnamefont {Giovanni}\
  \bibnamefont {Borghi}}, \bibinfo {author} {\bibfnamefont {Marco}\
  \bibnamefont {Polini}}, \bibinfo {author} {\bibfnamefont {Reza}\ \bibnamefont
  {Asgari}}, \ and\ \bibinfo {author} {\bibfnamefont {A.~H.}\ \bibnamefont
  {MacDonald}},\ }\bibfield  {title} {\enquote {\bibinfo {title} {Dynamical
  response functions and collective modes of bilayer graphene},}\ }\href@noop
  {} {\bibfield  {journal} {\bibinfo  {journal} {Phys. Rev. B}\ }\textbf
  {\bibinfo {volume} {80}},\ \bibinfo {pages} {241402} (\bibinfo {year}
  {2009})}\BibitemShut {NoStop}%
\bibitem [{\citenamefont {Abergel}\ \emph {et~al.}(2012)\citenamefont
  {Abergel}, \citenamefont {Rossi},\ and\ \citenamefont {{Das
  Sarma}}}]{abergel2012}%
  \BibitemOpen
  \bibfield  {author} {\bibinfo {author} {\bibfnamefont {D~S~L}\ \bibnamefont
  {Abergel}}, \bibinfo {author} {\bibfnamefont {E}~\bibnamefont {Rossi}}, \
  and\ \bibinfo {author} {\bibfnamefont {S}~\bibnamefont {{Das Sarma}}},\
  }\bibfield  {title} {\enquote {\bibinfo {title} {{Inhomogeneity and nonlinear
  screening in gapped bilayer graphene}},}\ }\href
  {https://link.aps.org/doi/10.1103/PhysRevB.86.155447} {\bibfield  {journal}
  {\bibinfo  {journal} {Physical Review B}\ }\textbf {\bibinfo {volume} {86}},\
  \bibinfo {pages} {155447} (\bibinfo {year} {2012})}\BibitemShut {NoStop}%
\bibitem [{\citenamefont {Triola}\ and\ \citenamefont
  {Rossi}(2012)}]{triola2012}%
  \BibitemOpen
  \bibfield  {author} {\bibinfo {author} {\bibfnamefont {C}~\bibnamefont
  {Triola}}\ and\ \bibinfo {author} {\bibfnamefont {E}~\bibnamefont {Rossi}},\
  }\bibfield  {title} {\enquote {\bibinfo {title} {{Screening and collective
  modes in gapped bilayer graphene}},}\ }\href
  {https://link.aps.org/doi/10.1103/PhysRevB.86.161408} {\bibfield  {journal}
  {\bibinfo  {journal} {Physical Review B}\ }\textbf {\bibinfo {volume} {86}},\
  \bibinfo {pages} {161408} (\bibinfo {year} {2012})}\BibitemShut {NoStop}%
\bibitem [{\citenamefont {Lu}\ \emph {et~al.}(2016)\citenamefont {Lu},
  \citenamefont {Rodriguez-Vega}, \citenamefont {Li}, \citenamefont
  {Luican-Mayer}, \citenamefont {Watanabe}, \citenamefont {Taniguchi},
  \citenamefont {Rossi},\ and\ \citenamefont {Andrei}}]{lu2016}%
  \BibitemOpen
  \bibfield  {author} {\bibinfo {author} {\bibfnamefont {Chih-Pin}\
  \bibnamefont {Lu}}, \bibinfo {author} {\bibfnamefont {Martin}\ \bibnamefont
  {Rodriguez-Vega}}, \bibinfo {author} {\bibfnamefont {Guohong}\ \bibnamefont
  {Li}}, \bibinfo {author} {\bibfnamefont {Adina}\ \bibnamefont
  {Luican-Mayer}}, \bibinfo {author} {\bibfnamefont {Kenji}\ \bibnamefont
  {Watanabe}}, \bibinfo {author} {\bibfnamefont {Takashi}\ \bibnamefont
  {Taniguchi}}, \bibinfo {author} {\bibfnamefont {Enrico}\ \bibnamefont
  {Rossi}}, \ and\ \bibinfo {author} {\bibfnamefont {Eva~Y}\ \bibnamefont
  {Andrei}},\ }\bibfield  {title} {\enquote {\bibinfo {title} {{Local, global,
  and nonlinear screening in twisted double-layer graphene}},}\ }\href
  {\doibase 10.1073/pnas.1606278113} {\bibfield  {journal} {\bibinfo  {journal}
  {Proceedings of the National Academy of Sciences}\ }\textbf {\bibinfo
  {volume} {113}},\ \bibinfo {pages} {6623--6628} (\bibinfo {year}
  {2016})}\BibitemShut {NoStop}%
\bibitem [{\citenamefont {Rodriguez-Vega}\ \emph {et~al.}(2017)\citenamefont
  {Rodriguez-Vega}, \citenamefont {Schwiete}, \citenamefont {Sinova},\ and\
  \citenamefont {Rossi}}]{rodriguez2017}%
  \BibitemOpen
  \bibfield  {author} {\bibinfo {author} {\bibfnamefont {M.}~\bibnamefont
  {Rodriguez-Vega}}, \bibinfo {author} {\bibfnamefont {G.}~\bibnamefont
  {Schwiete}}, \bibinfo {author} {\bibfnamefont {J.}~\bibnamefont {Sinova}}, \
  and\ \bibinfo {author} {\bibfnamefont {E.}~\bibnamefont {Rossi}},\ }\bibfield
   {title} {\enquote {\bibinfo {title} {Giant edelstein effect in
  topological-insulator--graphene heterostructures},}\ }\href
  {https://link.aps.org/doi/10.1103/PhysRevB.96.235419} {\bibfield  {journal}
  {\bibinfo  {journal} {Phys. Rev. B}\ }\textbf {\bibinfo {volume} {96}},\
  \bibinfo {pages} {235419} (\bibinfo {year} {2017})}\BibitemShut {NoStop}%
\bibitem [{\citenamefont {Rossi}\ and\ \citenamefont {{Das
  Sarma}}(2008)}]{rossi2008}%
  \BibitemOpen
  \bibfield  {author} {\bibinfo {author} {\bibfnamefont {Enrico}\ \bibnamefont
  {Rossi}}\ and\ \bibinfo {author} {\bibfnamefont {S.}~\bibnamefont {{Das
  Sarma}}},\ }\bibfield  {title} {\enquote {\bibinfo {title} {{Ground State of
  Graphene in the Presence of Random Charged Impurities}},}\ }\href
  {https://link.aps.org/doi/10.1103/PhysRevLett.101.166803} {\bibfield
  {journal} {\bibinfo  {journal} {Physical Review Letters}\ }\textbf {\bibinfo
  {volume} {101}},\ \bibinfo {pages} {166803} (\bibinfo {year}
  {2008})}\BibitemShut {NoStop}%
\bibitem [{\citenamefont {Culcer}\ \emph {et~al.}(2010)\citenamefont {Culcer},
  \citenamefont {Hwang}, \citenamefont {Stanescu},\ and\ \citenamefont
  {Das~Sarma}}]{culcer2010}%
  \BibitemOpen
  \bibfield  {author} {\bibinfo {author} {\bibfnamefont {Dimitrie}\
  \bibnamefont {Culcer}}, \bibinfo {author} {\bibfnamefont {E.~H.}\
  \bibnamefont {Hwang}}, \bibinfo {author} {\bibfnamefont {Tudor~D.}\
  \bibnamefont {Stanescu}}, \ and\ \bibinfo {author} {\bibfnamefont
  {S.}~\bibnamefont {Das~Sarma}},\ }\bibfield  {title} {\enquote {\bibinfo
  {title} {Two-dimensional surface charge transport in topological
  insulators},}\ }\href@noop {} {\bibfield  {journal} {\bibinfo  {journal}
  {Phys. Rev. B}\ }\textbf {\bibinfo {volume} {82}},\ \bibinfo {pages} {155457}
  (\bibinfo {year} {2010})}\BibitemShut {NoStop}%
\bibitem [{\citenamefont {Li}\ \emph {et~al.}(2012{\natexlab{b}})\citenamefont
  {Li}, \citenamefont {Rossi},\ and\ \citenamefont {{Das Sarma}}}]{li2012b}%
  \BibitemOpen
  \bibfield  {author} {\bibinfo {author} {\bibfnamefont {Qiuzi}\ \bibnamefont
  {Li}}, \bibinfo {author} {\bibfnamefont {E.}~\bibnamefont {Rossi}}, \ and\
  \bibinfo {author} {\bibfnamefont {S.}~\bibnamefont {{Das Sarma}}},\
  }\bibfield  {title} {\enquote {\bibinfo {title} {{Two-dimensional electronic
  transport on the surface of three-dimensional topological insulators}},}\
  }\href {\doibase 10.1103/PhysRevB.86.235443} {\bibfield  {journal} {\bibinfo
  {journal} {Physical Review B}\ }\textbf {\bibinfo {volume} {86}},\ \bibinfo
  {pages} {235443} (\bibinfo {year} {2012}{\natexlab{b}})}\BibitemShut
  {NoStop}%
\bibitem [{\citenamefont {Adam}\ \emph {et~al.}(2012)\citenamefont {Adam},
  \citenamefont {Hwang},\ and\ \citenamefont {Sarma}}]{adam2012}%
  \BibitemOpen
  \bibfield  {author} {\bibinfo {author} {\bibfnamefont {S.}~\bibnamefont
  {Adam}}, \bibinfo {author} {\bibfnamefont {E.~H.}\ \bibnamefont {Hwang}}, \
  and\ \bibinfo {author} {\bibfnamefont {S.~Das}\ \bibnamefont {Sarma}},\
  }\bibfield  {title} {\enquote {\bibinfo {title} {Two-dimensional transport
  and screening in topological insulator surface states},}\ }\href@noop {}
  {\bibfield  {journal} {\bibinfo  {journal} {Phys. Rev. B}\ }\textbf {\bibinfo
  {volume} {85}},\ \bibinfo {pages} {235413} (\bibinfo {year}
  {2012})}\BibitemShut {NoStop}%
\bibitem [{\citenamefont {Skinner}\ \emph {et~al.}(2012)\citenamefont
  {Skinner}, \citenamefont {Chen},\ and\ \citenamefont
  {Shklovskii}}]{skinner2012}%
  \BibitemOpen
  \bibfield  {author} {\bibinfo {author} {\bibfnamefont {B.}~\bibnamefont
  {Skinner}}, \bibinfo {author} {\bibfnamefont {T.~R.}\ \bibnamefont {Chen}}, \
  and\ \bibinfo {author} {\bibfnamefont {B.~I.}\ \bibnamefont {Shklovskii}},\
  }\bibfield  {title} {\enquote {\bibinfo {title} {Why {Is} the {Bulk}
  {Resistivity} of {Topological} {Insulators} {So} small?}}\ }\href@noop {}
  {\bibfield  {journal} {\bibinfo  {journal} {Phys. Rev. Lett.}\ }\textbf
  {\bibinfo {volume} {109}},\ \bibinfo {pages} {176801} (\bibinfo {year}
  {2012})}\BibitemShut {NoStop}%
\bibitem [{\citenamefont {Skinner}\ and\ \citenamefont
  {Shklovskii}(2013)}]{skinner2013}%
  \BibitemOpen
  \bibfield  {author} {\bibinfo {author} {\bibfnamefont {B.}~\bibnamefont
  {Skinner}}\ and\ \bibinfo {author} {\bibfnamefont {B.~I.}\ \bibnamefont
  {Shklovskii}},\ }\bibfield  {title} {\enquote {\bibinfo {title} {Theory of
  the random potential and conductivity at the surface of a topological
  insulator},}\ }\href@noop {} {\bibfield  {journal} {\bibinfo  {journal}
  {Phys. Rev. B}\ }\textbf {\bibinfo {volume} {87}},\ \bibinfo {pages} {075454}
  (\bibinfo {year} {2013})}\BibitemShut {NoStop}%
\bibitem [{\citenamefont {Butch}\ \emph {et~al.}(2010)\citenamefont {Butch},
  \citenamefont {Kirshenbaum}, \citenamefont {Syers}, \citenamefont {Sushkov},
  \citenamefont {Jenkins}, \citenamefont {Drew},\ and\ \citenamefont
  {Paglione}}]{butch2010}%
  \BibitemOpen
  \bibfield  {author} {\bibinfo {author} {\bibfnamefont {N.~P.}\ \bibnamefont
  {Butch}}, \bibinfo {author} {\bibfnamefont {K.}~\bibnamefont {Kirshenbaum}},
  \bibinfo {author} {\bibfnamefont {P.}~\bibnamefont {Syers}}, \bibinfo
  {author} {\bibfnamefont {A.~B.}\ \bibnamefont {Sushkov}}, \bibinfo {author}
  {\bibfnamefont {G.~S.}\ \bibnamefont {Jenkins}}, \bibinfo {author}
  {\bibfnamefont {H.~D.}\ \bibnamefont {Drew}}, \ and\ \bibinfo {author}
  {\bibfnamefont {J.}~\bibnamefont {Paglione}},\ }\bibfield  {title} {\enquote
  {\bibinfo {title} {Strong surface scattering in ultrahigh-mobility {Bi2Se3}
  topological insulator crystals},}\ }\href@noop {} {\bibfield  {journal}
  {\bibinfo  {journal} {Phys. Rev. B}\ }\textbf {\bibinfo {volume} {81}},\
  \bibinfo {pages} {241301} (\bibinfo {year} {2010})}\BibitemShut {NoStop}%
\bibitem [{\citenamefont {Beidenkopf}\ \emph {et~al.}(2011)\citenamefont
  {Beidenkopf}, \citenamefont {Roushan}, \citenamefont {Seo}, \citenamefont
  {Gorman}, \citenamefont {Drozdov}, \citenamefont {Hor}, \citenamefont
  {Cava},\ and\ \citenamefont {Yazdani}}]{beidenkopf2011}%
  \BibitemOpen
  \bibfield  {author} {\bibinfo {author} {\bibfnamefont {Haim}\ \bibnamefont
  {Beidenkopf}}, \bibinfo {author} {\bibfnamefont {Pedram}\ \bibnamefont
  {Roushan}}, \bibinfo {author} {\bibfnamefont {Jungpil}\ \bibnamefont {Seo}},
  \bibinfo {author} {\bibfnamefont {Lindsay}\ \bibnamefont {Gorman}}, \bibinfo
  {author} {\bibfnamefont {Ilya}\ \bibnamefont {Drozdov}}, \bibinfo {author}
  {\bibfnamefont {Yew~San}\ \bibnamefont {Hor}}, \bibinfo {author}
  {\bibfnamefont {R.~J.}\ \bibnamefont {Cava}}, \ and\ \bibinfo {author}
  {\bibfnamefont {Ali}\ \bibnamefont {Yazdani}},\ }\bibfield  {title} {\enquote
  {\bibinfo {title} {Spatial fluctuations of helical dirac fermions on the
  surface of topological insulators},}\ }\href@noop {} {\bibfield  {journal}
  {\bibinfo  {journal} {Nat Phys}\ }\textbf {\bibinfo {volume} {7}},\ \bibinfo
  {pages} {939--943} (\bibinfo {year} {2011})}\BibitemShut {NoStop}%
\bibitem [{\citenamefont {Kim}\ \emph {et~al.}(2012)\citenamefont {Kim},
  \citenamefont {Cho}, \citenamefont {Butch}, \citenamefont {Syers},
  \citenamefont {Kirshenbaum}, \citenamefont {Adam}, \citenamefont {Paglione},\
  and\ \citenamefont {Fuhrer}}]{kim2012b}%
  \BibitemOpen
  \bibfield  {author} {\bibinfo {author} {\bibfnamefont {D.}~\bibnamefont
  {Kim}}, \bibinfo {author} {\bibfnamefont {S.}~\bibnamefont {Cho}}, \bibinfo
  {author} {\bibfnamefont {N.~P.}\ \bibnamefont {Butch}}, \bibinfo {author}
  {\bibfnamefont {P.}~\bibnamefont {Syers}}, \bibinfo {author} {\bibfnamefont
  {K.}~\bibnamefont {Kirshenbaum}}, \bibinfo {author} {\bibfnamefont
  {S.}~\bibnamefont {Adam}}, \bibinfo {author} {\bibfnamefont {J.}~\bibnamefont
  {Paglione}}, \ and\ \bibinfo {author} {\bibfnamefont {M.~S.}\ \bibnamefont
  {Fuhrer}},\ }\bibfield  {title} {\enquote {\bibinfo {title} {Surface
  conduction of topological {Dirac} electrons in bulk insulating bi2se3},}\
  }\href@noop {} {\bibfield  {journal} {\bibinfo  {journal} {Nature Phys.}\
  }\textbf {\bibinfo {volume} {8}},\ \bibinfo {pages} {458} (\bibinfo {year}
  {2012})}\BibitemShut {NoStop}%
\bibitem [{\citenamefont {Rodriguez-Vega}\ \emph {et~al.}(2014)\citenamefont
  {Rodriguez-Vega}, \citenamefont {Fischer}, \citenamefont {Das~Sarma},\ and\
  \citenamefont {Rossi}}]{rodriguez2014}%
  \BibitemOpen
  \bibfield  {author} {\bibinfo {author} {\bibfnamefont {Martin}\ \bibnamefont
  {Rodriguez-Vega}}, \bibinfo {author} {\bibfnamefont {Jonathan}\ \bibnamefont
  {Fischer}}, \bibinfo {author} {\bibfnamefont {S.}~\bibnamefont {Das~Sarma}},
  \ and\ \bibinfo {author} {\bibfnamefont {E.}~\bibnamefont {Rossi}},\
  }\bibfield  {title} {\enquote {\bibinfo {title} {Ground state of graphene
  heterostructures in the presence of random charged impurities},}\ }\href
  {http://link.aps.org/doi/10.1103/PhysRevB.90.035406} {\bibfield  {journal}
  {\bibinfo  {journal} {Phys. Rev. B}\ }\textbf {\bibinfo {volume} {90}},\
  \bibinfo {pages} {035406} (\bibinfo {year} {2014})}\BibitemShut {NoStop}%
\bibitem [{\citenamefont {Dyakonov}\ and\ \citenamefont
  {Perel}(1971)}]{dyakonov1971}%
  \BibitemOpen
  \bibfield  {author} {\bibinfo {author} {\bibfnamefont {M.~I.}\ \bibnamefont
  {Dyakonov}}\ and\ \bibinfo {author} {\bibfnamefont {V.~I.}\ \bibnamefont
  {Perel}},\ }\bibfield  {title} {\enquote {\bibinfo {title} {Current-induced
  spin orientation of electrons in semiconductors},}\ }\href@noop {} {\bibfield
   {journal} {\bibinfo  {journal} {Phys. Lett. A}\ }\textbf {\bibinfo {volume}
  {A 35}},\ \bibinfo {pages} {459} (\bibinfo {year} {1971})}\BibitemShut
  {NoStop}%
\bibitem [{\citenamefont {Edelstein}(1990)}]{edelstein1990}%
  \BibitemOpen
  \bibfield  {author} {\bibinfo {author} {\bibfnamefont {V.~M.}\ \bibnamefont
  {Edelstein}},\ }\bibfield  {title} {\enquote {\bibinfo {title} {Spin
  polarization of conduction electrons induced by electric-current in
  2-dimensional asymmetric electron-systems},}\ }\href@noop {} {\bibfield
  {journal} {\bibinfo  {journal} {Solid State Comm.}\ }\textbf {\bibinfo
  {volume} {73}},\ \bibinfo {pages} {233--235} (\bibinfo {year}
  {1990})}\BibitemShut {NoStop}%
\bibitem [{\citenamefont {Garate}\ and\ \citenamefont
  {Franz}(2010)}]{garate2010}%
  \BibitemOpen
  \bibfield  {author} {\bibinfo {author} {\bibfnamefont {I.}~\bibnamefont
  {Garate}}\ and\ \bibinfo {author} {\bibfnamefont {M.}~\bibnamefont {Franz}},\
  }\bibfield  {title} {\enquote {\bibinfo {title} {Inverse {Spin-Galvanic}
  {Effect} in the {Interface} between a {Topological} {Insulator} and a
  ferromagnet},}\ }\href@noop {} {\bibfield  {journal} {\bibinfo  {journal}
  {Phys. Rev. Lett.}\ }\textbf {\bibinfo {volume} {104}},\ \bibinfo {pages}
  {146802} (\bibinfo {year} {2010})}\BibitemShut {NoStop}%
\bibitem [{\citenamefont {Yokoyama}\ \emph {et~al.}(2010)\citenamefont
  {Yokoyama}, \citenamefont {Tanaka},\ and\ \citenamefont
  {Nagaosa}}]{yokoyama2010}%
  \BibitemOpen
  \bibfield  {author} {\bibinfo {author} {\bibfnamefont {T.}~\bibnamefont
  {Yokoyama}}, \bibinfo {author} {\bibfnamefont {Y.}~\bibnamefont {Tanaka}}, \
  and\ \bibinfo {author} {\bibfnamefont {N.}~\bibnamefont {Nagaosa}},\
  }\bibfield  {title} {\enquote {\bibinfo {title} {Anomalous magnetoresistance
  of a two-dimensional ferromagnet/ferromagnet junction on the surface of a
  topological insulator},}\ }\href@noop {} {\bibfield  {journal} {\bibinfo
  {journal} {Phys. Rev. B}\ }\textbf {\bibinfo {volume} {81}},\ \bibinfo
  {pages} {121401} (\bibinfo {year} {2010})}\BibitemShut {NoStop}%
\bibitem [{\citenamefont {Sakai}\ and\ \citenamefont
  {Kohno}(2014)}]{sakai2014}%
  \BibitemOpen
  \bibfield  {author} {\bibinfo {author} {\bibfnamefont {Akio}\ \bibnamefont
  {Sakai}}\ and\ \bibinfo {author} {\bibfnamefont {Hiroshi}\ \bibnamefont
  {Kohno}},\ }\bibfield  {title} {\enquote {\bibinfo {title} {Spin torques and
  charge transport on the surface of topological insulator},}\ }\href@noop {}
  {\bibfield  {journal} {\bibinfo  {journal} {Phys. Rev. B}\ }\textbf {\bibinfo
  {volume} {89}},\ \bibinfo {pages} {165307} (\bibinfo {year}
  {2014})}\BibitemShut {NoStop}%
\bibitem [{\citenamefont {{Fischer}}\ \emph {et~al.}(2013)\citenamefont
  {{Fischer}}, \citenamefont {{Vaezi}}, \citenamefont {{Manchon}},\ and\
  \citenamefont {{Kim}}}]{fischer2013}%
  \BibitemOpen
  \bibfield  {author} {\bibinfo {author} {\bibfnamefont {M.~H}\ \bibnamefont
  {{Fischer}}}, \bibinfo {author} {\bibfnamefont {A.}~\bibnamefont {{Vaezi}}},
  \bibinfo {author} {\bibfnamefont {A.}~\bibnamefont {{Manchon}}}, \ and\
  \bibinfo {author} {\bibfnamefont {E.-A.}\ \bibnamefont {{Kim}}},\ }\bibfield
  {title} {\enquote {\bibinfo {title} {{Large Spin Torque in Topological
  Insulator/Ferromagnetic Metal Bilayers}},}\ }\href@noop {} {\bibfield
  {journal} {\bibinfo  {journal} {ArXiv e-prints}\ } (\bibinfo {year}
  {2013})}\BibitemShut {NoStop}%
\bibitem [{\citenamefont {Mellnik}\ \emph {et~al.}(2014)\citenamefont {Mellnik}
  \emph {et~al.}}]{mellnik2014b}%
  \BibitemOpen
  \bibfield  {author} {\bibinfo {author} {\bibfnamefont {A.~R.}\ \bibnamefont
  {Mellnik}} \emph {et~al.},\ }\bibfield  {title} {\enquote {\bibinfo {title}
  {Spin-transfer torque generated by a topological insulator},}\ }\href@noop {}
  {\bibfield  {journal} {\bibinfo  {journal} {Nature}\ }\textbf {\bibinfo
  {volume} {511}},\ \bibinfo {pages} {449--451} (\bibinfo {year}
  {2014})}\BibitemShut {NoStop}%
\bibitem [{\citenamefont {Fan}\ \emph {et~al.}(2016)\citenamefont {Fan} \emph
  {et~al.}}]{fan2016}%
  \BibitemOpen
  \bibfield  {author} {\bibinfo {author} {\bibfnamefont {Yabin}\ \bibnamefont
  {Fan}} \emph {et~al.},\ }\bibfield  {title} {\enquote {\bibinfo {title}
  {{Electric-field control of spin–orbit torque in a magnetically doped
  topological insulator}},}\ }\href@noop {} {\bibfield  {journal} {\bibinfo
  {journal} {Nature Nanotechnology}\ }\textbf {\bibinfo {volume} {11}},\
  \bibinfo {pages} {352--359} (\bibinfo {year} {2016})}\BibitemShut {NoStop}%
\bibitem [{\citenamefont {Rodriguez-Vega}\ \emph {et~al.}(2019)\citenamefont
  {Rodriguez-Vega}, \citenamefont {Schwiete},\ and\ \citenamefont
  {Rossi}}]{rodriguez2019}%
  \BibitemOpen
  \bibfield  {author} {\bibinfo {author} {\bibfnamefont {M}~\bibnamefont
  {Rodriguez-Vega}}, \bibinfo {author} {\bibfnamefont {G}~\bibnamefont
  {Schwiete}}, \ and\ \bibinfo {author} {\bibfnamefont {Enrico}\ \bibnamefont
  {Rossi}},\ }\bibfield  {title} {\enquote {\bibinfo {title} {{Spin-charge
  coupled transport in van der {Waals} systems with random tunneling}},}\
  }\href {\doibase 10.1103/PhysRevResearch.1.033085} {\bibfield  {journal}
  {\bibinfo  {journal} {Physical Review Research}\ }\textbf {\bibinfo {volume}
  {1}},\ \bibinfo {pages} {033085} (\bibinfo {year} {2019})}\BibitemShut
  {NoStop}%
\bibitem [{\citenamefont {Xiao}\ \emph {et~al.}(2012)\citenamefont {Xiao},
  \citenamefont {Liu}, \citenamefont {Feng}, \citenamefont {Xu},\ and\
  \citenamefont {Yao}}]{xiao2012}%
  \BibitemOpen
  \bibfield  {author} {\bibinfo {author} {\bibfnamefont {Di}~\bibnamefont
  {Xiao}}, \bibinfo {author} {\bibfnamefont {Gui-Bin}\ \bibnamefont {Liu}},
  \bibinfo {author} {\bibfnamefont {Wanxiang}\ \bibnamefont {Feng}}, \bibinfo
  {author} {\bibfnamefont {Xiaodong}\ \bibnamefont {Xu}}, \ and\ \bibinfo
  {author} {\bibfnamefont {Wang}\ \bibnamefont {Yao}},\ }\bibfield  {title}
  {\enquote {\bibinfo {title} {Coupled spin and valley physics in monolayers of
  {MoS$_2$} and other {Group-VI} dichalcogenides},}\ }\href
  {http://link.aps.org/doi/10.1103/PhysRevLett.108.196802} {\bibfield
  {journal} {\bibinfo  {journal} {Phys. Rev. Lett.}\ }\textbf {\bibinfo
  {volume} {108}},\ \bibinfo {pages} {196802} (\bibinfo {year}
  {2012})}\BibitemShut {NoStop}%
\bibitem [{\citenamefont {Ramasubramaniam}(2012)}]{ashwin2012}%
  \BibitemOpen
  \bibfield  {author} {\bibinfo {author} {\bibfnamefont {Ashwin}\ \bibnamefont
  {Ramasubramaniam}},\ }\bibfield  {title} {\enquote {\bibinfo {title} {Large
  excitonic effects in monolayers of molybdenum and tungsten
  dichalcogenides},}\ }\href {\doibase 10.1103/PhysRevB.86.115409} {\bibfield
  {journal} {\bibinfo  {journal} {Phys. Rev. B}\ }\textbf {\bibinfo {volume}
  {86}},\ \bibinfo {pages} {115409} (\bibinfo {year} {2012})}\BibitemShut
  {NoStop}%
\bibitem [{\citenamefont {Wang}\ \emph {et~al.}(2012)\citenamefont {Wang},
  \citenamefont {Kalantar-Zadeh}, \citenamefont {Kis}, \citenamefont
  {Coleman},\ and\ \citenamefont {Strano}}]{strano2012}%
  \BibitemOpen
  \bibfield  {author} {\bibinfo {author} {\bibfnamefont {Qing~Hua}\
  \bibnamefont {Wang}}, \bibinfo {author} {\bibfnamefont {Kourosh}\
  \bibnamefont {Kalantar-Zadeh}}, \bibinfo {author} {\bibfnamefont {Andras}\
  \bibnamefont {Kis}}, \bibinfo {author} {\bibfnamefont {Jonathan~N.}\
  \bibnamefont {Coleman}}, \ and\ \bibinfo {author} {\bibfnamefont
  {Michael~S.}\ \bibnamefont {Strano}},\ }\bibfield  {title} {\enquote
  {\bibinfo {title} {Electronics and optoelectronics of two-dimensional
  transition metal dichalcogenides},}\ }\href {\doibase 10.1038/nnano.2012.193}
  {\bibfield  {journal} {\bibinfo  {journal} {Nature Nanotechnology}\ }\textbf
  {\bibinfo {volume} {7}},\ \bibinfo {pages} {699--712} (\bibinfo {year}
  {2012})}\BibitemShut {NoStop}%
\bibitem [{\citenamefont {Xu}\ \emph {et~al.}(2014{\natexlab{b}})\citenamefont
  {Xu}, \citenamefont {Yao}, \citenamefont {Xiao},\ and\ \citenamefont
  {Heinz}}]{xiaodong2014}%
  \BibitemOpen
  \bibfield  {author} {\bibinfo {author} {\bibfnamefont {Xiaodong}\
  \bibnamefont {Xu}}, \bibinfo {author} {\bibfnamefont {Wang}\ \bibnamefont
  {Yao}}, \bibinfo {author} {\bibfnamefont {Di}~\bibnamefont {Xiao}}, \ and\
  \bibinfo {author} {\bibfnamefont {Tony~F.}\ \bibnamefont {Heinz}},\
  }\bibfield  {title} {\enquote {\bibinfo {title} {Spin and pseudospins in
  layered transition metal dichalcogenides},}\ }\href {\doibase
  10.1038/nphys2942} {\bibfield  {journal} {\bibinfo  {journal} {Nature
  Physics}\ }\textbf {\bibinfo {volume} {10}},\ \bibinfo {pages} {343--350}
  (\bibinfo {year} {2014}{\natexlab{b}})}\BibitemShut {NoStop}%
\bibitem [{\citenamefont {Latzke}\ \emph {et~al.}(2015)\citenamefont {Latzke},
  \citenamefont {Zhang}, \citenamefont {Suslu}, \citenamefont {Chang},
  \citenamefont {Lin}, \citenamefont {Jeng}, \citenamefont {Tongay},
  \citenamefont {Wu}, \citenamefont {Bansil},\ and\ \citenamefont
  {Lanzara}}]{latzke2015}%
  \BibitemOpen
  \bibfield  {author} {\bibinfo {author} {\bibfnamefont {Drew~W.}\ \bibnamefont
  {Latzke}}, \bibinfo {author} {\bibfnamefont {Wentao}\ \bibnamefont {Zhang}},
  \bibinfo {author} {\bibfnamefont {Aslihan}\ \bibnamefont {Suslu}}, \bibinfo
  {author} {\bibfnamefont {Tay-Rong}\ \bibnamefont {Chang}}, \bibinfo {author}
  {\bibfnamefont {Hsin}\ \bibnamefont {Lin}}, \bibinfo {author} {\bibfnamefont
  {Horng-Tay}\ \bibnamefont {Jeng}}, \bibinfo {author} {\bibfnamefont
  {Sefaattin}\ \bibnamefont {Tongay}}, \bibinfo {author} {\bibfnamefont
  {Junqiao}\ \bibnamefont {Wu}}, \bibinfo {author} {\bibfnamefont {Arun}\
  \bibnamefont {Bansil}}, \ and\ \bibinfo {author} {\bibfnamefont {Alessandra}\
  \bibnamefont {Lanzara}},\ }\bibfield  {title} {\enquote {\bibinfo {title}
  {Electronic structure, spin-orbit coupling, and interlayer interaction in
  bulk ${\mathrm{mos}}_{2}$ and ${\mathrm{ws}}_{2}$},}\ }\href {\doibase
  10.1103/PhysRevB.91.235202} {\bibfield  {journal} {\bibinfo  {journal} {Phys.
  Rev. B}\ }\textbf {\bibinfo {volume} {91}},\ \bibinfo {pages} {235202}
  (\bibinfo {year} {2015})}\BibitemShut {NoStop}%
\bibitem [{\citenamefont {Gmitra}\ \emph {et~al.}(2009)\citenamefont {Gmitra},
  \citenamefont {Konschuh}, \citenamefont {Ertler}, \citenamefont
  {Ambrosch-Draxl},\ and\ \citenamefont {Fabian}}]{gmitra2009}%
  \BibitemOpen
  \bibfield  {author} {\bibinfo {author} {\bibfnamefont {M.}~\bibnamefont
  {Gmitra}}, \bibinfo {author} {\bibfnamefont {S.}~\bibnamefont {Konschuh}},
  \bibinfo {author} {\bibfnamefont {C.}~\bibnamefont {Ertler}}, \bibinfo
  {author} {\bibfnamefont {C.}~\bibnamefont {Ambrosch-Draxl}}, \ and\ \bibinfo
  {author} {\bibfnamefont {J.}~\bibnamefont {Fabian}},\ }\bibfield  {title}
  {\enquote {\bibinfo {title} {Band-structure topologies of graphene:
  {Spin-orbit} coupling effects from first principles},}\ }\href@noop {}
  {\bibfield  {journal} {\bibinfo  {journal} {Phys. Rev. B}\ }\textbf {\bibinfo
  {volume} {80}},\ \bibinfo {pages} {235431} (\bibinfo {year}
  {2009})}\BibitemShut {NoStop}%
\bibitem [{\citenamefont {Gmitra}\ and\ \citenamefont
  {Fabian}(2015)}]{Gmitra2015}%
  \BibitemOpen
  \bibfield  {author} {\bibinfo {author} {\bibfnamefont {Martin}\ \bibnamefont
  {Gmitra}}\ and\ \bibinfo {author} {\bibfnamefont {Jaroslav}\ \bibnamefont
  {Fabian}},\ }\bibfield  {title} {\enquote {\bibinfo {title} {{Graphene on
  transition-metal dichalcogenides: A platform for proximity spin-orbit physics
  and optospintronics}},}\ }\href {\doibase 10.1103/PhysRevB.92.155403}
  {\bibfield  {journal} {\bibinfo  {journal} {Physical Review B - Condensed
  Matter and Materials Physics}\ }\textbf {\bibinfo {volume} {92}},\ \bibinfo
  {pages} {1--6} (\bibinfo {year} {2015})}\BibitemShut {NoStop}%
\bibitem [{\citenamefont {Gmitra}\ \emph {et~al.}(2016)\citenamefont {Gmitra},
  \citenamefont {Kochan}, \citenamefont {H{\"{o}}gl},\ and\ \citenamefont
  {Fabian}}]{Gmitra2016}%
  \BibitemOpen
  \bibfield  {author} {\bibinfo {author} {\bibfnamefont {Martin}\ \bibnamefont
  {Gmitra}}, \bibinfo {author} {\bibfnamefont {Denis}\ \bibnamefont {Kochan}},
  \bibinfo {author} {\bibfnamefont {Petra}\ \bibnamefont {H{\"{o}}gl}}, \ and\
  \bibinfo {author} {\bibfnamefont {Jaroslav}\ \bibnamefont {Fabian}},\
  }\bibfield  {title} {\enquote {\bibinfo {title} {{Trivial and inverted Dirac
  bands and the emergence of quantum spin Hall states in graphene on
  transition-metal dichalcogenides}},}\ }\href {\doibase
  10.1103/PhysRevB.93.155104} {\bibfield  {journal} {\bibinfo  {journal}
  {Physical Review B}\ }\textbf {\bibinfo {volume} {93}},\ \bibinfo {pages}
  {1--10} (\bibinfo {year} {2016})}\BibitemShut {NoStop}%
\bibitem [{\citenamefont {Gmitra}\ and\ \citenamefont
  {Fabian}(2017)}]{Gmitra2017}%
  \BibitemOpen
  \bibfield  {author} {\bibinfo {author} {\bibfnamefont {Martin}\ \bibnamefont
  {Gmitra}}\ and\ \bibinfo {author} {\bibfnamefont {Jaroslav}\ \bibnamefont
  {Fabian}},\ }\bibfield  {title} {\enquote {\bibinfo {title} {{Proximity
  Effects in Bilayer Graphene on Monolayer WSe2: Field-Effect Spin Valley
  Locking, Spin-Orbit Valve, and Spin Transistor}},}\ }\href {\doibase
  10.1103/PhysRevLett.119.146401} {\bibfield  {journal} {\bibinfo  {journal}
  {Physical Review Letters}\ }\textbf {\bibinfo {volume} {119}},\ \bibinfo
  {pages} {2--6} (\bibinfo {year} {2017})}\BibitemShut {NoStop}%
\bibitem [{\citenamefont {Alsharari}\ \emph {et~al.}(2016)\citenamefont
  {Alsharari}, \citenamefont {Asmar},\ and\ \citenamefont
  {Ulloa}}]{alsharari2016}%
  \BibitemOpen
  \bibfield  {author} {\bibinfo {author} {\bibfnamefont {Abdulrhman~M.}\
  \bibnamefont {Alsharari}}, \bibinfo {author} {\bibfnamefont {Mahmoud~M.}\
  \bibnamefont {Asmar}}, \ and\ \bibinfo {author} {\bibfnamefont {Sergio~E.}\
  \bibnamefont {Ulloa}},\ }\bibfield  {title} {\enquote {\bibinfo {title} {Mass
  inversion in graphene by proximity to dichalcogenide monolayer},}\ }\href
  {\doibase 10.1103/PhysRevB.94.241106} {\bibfield  {journal} {\bibinfo
  {journal} {Phys. Rev. B}\ }\textbf {\bibinfo {volume} {94}},\ \bibinfo
  {pages} {241106} (\bibinfo {year} {2016})}\BibitemShut {NoStop}%
\bibitem [{\citenamefont {Alsharari}\ \emph
  {et~al.}(2018{\natexlab{a}})\citenamefont {Alsharari}, \citenamefont
  {Asmar},\ and\ \citenamefont {Ulloa}}]{alsharari2018}%
  \BibitemOpen
  \bibfield  {author} {\bibinfo {author} {\bibfnamefont {Abdulrhman~M.}\
  \bibnamefont {Alsharari}}, \bibinfo {author} {\bibfnamefont {Mahmoud~M.}\
  \bibnamefont {Asmar}}, \ and\ \bibinfo {author} {\bibfnamefont {Sergio~E.}\
  \bibnamefont {Ulloa}},\ }\bibfield  {title} {\enquote {\bibinfo {title}
  {Topological phases and twisting of graphene on a dichalcogenide
  monolayer},}\ }\href {\doibase 10.1103/PhysRevB.98.195129} {\bibfield
  {journal} {\bibinfo  {journal} {Phys. Rev. B}\ }\textbf {\bibinfo {volume}
  {98}},\ \bibinfo {pages} {195129} (\bibinfo {year}
  {2018}{\natexlab{a}})}\BibitemShut {NoStop}%
\bibitem [{\citenamefont {Alsharari}\ \emph
  {et~al.}(2018{\natexlab{b}})\citenamefont {Alsharari}, \citenamefont
  {Asmar},\ and\ \citenamefont {Ulloa}}]{alsharari2018a}%
  \BibitemOpen
  \bibfield  {author} {\bibinfo {author} {\bibfnamefont {Abdulrhman~M.}\
  \bibnamefont {Alsharari}}, \bibinfo {author} {\bibfnamefont {Mahmoud~M.}\
  \bibnamefont {Asmar}}, \ and\ \bibinfo {author} {\bibfnamefont {Sergio~E.}\
  \bibnamefont {Ulloa}},\ }\bibfield  {title} {\enquote {\bibinfo {title}
  {Proximity-induced topological phases in bilayer graphene},}\ }\href
  {\doibase 10.1103/PhysRevB.97.241104} {\bibfield  {journal} {\bibinfo
  {journal} {Phys. Rev. B}\ }\textbf {\bibinfo {volume} {97}},\ \bibinfo
  {pages} {241104} (\bibinfo {year} {2018}{\natexlab{b}})}\BibitemShut
  {NoStop}%
\bibitem [{\citenamefont {Wang}\ \emph {et~al.}(2015)\citenamefont {Wang},
  \citenamefont {Ki}, \citenamefont {Chen}, \citenamefont {Berger},
  \citenamefont {MacDonald},\ and\ \citenamefont {Morpurgo}}]{wang2015}%
  \BibitemOpen
  \bibfield  {author} {\bibinfo {author} {\bibfnamefont {Z.}~\bibnamefont
  {Wang}}, \bibinfo {author} {\bibfnamefont {DongKeun}\ \bibnamefont {Ki}},
  \bibinfo {author} {\bibfnamefont {H.}~\bibnamefont {Chen}}, \bibinfo {author}
  {\bibfnamefont {H}~\bibnamefont {Berger}}, \bibinfo {author} {\bibfnamefont
  {A.~H.}\ \bibnamefont {MacDonald}}, \ and\ \bibinfo {author} {\bibfnamefont
  {A.~F.}\ \bibnamefont {Morpurgo}},\ }\bibfield  {title} {\enquote {\bibinfo
  {title} {Strong interface-induced spin-orbit interaction in graphene on
  ws2},}\ }\href@noop {} {\bibfield  {journal} {\bibinfo  {journal} {Nature
  Comm.}\ }\textbf {\bibinfo {volume} {6}},\ \bibinfo {pages} {8339} (\bibinfo
  {year} {2015})}\BibitemShut {NoStop}%
\bibitem [{\citenamefont {Wang}\ \emph {et~al.}(2016)\citenamefont {Wang},
  \citenamefont {Ki}, \citenamefont {Yong}, \citenamefont {Mauro},
  \citenamefont {Berger}, \citenamefont {L.S.},\ and\ \citenamefont
  {Morpurgo}}]{wang2016}%
  \BibitemOpen
  \bibfield  {author} {\bibinfo {author} {\bibfnamefont {Z.}~\bibnamefont
  {Wang}}, \bibinfo {author} {\bibfnamefont {D.}~\bibnamefont {Ki}}, \bibinfo
  {author} {\bibfnamefont {J.~Y.}\ \bibnamefont {Yong}, \bibfnamefont {Khoo}},
  \bibinfo {author} {\bibfnamefont {D.}~\bibnamefont {Mauro}}, \bibinfo
  {author} {\bibfnamefont {H.}~\bibnamefont {Berger}}, \bibinfo {author}
  {\bibfnamefont {Levitov}\ \bibnamefont {L.S.}}, \ and\ \bibinfo {author}
  {\bibfnamefont {A.~F.}\ \bibnamefont {Morpurgo}},\ }\bibfield  {title}
  {\enquote {\bibinfo {title} {Origin and {Magnitude} of {'Designer'}
  {Spin-Orbit} {Interaction} in {Graphene} on {Semiconducting} {Transition}
  {Metal} dichalcogenides},}\ }\href@noop {} {\bibfield  {journal} {\bibinfo
  {journal} {Phys. Rev. X}\ }\textbf {\bibinfo {volume} {6}},\ \bibinfo {pages}
  {041020} (\bibinfo {year} {2016})}\BibitemShut {NoStop}%
\bibitem [{\citenamefont {Yang}\ \emph {et~al.}(2016)\citenamefont {Yang},
  \citenamefont {Tu}, \citenamefont {Kim}, \citenamefont {Wu}, \citenamefont
  {Wang}, \citenamefont {Alicea}, \citenamefont {Wu}, \citenamefont
  {Bockrath},\ and\ \citenamefont {Shi}}]{yang2016}%
  \BibitemOpen
  \bibfield  {author} {\bibinfo {author} {\bibfnamefont {Bowen}\ \bibnamefont
  {Yang}}, \bibinfo {author} {\bibfnamefont {Min-Feng}\ \bibnamefont {Tu}},
  \bibinfo {author} {\bibfnamefont {Jeongwoo}\ \bibnamefont {Kim}}, \bibinfo
  {author} {\bibfnamefont {Yong}\ \bibnamefont {Wu}}, \bibinfo {author}
  {\bibfnamefont {Hui}\ \bibnamefont {Wang}}, \bibinfo {author} {\bibfnamefont
  {Jason}\ \bibnamefont {Alicea}}, \bibinfo {author} {\bibfnamefont {Ruqian}\
  \bibnamefont {Wu}}, \bibinfo {author} {\bibfnamefont {Marc}\ \bibnamefont
  {Bockrath}}, \ and\ \bibinfo {author} {\bibfnamefont {Jing}\ \bibnamefont
  {Shi}},\ }\bibfield  {title} {\enquote {\bibinfo {title} {Tunable
  spin{\textendash}orbit coupling and symmetry-protected edge states in
  graphene/{WS}2},}\ }\href {\doibase 10.1088/2053-1583/3/3/031012} {\bibfield
  {journal} {\bibinfo  {journal} {2D Materials}\ }\textbf {\bibinfo {volume}
  {3}},\ \bibinfo {pages} {031012} (\bibinfo {year} {2016})}\BibitemShut
  {NoStop}%
\bibitem [{\citenamefont {Yang}\ \emph {et~al.}(2017)\citenamefont {Yang},
  \citenamefont {Lohmann}, \citenamefont {Barroso}, \citenamefont {Liao},
  \citenamefont {Lin}, \citenamefont {Liu}, \citenamefont {Bartels},
  \citenamefont {Watanabe}, \citenamefont {Taniguchi},\ and\ \citenamefont
  {Shi}}]{yang2017}%
  \BibitemOpen
  \bibfield  {author} {\bibinfo {author} {\bibfnamefont {Bowen}\ \bibnamefont
  {Yang}}, \bibinfo {author} {\bibfnamefont {Mark}\ \bibnamefont {Lohmann}},
  \bibinfo {author} {\bibfnamefont {David}\ \bibnamefont {Barroso}}, \bibinfo
  {author} {\bibfnamefont {Ingrid}\ \bibnamefont {Liao}}, \bibinfo {author}
  {\bibfnamefont {Zhisheng}\ \bibnamefont {Lin}}, \bibinfo {author}
  {\bibfnamefont {Yawen}\ \bibnamefont {Liu}}, \bibinfo {author} {\bibfnamefont
  {Ludwig}\ \bibnamefont {Bartels}}, \bibinfo {author} {\bibfnamefont {Kenji}\
  \bibnamefont {Watanabe}}, \bibinfo {author} {\bibfnamefont {Takashi}\
  \bibnamefont {Taniguchi}}, \ and\ \bibinfo {author} {\bibfnamefont {Jing}\
  \bibnamefont {Shi}},\ }\bibfield  {title} {\enquote {\bibinfo {title} {Strong
  electron-hole symmetric rashba spin-orbit coupling in graphene/monolayer
  transition metal dichalcogenide heterostructures},}\ }\href {\doibase
  10.1103/PhysRevB.96.041409} {\bibfield  {journal} {\bibinfo  {journal} {Phys.
  Rev. B}\ }\textbf {\bibinfo {volume} {96}},\ \bibinfo {pages} {041409}
  (\bibinfo {year} {2017})}\BibitemShut {NoStop}%
\bibitem [{\citenamefont {V\"olkl}\ \emph {et~al.}(2017)\citenamefont
  {V\"olkl}, \citenamefont {Rockinger}, \citenamefont {Drienovsky},
  \citenamefont {Watanabe}, \citenamefont {Taniguchi}, \citenamefont {Weiss},\
  and\ \citenamefont {Eroms}}]{voekl2017}%
  \BibitemOpen
  \bibfield  {author} {\bibinfo {author} {\bibfnamefont {Tobias}\ \bibnamefont
  {V\"olkl}}, \bibinfo {author} {\bibfnamefont {Tobias}\ \bibnamefont
  {Rockinger}}, \bibinfo {author} {\bibfnamefont {Martin}\ \bibnamefont
  {Drienovsky}}, \bibinfo {author} {\bibfnamefont {Kenji}\ \bibnamefont
  {Watanabe}}, \bibinfo {author} {\bibfnamefont {Takashi}\ \bibnamefont
  {Taniguchi}}, \bibinfo {author} {\bibfnamefont {Dieter}\ \bibnamefont
  {Weiss}}, \ and\ \bibinfo {author} {\bibfnamefont {Jonathan}\ \bibnamefont
  {Eroms}},\ }\bibfield  {title} {\enquote {\bibinfo {title} {Magnetotransport
  in heterostructures of transition metal dichalcogenides and graphene},}\
  }\href {\doibase 10.1103/PhysRevB.96.125405} {\bibfield  {journal} {\bibinfo
  {journal} {Phys. Rev. B}\ }\textbf {\bibinfo {volume} {96}},\ \bibinfo
  {pages} {125405} (\bibinfo {year} {2017})}\BibitemShut {NoStop}%
\bibitem [{\citenamefont {Wakamura}\ \emph {et~al.}(2018)\citenamefont
  {Wakamura}, \citenamefont {Reale}, \citenamefont {Palczynski}, \citenamefont
  {Gueron}, \citenamefont {Mattevi},\ and\ \citenamefont
  {Bouchiat}}]{wakamura2018}%
  \BibitemOpen
  \bibfield  {author} {\bibinfo {author} {\bibfnamefont {T.}~\bibnamefont
  {Wakamura}}, \bibinfo {author} {\bibfnamefont {F}~\bibnamefont {Reale}},
  \bibinfo {author} {\bibfnamefont {P.}~\bibnamefont {Palczynski}}, \bibinfo
  {author} {\bibfnamefont {S.}~\bibnamefont {Gueron}}, \bibinfo {author}
  {\bibfnamefont {C.}~\bibnamefont {Mattevi}}, \ and\ \bibinfo {author}
  {\bibfnamefont {H}~\bibnamefont {Bouchiat}},\ }\bibfield  {title} {\enquote
  {\bibinfo {title} {Strong {Anisotropic} {Spin-Orbit} {Interaction} {Induced}
  in {Graphene} by {Monolayer} ws2},}\ }\href@noop {} {\bibfield  {journal}
  {\bibinfo  {journal} {Phys. Rev. Lett.}\ }\textbf {\bibinfo {volume} {120}},\
  \bibinfo {pages} {106802} (\bibinfo {year} {2018})}\BibitemShut {NoStop}%
\bibitem [{\citenamefont {Zihlmann}\ \emph {et~al.}(2018)\citenamefont
  {Zihlmann}, \citenamefont {Cummings}, \citenamefont {Garcia}, \citenamefont
  {Kedves}, \citenamefont {Watanabe}, \citenamefont {Taniguchi}, \citenamefont
  {Sch\"onenberger},\ and\ \citenamefont {Makk}}]{zihlman2018}%
  \BibitemOpen
  \bibfield  {author} {\bibinfo {author} {\bibfnamefont {Simon}\ \bibnamefont
  {Zihlmann}}, \bibinfo {author} {\bibfnamefont {Aron~W.}\ \bibnamefont
  {Cummings}}, \bibinfo {author} {\bibfnamefont {Jose~H.}\ \bibnamefont
  {Garcia}}, \bibinfo {author} {\bibfnamefont {M\'at\'e}\ \bibnamefont
  {Kedves}}, \bibinfo {author} {\bibfnamefont {Kenji}\ \bibnamefont
  {Watanabe}}, \bibinfo {author} {\bibfnamefont {Takashi}\ \bibnamefont
  {Taniguchi}}, \bibinfo {author} {\bibfnamefont {Christian}\ \bibnamefont
  {Sch\"onenberger}}, \ and\ \bibinfo {author} {\bibfnamefont {P\'eter}\
  \bibnamefont {Makk}},\ }\bibfield  {title} {\enquote {\bibinfo {title} {Large
  spin relaxation anisotropy and valley-zeeman spin-orbit coupling in
  ${\mathrm{wse}}_{2}$/graphene/$h$-bn heterostructures},}\ }\href {\doibase
  10.1103/PhysRevB.97.075434} {\bibfield  {journal} {\bibinfo  {journal} {Phys.
  Rev. B}\ }\textbf {\bibinfo {volume} {97}},\ \bibinfo {pages} {075434}
  (\bibinfo {year} {2018})}\BibitemShut {NoStop}%
\bibitem [{\citenamefont {Avsar}\ \emph {et~al.}(2014)\citenamefont {Avsar},
  \citenamefont {Tan}, \citenamefont {Taychatanapat}, \citenamefont
  {Balakrishnan}, \citenamefont {Koon}, \citenamefont {Yeo}, \citenamefont
  {Lahiri}, \citenamefont {Carvalho}, \citenamefont {Rodin}, \citenamefont
  {O'Farrell}, \citenamefont {Eda}, \citenamefont {{Castro Neto}},\ and\
  \citenamefont {{\"{O}}zyilmaz}}]{Avsar2014}%
  \BibitemOpen
  \bibfield  {author} {\bibinfo {author} {\bibfnamefont {A.}~\bibnamefont
  {Avsar}}, \bibinfo {author} {\bibfnamefont {J.~Y.}\ \bibnamefont {Tan}},
  \bibinfo {author} {\bibfnamefont {T.}~\bibnamefont {Taychatanapat}}, \bibinfo
  {author} {\bibfnamefont {J.}~\bibnamefont {Balakrishnan}}, \bibinfo {author}
  {\bibfnamefont {G.K.W.}\ \bibnamefont {Koon}}, \bibinfo {author}
  {\bibfnamefont {Y.}~\bibnamefont {Yeo}}, \bibinfo {author} {\bibfnamefont
  {J.}~\bibnamefont {Lahiri}}, \bibinfo {author} {\bibfnamefont
  {A.}~\bibnamefont {Carvalho}}, \bibinfo {author} {\bibfnamefont {A.~S.}\
  \bibnamefont {Rodin}}, \bibinfo {author} {\bibfnamefont {E.C.T.}\
  \bibnamefont {O'Farrell}}, \bibinfo {author} {\bibfnamefont {G.}~\bibnamefont
  {Eda}}, \bibinfo {author} {\bibfnamefont {A.~H.}\ \bibnamefont {{Castro
  Neto}}}, \ and\ \bibinfo {author} {\bibfnamefont {B.}~\bibnamefont
  {{\"{O}}zyilmaz}},\ }\bibfield  {title} {\enquote {\bibinfo {title}
  {{Spin–orbit proximity effect in graphene}},}\ }\href {\doibase
  10.1038/ncomms5875} {\bibfield  {journal} {\bibinfo  {journal} {Nature
  Communications}\ }\textbf {\bibinfo {volume} {5}},\ \bibinfo {pages} {4875}
  (\bibinfo {year} {2014})}\BibitemShut {NoStop}%
\bibitem [{\citenamefont {Dankert}\ and\ \citenamefont
  {Dash}(2017)}]{dankert2017}%
  \BibitemOpen
  \bibfield  {author} {\bibinfo {author} {\bibfnamefont {A}~\bibnamefont
  {Dankert}}\ and\ \bibinfo {author} {\bibfnamefont {S.~P.}\ \bibnamefont
  {Dash}},\ }\bibfield  {title} {\enquote {\bibinfo {title} {Electrical gate
  control of spin current in van der {Waals} heterostructures at room
  temperature},}\ }\href@noop {} {\bibfield  {journal} {\bibinfo  {journal}
  {Nature Comm.}\ }\textbf {\bibinfo {volume} {8}},\ \bibinfo {pages} {16093}
  (\bibinfo {year} {2017})}\BibitemShut {NoStop}%
\bibitem [{\citenamefont {Ghiasi}\ \emph {et~al.}(2017)\citenamefont {Ghiasi},
  \citenamefont {Ingla-Ayn{\'{e}}s}, \citenamefont {Kaverzin},\ and\
  \citenamefont {van Wees}}]{Ghiasi2017a}%
  \BibitemOpen
  \bibfield  {author} {\bibinfo {author} {\bibfnamefont {Talieh~S.}\
  \bibnamefont {Ghiasi}}, \bibinfo {author} {\bibfnamefont {Josep}\
  \bibnamefont {Ingla-Ayn{\'{e}}s}}, \bibinfo {author} {\bibfnamefont
  {Alexey~A.}\ \bibnamefont {Kaverzin}}, \ and\ \bibinfo {author}
  {\bibfnamefont {Bart~J.}\ \bibnamefont {van Wees}},\ }\bibfield  {title}
  {\enquote {\bibinfo {title} {{Large Proximity-Induced Spin Lifetime
  Anisotropy in Transition-Metal Dichalcogenide/Graphene Heterostructures}},}\
  }\href {\doibase 10.1021/acs.nanolett.7b03460} {\bibfield  {journal}
  {\bibinfo  {journal} {Nano Letters}\ }\textbf {\bibinfo {volume} {17}},\
  \bibinfo {pages} {7528--7532} (\bibinfo {year} {2017})}\BibitemShut {NoStop}%
\bibitem [{\citenamefont {Omar}\ and\ \citenamefont {van
  Wees}(2018)}]{omar2018}%
  \BibitemOpen
  \bibfield  {author} {\bibinfo {author} {\bibfnamefont {S.}~\bibnamefont
  {Omar}}\ and\ \bibinfo {author} {\bibfnamefont {B.~J.}\ \bibnamefont {van
  Wees}},\ }\bibfield  {title} {\enquote {\bibinfo {title} {Spin transport in
  high-mobility graphene on {WS2} substrate with electric-field tunable
  proximity spin-orbit interaction},}\ }\href@noop {} {\bibfield  {journal}
  {\bibinfo  {journal} {Phys. Rev. B}\ }\textbf {\bibinfo {volume} {97}},\
  \bibinfo {pages} {045414} (\bibinfo {year} {2018})}\BibitemShut {NoStop}%
\bibitem [{\citenamefont {Ben{\'{i}}tez}\ \emph {et~al.}(2018)\citenamefont
  {Ben{\'{i}}tez}, \citenamefont {Sierra}, \citenamefont {{Savero Torres}},
  \citenamefont {Arrighi}, \citenamefont {Bonell}, \citenamefont {Costache},\
  and\ \citenamefont {Valenzuela}}]{Benitez2018a}%
  \BibitemOpen
  \bibfield  {author} {\bibinfo {author} {\bibfnamefont {L.~Antonio}\
  \bibnamefont {Ben{\'{i}}tez}}, \bibinfo {author} {\bibfnamefont {Juan~F.}\
  \bibnamefont {Sierra}}, \bibinfo {author} {\bibfnamefont {Williams}\
  \bibnamefont {{Savero Torres}}}, \bibinfo {author} {\bibfnamefont
  {Alo{\"{i}}s}\ \bibnamefont {Arrighi}}, \bibinfo {author} {\bibfnamefont
  {Fr{\'{e}}d{\'{e}}ric}\ \bibnamefont {Bonell}}, \bibinfo {author}
  {\bibfnamefont {Marius~V.}\ \bibnamefont {Costache}}, \ and\ \bibinfo
  {author} {\bibfnamefont {Sergio~O.}\ \bibnamefont {Valenzuela}},\ }\bibfield
  {title} {\enquote {\bibinfo {title} {{Strongly anisotropic spin relaxation in
  graphene–transition metal dichalcogenide heterostructures at room
  temperature}},}\ }\href {\doibase 10.1038/s41567-017-0019-2} {\bibfield
  {journal} {\bibinfo  {journal} {Nature Physics}\ }\textbf {\bibinfo {volume}
  {14}},\ \bibinfo {pages} {303--308} (\bibinfo {year} {2018})}\BibitemShut
  {NoStop}%
\bibitem [{\citenamefont {Li}\ and\ \citenamefont {Koshino}(2019)}]{Li2019}%
  \BibitemOpen
  \bibfield  {author} {\bibinfo {author} {\bibfnamefont {Yang}\ \bibnamefont
  {Li}}\ and\ \bibinfo {author} {\bibfnamefont {Mikito}\ \bibnamefont
  {Koshino}},\ }\bibfield  {title} {\enquote {\bibinfo {title} {{Twist-angle
  dependence of the proximity spin-orbit coupling in graphene on
  transition-metal dichalcogenides}},}\ }\href {\doibase
  10.1103/PhysRevB.99.075438} {\bibfield  {journal} {\bibinfo  {journal}
  {Physical Review B}\ }\textbf {\bibinfo {volume} {99}},\ \bibinfo {pages}
  {075438} (\bibinfo {year} {2019})}\BibitemShut {NoStop}%
\bibitem [{\citenamefont {David}\ \emph {et~al.}(2019)\citenamefont {David},
  \citenamefont {Rakyta}, \citenamefont {Korm\'anyos},\ and\ \citenamefont
  {Burkard}}]{david2019}%
  \BibitemOpen
  \bibfield  {author} {\bibinfo {author} {\bibfnamefont {Alessandro}\
  \bibnamefont {David}}, \bibinfo {author} {\bibfnamefont {P\'eter}\
  \bibnamefont {Rakyta}}, \bibinfo {author} {\bibfnamefont {Andor}\
  \bibnamefont {Korm\'anyos}}, \ and\ \bibinfo {author} {\bibfnamefont {Guido}\
  \bibnamefont {Burkard}},\ }\bibfield  {title} {\enquote {\bibinfo {title}
  {Induced spin-orbit coupling in twisted graphene--transition metal
  dichalcogenide heterobilayers: Twistronics meets spintronics},}\ }\href
  {\doibase 10.1103/PhysRevB.100.085412} {\bibfield  {journal} {\bibinfo
  {journal} {Phys. Rev. B}\ }\textbf {\bibinfo {volume} {100}},\ \bibinfo
  {pages} {085412} (\bibinfo {year} {2019})}\BibitemShut {NoStop}%
\bibitem [{\citenamefont {Gani}\ \emph {et~al.}(2020)\citenamefont {Gani},
  \citenamefont {Walter},\ and\ \citenamefont {Rossi}}]{Gani2019b}%
  \BibitemOpen
  \bibfield  {author} {\bibinfo {author} {\bibfnamefont {Yohanes~S.}\
  \bibnamefont {Gani}}, \bibinfo {author} {\bibfnamefont {Eric~J.}\
  \bibnamefont {Walter}}, \ and\ \bibinfo {author} {\bibfnamefont {Enrico}\
  \bibnamefont {Rossi}},\ }\bibfield  {title} {\enquote {\bibinfo {title}
  {{Proximity-induced spin-orbit splitting in graphene nanoribbons on
  transition-metal dichalcogenides}},}\ }\href
  {https://link.aps.org/doi/10.1103/PhysRevB.101.195416} {\bibfield  {journal}
  {\bibinfo  {journal} {Physical Review B}\ }\textbf {\bibinfo {volume}
  {101}},\ \bibinfo {pages} {195416} (\bibinfo {year} {2020})}\BibitemShut
  {NoStop}%
\bibitem [{\citenamefont {Offidani}\ \emph {et~al.}(2017)\citenamefont
  {Offidani}, \citenamefont {Milletar{\`{i}}}, \citenamefont {Raimondi},\ and\
  \citenamefont {Ferreira}}]{Offidani2017}%
  \BibitemOpen
  \bibfield  {author} {\bibinfo {author} {\bibfnamefont {Manuel}\ \bibnamefont
  {Offidani}}, \bibinfo {author} {\bibfnamefont {Mirco}\ \bibnamefont
  {Milletar{\`{i}}}}, \bibinfo {author} {\bibfnamefont {Roberto}\ \bibnamefont
  {Raimondi}}, \ and\ \bibinfo {author} {\bibfnamefont {Aires}\ \bibnamefont
  {Ferreira}},\ }\bibfield  {title} {\enquote {\bibinfo {title} {{Optimal
  Charge-to-Spin Conversion in Graphene on Transition-Metal
  Dichalcogenides}},}\ }\href {\doibase 10.1103/PhysRevLett.119.196801}
  {\bibfield  {journal} {\bibinfo  {journal} {Physical Review Letters}\
  }\textbf {\bibinfo {volume} {119}},\ \bibinfo {pages} {1--5} (\bibinfo {year}
  {2017})}\BibitemShut {NoStop}%
\bibitem [{\citenamefont {Garcia}\ \emph {et~al.}(2017)\citenamefont {Garcia},
  \citenamefont {Cummings},\ and\ \citenamefont {Roche}}]{garcia2017}%
  \BibitemOpen
  \bibfield  {author} {\bibinfo {author} {\bibfnamefont {Jose~H.}\ \bibnamefont
  {Garcia}}, \bibinfo {author} {\bibfnamefont {Aron~W.}\ \bibnamefont
  {Cummings}}, \ and\ \bibinfo {author} {\bibfnamefont {Stephan}\ \bibnamefont
  {Roche}},\ }\bibfield  {title} {\enquote {\bibinfo {title} {Spin hall effect
  and weak antilocalization in graphene/transition metal dichalcogenide
  heterostructures},}\ }\href {\doibase 10.1021/acs.nanolett.7b02364}
  {\bibfield  {journal} {\bibinfo  {journal} {Nano Letters}\ }\textbf {\bibinfo
  {volume} {17}},\ \bibinfo {pages} {5078--5083} (\bibinfo {year}
  {2017})}\BibitemShut {NoStop}%
\bibitem [{\citenamefont {Yan}\ \emph {et~al.}(2016)\citenamefont {Yan},
  \citenamefont {Txoperena}, \citenamefont {Llopis}, \citenamefont {Dery},
  \citenamefont {Hueso},\ and\ \citenamefont {Casanova}}]{yan2016}%
  \BibitemOpen
  \bibfield  {author} {\bibinfo {author} {\bibfnamefont {Wenjing}\ \bibnamefont
  {Yan}}, \bibinfo {author} {\bibfnamefont {Oihana}\ \bibnamefont {Txoperena}},
  \bibinfo {author} {\bibfnamefont {Roger}\ \bibnamefont {Llopis}}, \bibinfo
  {author} {\bibfnamefont {Hanan}\ \bibnamefont {Dery}}, \bibinfo {author}
  {\bibfnamefont {Luis~E.}\ \bibnamefont {Hueso}}, \ and\ \bibinfo {author}
  {\bibfnamefont {F{\`e}lix}\ \bibnamefont {Casanova}},\ }\bibfield  {title}
  {\enquote {\bibinfo {title} {A two-dimensional spin field-effect switch},}\
  }\href {\doibase 10.1038/ncomms13372} {\bibfield  {journal} {\bibinfo
  {journal} {Nature Communications}\ }\textbf {\bibinfo {volume} {7}},\
  \bibinfo {pages} {13372} (\bibinfo {year} {2016})}\BibitemShut {NoStop}%
\bibitem [{\citenamefont {Safeer}\ \emph {et~al.}(2019)\citenamefont {Safeer},
  \citenamefont {Ingla-Aynés}, \citenamefont {Herling}, \citenamefont
  {Garcia}, \citenamefont {Vila}, \citenamefont {Ontoso}, \citenamefont
  {Calvo}, \citenamefont {Roche}, \citenamefont {Hueso},\ and\ \citenamefont
  {Casanova}}]{safeer2019}%
  \BibitemOpen
  \bibfield  {author} {\bibinfo {author} {\bibfnamefont {C.~K.}\ \bibnamefont
  {Safeer}}, \bibinfo {author} {\bibfnamefont {Josep}\ \bibnamefont
  {Ingla-Aynés}}, \bibinfo {author} {\bibfnamefont {Franz}\ \bibnamefont
  {Herling}}, \bibinfo {author} {\bibfnamefont {José~H.}\ \bibnamefont
  {Garcia}}, \bibinfo {author} {\bibfnamefont {Marc}\ \bibnamefont {Vila}},
  \bibinfo {author} {\bibfnamefont {Nerea}\ \bibnamefont {Ontoso}}, \bibinfo
  {author} {\bibfnamefont {M.~Reyes}\ \bibnamefont {Calvo}}, \bibinfo {author}
  {\bibfnamefont {Stephan}\ \bibnamefont {Roche}}, \bibinfo {author}
  {\bibfnamefont {Luis~E.}\ \bibnamefont {Hueso}}, \ and\ \bibinfo {author}
  {\bibfnamefont {Fèlix}\ \bibnamefont {Casanova}},\ }\bibfield  {title}
  {\enquote {\bibinfo {title} {Room-temperature spin hall effect in
  graphene/mos2 van der {Waals} heterostructures},}\ }\href {\doibase
  10.1021/acs.nanolett.8b04368} {\bibfield  {journal} {\bibinfo  {journal}
  {Nano Letters}\ }\textbf {\bibinfo {volume} {19}},\ \bibinfo {pages}
  {1074--1082} (\bibinfo {year} {2019})}\BibitemShut {NoStop}%
\bibitem [{\citenamefont {Ghiasi}\ \emph {et~al.}(2019)\citenamefont {Ghiasi},
  \citenamefont {Kaverzin}, \citenamefont {Blah},\ and\ \citenamefont {van
  Wees}}]{ghiasi2019}%
  \BibitemOpen
  \bibfield  {author} {\bibinfo {author} {\bibfnamefont {Talieh~S.}\
  \bibnamefont {Ghiasi}}, \bibinfo {author} {\bibfnamefont {Alexey~A.}\
  \bibnamefont {Kaverzin}}, \bibinfo {author} {\bibfnamefont {Patrick~J.}\
  \bibnamefont {Blah}}, \ and\ \bibinfo {author} {\bibfnamefont {Bart~J.}\
  \bibnamefont {van Wees}},\ }\bibfield  {title} {\enquote {\bibinfo {title}
  {Charge-to-spin conversion by the rashba–edelstein effect in
  two-dimensional van der {Waals} heterostructures up to room temperature},}\
  }\href {\doibase 10.1021/acs.nanolett.9b01611} {\bibfield  {journal}
  {\bibinfo  {journal} {Nano Letters}\ }\textbf {\bibinfo {volume} {19}},\
  \bibinfo {pages} {5959--5966} (\bibinfo {year} {2019})}\BibitemShut {NoStop}%
\bibitem [{\citenamefont {Benitez}\ \emph {et~al.}(2019)\citenamefont
  {Benitez}, \citenamefont {Torres}, \citenamefont {Sierra}, \citenamefont
  {Timmermans}, \citenamefont {Garcia}, \citenamefont {Roche}, \citenamefont
  {Costache},\ and\ \citenamefont {Valenzuela}}]{benitez2019tunable}%
  \BibitemOpen
  \bibfield  {author} {\bibinfo {author} {\bibfnamefont {L~Antonio}\
  \bibnamefont {Benitez}}, \bibinfo {author} {\bibfnamefont {Williams~Savero}\
  \bibnamefont {Torres}}, \bibinfo {author} {\bibfnamefont {Juan~F}\
  \bibnamefont {Sierra}}, \bibinfo {author} {\bibfnamefont {Matias}\
  \bibnamefont {Timmermans}}, \bibinfo {author} {\bibfnamefont {Jose~H}\
  \bibnamefont {Garcia}}, \bibinfo {author} {\bibfnamefont {Stephan}\
  \bibnamefont {Roche}}, \bibinfo {author} {\bibfnamefont {Marius~V}\
  \bibnamefont {Costache}}, \ and\ \bibinfo {author} {\bibfnamefont {Sergio~O}\
  \bibnamefont {Valenzuela}},\ }\bibfield  {title} {\enquote {\bibinfo {title}
  {Tunable room-temperature spin galvanic and spin hall effects in van der
  {Waals} heterostructures},}\ }\href@noop {} {\bibfield  {journal} {\bibinfo
  {journal} {arXiv preprint arXiv:1908.07868}\ } (\bibinfo {year}
  {2019})}\BibitemShut {NoStop}%
\bibitem [{\citenamefont {Hoque}\ \emph {et~al.}(2019)\citenamefont {Hoque},
  \citenamefont {Khokhriakov}, \citenamefont {Karpiak},\ and\ \citenamefont
  {Dash}}]{hoque2019all}%
  \BibitemOpen
  \bibfield  {author} {\bibinfo {author} {\bibfnamefont {Anamul~Md}\
  \bibnamefont {Hoque}}, \bibinfo {author} {\bibfnamefont {Dmitrii}\
  \bibnamefont {Khokhriakov}}, \bibinfo {author} {\bibfnamefont {Bogdan}\
  \bibnamefont {Karpiak}}, \ and\ \bibinfo {author} {\bibfnamefont {Saroj~P}\
  \bibnamefont {Dash}},\ }\bibfield  {title} {\enquote {\bibinfo {title}
  {All-electrical creation and control of giant spin-galvanic effect in
  1t-mote2/graphene heterostructures at room temperature},}\ }\href@noop {}
  {\bibfield  {journal} {\bibinfo  {journal} {arXiv preprint arXiv:1908.09367}\
  } (\bibinfo {year} {2019})}\BibitemShut {NoStop}%
\bibitem [{\citenamefont {Li}\ \emph {et~al.}(2019)\citenamefont {Li},
  \citenamefont {Zhang}, \citenamefont {Myeong}, \citenamefont {Shin},
  \citenamefont {Lim}, \citenamefont {Kim}, \citenamefont {Kim}, \citenamefont
  {Jin}, \citenamefont {Kim}, \citenamefont {Kim} \emph
  {et~al.}}]{li2019electrical}%
  \BibitemOpen
  \bibfield  {author} {\bibinfo {author} {\bibfnamefont {Lijun}\ \bibnamefont
  {Li}}, \bibinfo {author} {\bibfnamefont {Jin}\ \bibnamefont {Zhang}},
  \bibinfo {author} {\bibfnamefont {Gyuho}\ \bibnamefont {Myeong}}, \bibinfo
  {author} {\bibfnamefont {Wongil}\ \bibnamefont {Shin}}, \bibinfo {author}
  {\bibfnamefont {Hongsik}\ \bibnamefont {Lim}}, \bibinfo {author}
  {\bibfnamefont {Boram}\ \bibnamefont {Kim}}, \bibinfo {author} {\bibfnamefont
  {Seungho}\ \bibnamefont {Kim}}, \bibinfo {author} {\bibfnamefont {Taehyeok}\
  \bibnamefont {Jin}}, \bibinfo {author} {\bibfnamefont {Bumseo}\ \bibnamefont
  {Kim}}, \bibinfo {author} {\bibfnamefont {Changyoung}\ \bibnamefont {Kim}},
  \emph {et~al.},\ }\bibfield  {title} {\enquote {\bibinfo {title} {Electrical
  control of the rashba-edelstein effect in a graphene/2h-tas2 van der {Waals}
  heterostructure at room temperature},}\ }\href@noop {} {\bibfield  {journal}
  {\bibinfo  {journal} {arXiv preprint arXiv:1906.10702}\ } (\bibinfo {year}
  {2019})}\BibitemShut {NoStop}%
\bibitem [{\citenamefont {Island}\ \emph {et~al.}(2019)\citenamefont {Island},
  \citenamefont {Cui}, \citenamefont {Lewandowski}, \citenamefont {Khoo},
  \citenamefont {Spanton}, \citenamefont {Zhou}, \citenamefont {Rhodes},
  \citenamefont {Hone}, \citenamefont {Taniguchi}, \citenamefont {Watanabe},
  \citenamefont {Levitov}, \citenamefont {Zaletel},\ and\ \citenamefont
  {Young}}]{Island2019}%
  \BibitemOpen
  \bibfield  {author} {\bibinfo {author} {\bibfnamefont {J~O}\ \bibnamefont
  {Island}}, \bibinfo {author} {\bibfnamefont {X}~\bibnamefont {Cui}}, \bibinfo
  {author} {\bibfnamefont {C}~\bibnamefont {Lewandowski}}, \bibinfo {author}
  {\bibfnamefont {J~Y}\ \bibnamefont {Khoo}}, \bibinfo {author} {\bibfnamefont
  {E~M}\ \bibnamefont {Spanton}}, \bibinfo {author} {\bibfnamefont
  {H}~\bibnamefont {Zhou}}, \bibinfo {author} {\bibfnamefont {D}~\bibnamefont
  {Rhodes}}, \bibinfo {author} {\bibfnamefont {J~C}\ \bibnamefont {Hone}},
  \bibinfo {author} {\bibfnamefont {T}~\bibnamefont {Taniguchi}}, \bibinfo
  {author} {\bibfnamefont {K}~\bibnamefont {Watanabe}}, \bibinfo {author}
  {\bibfnamefont {L~S}\ \bibnamefont {Levitov}}, \bibinfo {author}
  {\bibfnamefont {M~P}\ \bibnamefont {Zaletel}}, \ and\ \bibinfo {author}
  {\bibfnamefont {A~F}\ \bibnamefont {Young}},\ }\bibfield  {title} {\enquote
  {\bibinfo {title} {{Spin–orbit-driven band inversion in bilayer graphene by
  the van der {Waals} proximity effect}},}\ }\href {\doibase
  10.1038/s41586-019-1304-2} {\bibfield  {journal} {\bibinfo  {journal}
  {Nature}\ }\textbf {\bibinfo {volume} {571}},\ \bibinfo {pages} {85}
  (\bibinfo {year} {2019})}\BibitemShut {NoStop}%
\bibitem [{\citenamefont {Fu}\ and\ \citenamefont {Kane}(2008)}]{fu2008}%
  \BibitemOpen
  \bibfield  {author} {\bibinfo {author} {\bibfnamefont {L.}~\bibnamefont
  {Fu}}\ and\ \bibinfo {author} {\bibfnamefont {C.~L.}\ \bibnamefont {Kane}},\
  }\bibfield  {title} {\enquote {\bibinfo {title} {Superconducting proximity
  effect and {Majorana} fermions at the surface of a topological insulator},}\
  }\href@noop {} {\bibfield  {journal} {\bibinfo  {journal} {Phys. Rev. Lett.}\
  }\textbf {\bibinfo {volume} {100}},\ \bibinfo {pages} {096407} (\bibinfo
  {year} {2008})}\BibitemShut {NoStop}%
\bibitem [{\citenamefont {Sau}\ \emph {et~al.}(2010)\citenamefont {Sau},
  \citenamefont {Lutchyn}, \citenamefont {Tewari},\ and\ \citenamefont
  {Sarma}}]{sau2010}%
  \BibitemOpen
  \bibfield  {author} {\bibinfo {author} {\bibfnamefont {J.~D.}\ \bibnamefont
  {Sau}}, \bibinfo {author} {\bibfnamefont {R.~M.}\ \bibnamefont {Lutchyn}},
  \bibinfo {author} {\bibfnamefont {S.}~\bibnamefont {Tewari}}, \ and\ \bibinfo
  {author} {\bibfnamefont {S.~Das}\ \bibnamefont {Sarma}},\ }\bibfield  {title}
  {\enquote {\bibinfo {title} {Generic {New} {Platform} for {Topological}
  {Quantum} {Computation} {Using} {Semiconductor} heterostructures},}\
  }\href@noop {} {\bibfield  {journal} {\bibinfo  {journal} {Phys. Rev. Lett.}\
  }\textbf {\bibinfo {volume} {104}},\ \bibinfo {pages} {040502} (\bibinfo
  {year} {2010})}\BibitemShut {NoStop}%
\bibitem [{\citenamefont {Lutchyn}\ \emph {et~al.}(2010)\citenamefont
  {Lutchyn}, \citenamefont {Sau},\ and\ \citenamefont {Sarma}}]{lutchyn2010}%
  \BibitemOpen
  \bibfield  {author} {\bibinfo {author} {\bibfnamefont {R.~M.}\ \bibnamefont
  {Lutchyn}}, \bibinfo {author} {\bibfnamefont {J.~D.}\ \bibnamefont {Sau}}, \
  and\ \bibinfo {author} {\bibfnamefont {S.~Das}\ \bibnamefont {Sarma}},\
  }\bibfield  {title} {\enquote {\bibinfo {title} {Majorana fermions and a
  topological phase transition in semiconductor-superconductor
  heterostructures},}\ }\href@noop {} {\bibfield  {journal} {\bibinfo
  {journal} {Phys. Rev. Lett.}\ }\textbf {\bibinfo {volume} {105}},\ \bibinfo
  {pages} {077001} (\bibinfo {year} {2010})}\BibitemShut {NoStop}%
\bibitem [{\citenamefont {Oreg}\ \emph {et~al.}(2010)\citenamefont {Oreg},
  \citenamefont {Refael},\ and\ \citenamefont {von Oppen}}]{oreg2010}%
  \BibitemOpen
  \bibfield  {author} {\bibinfo {author} {\bibfnamefont {Y.}~\bibnamefont
  {Oreg}}, \bibinfo {author} {\bibfnamefont {G.}~\bibnamefont {Refael}}, \ and\
  \bibinfo {author} {\bibfnamefont {F.}~\bibnamefont {von Oppen}},\ }\bibfield
  {title} {\enquote {\bibinfo {title} {Helical {Liquids} and {Majorana} {Bound}
  {States} in {Quantum} wires},}\ }\href@noop {} {\bibfield  {journal}
  {\bibinfo  {journal} {Phys. Rev. Lett.}\ }\textbf {\bibinfo {volume} {105}},\
  \bibinfo {pages} {177002} (\bibinfo {year} {2010})}\BibitemShut {NoStop}%
\bibitem [{\citenamefont {Antipov}\ \emph {et~al.}(2018)\citenamefont
  {Antipov}, \citenamefont {Bargerbos}, \citenamefont {Winkler}, \citenamefont
  {Bauer}, \citenamefont {Rossi},\ and\ \citenamefont {Lutchyn}}]{antipov2018}%
  \BibitemOpen
  \bibfield  {author} {\bibinfo {author} {\bibfnamefont {Andrey~E.}\
  \bibnamefont {Antipov}}, \bibinfo {author} {\bibfnamefont {Arno}\
  \bibnamefont {Bargerbos}}, \bibinfo {author} {\bibfnamefont {Georg~W.}\
  \bibnamefont {Winkler}}, \bibinfo {author} {\bibfnamefont {Bela}\
  \bibnamefont {Bauer}}, \bibinfo {author} {\bibfnamefont {Enrico}\
  \bibnamefont {Rossi}}, \ and\ \bibinfo {author} {\bibfnamefont {Roman~M.}\
  \bibnamefont {Lutchyn}},\ }\bibfield  {title} {\enquote {\bibinfo {title}
  {Effects of gate-induced electric fields on semiconductor majorana
  nanowires},}\ }\href {\doibase 10.1103/PhysRevX.8.031041} {\bibfield
  {journal} {\bibinfo  {journal} {Phys. Rev. X}\ }\textbf {\bibinfo {volume}
  {8}},\ \bibinfo {pages} {031041} (\bibinfo {year} {2018})}\BibitemShut
  {NoStop}%
\bibitem [{\citenamefont {Bergeret}\ \emph {et~al.}(2005)\citenamefont
  {Bergeret}, \citenamefont {Volkov},\ and\ \citenamefont
  {Efetov}}]{bergeret2005}%
  \BibitemOpen
  \bibfield  {author} {\bibinfo {author} {\bibfnamefont {F.~S.}\ \bibnamefont
  {Bergeret}}, \bibinfo {author} {\bibfnamefont {A.~F.}\ \bibnamefont
  {Volkov}}, \ and\ \bibinfo {author} {\bibfnamefont {K.~B.}\ \bibnamefont
  {Efetov}},\ }\bibfield  {title} {\enquote {\bibinfo {title} {Odd triplet
  superconductivity and related phenomena in superconductor-ferromagnet
  structures},}\ }\href@noop {} {\bibfield  {journal} {\bibinfo  {journal}
  {Rev. Modern Phys.}\ }\textbf {\bibinfo {volume} {77}},\ \bibinfo {pages}
  {1321--1373} (\bibinfo {year} {2005})}\BibitemShut {NoStop}%
\bibitem [{\citenamefont {{Linder}}\ and\ \citenamefont
  {{Balatsky}}(2017)}]{linder2017}%
  \BibitemOpen
  \bibfield  {author} {\bibinfo {author} {\bibfnamefont {J.}~\bibnamefont
  {{Linder}}}\ and\ \bibinfo {author} {\bibfnamefont {A.~V.}\ \bibnamefont
  {{Balatsky}}},\ }\bibfield  {title} {\enquote {\bibinfo {title}
  {{Odd-frequency superconductivity}},}\ }\href@noop {} {\bibfield  {journal}
  {\bibinfo  {journal} {ArXiv e-prints}\ } (\bibinfo {year}
  {2017})}\BibitemShut {NoStop}%
\bibitem [{\citenamefont {{Berezinski{\v i}}}(1974)}]{berezinskii1974}%
  \BibitemOpen
  \bibfield  {author} {\bibinfo {author} {\bibfnamefont {V.~L.}\ \bibnamefont
  {{Berezinski{\v i}}}},\ }\bibfield  {title} {\enquote {\bibinfo {title} {{New
  model of the anisotropic phase of superfluid He$^{3}$}},}\ }\href@noop {}
  {\bibfield  {journal} {\bibinfo  {journal} {Soviet Journal of Experimental
  and Theoretical Physics Letters}\ }\textbf {\bibinfo {volume} {20}},\
  \bibinfo {pages} {287} (\bibinfo {year} {1974})}\BibitemShut {NoStop}%
\bibitem [{\citenamefont {Kirkpatrick}\ and\ \citenamefont
  {Belitz}(1991)}]{kirkpatrick_1991_prl}%
  \BibitemOpen
  \bibfield  {author} {\bibinfo {author} {\bibfnamefont {T.~R.}\ \bibnamefont
  {Kirkpatrick}}\ and\ \bibinfo {author} {\bibfnamefont {D.}~\bibnamefont
  {Belitz}},\ }\bibfield  {title} {\enquote {\bibinfo {title} {Disorder-induced
  triplet superconductivity},}\ }\href {\doibase 10.1103/PhysRevLett.66.1533}
  {\bibfield  {journal} {\bibinfo  {journal} {Phys. Rev. Lett.}\ }\textbf
  {\bibinfo {volume} {66}},\ \bibinfo {pages} {1533--1536} (\bibinfo {year}
  {1991})}\BibitemShut {NoStop}%
\bibitem [{\citenamefont {Belitz}\ and\ \citenamefont
  {Kirkpatrick}(1992)}]{belitz_1992_prb}%
  \BibitemOpen
  \bibfield  {author} {\bibinfo {author} {\bibfnamefont {D.}~\bibnamefont
  {Belitz}}\ and\ \bibinfo {author} {\bibfnamefont {T.~R.}\ \bibnamefont
  {Kirkpatrick}},\ }\bibfield  {title} {\enquote {\bibinfo {title} {Even-parity
  spin-triplet superconductivity in disordered electronic systems},}\ }\href
  {\doibase 10.1103/PhysRevB.46.8393} {\bibfield  {journal} {\bibinfo
  {journal} {Phys. Rev. B}\ }\textbf {\bibinfo {volume} {46}},\ \bibinfo
  {pages} {8393--8408} (\bibinfo {year} {1992})}\BibitemShut {NoStop}%
\bibitem [{\citenamefont {{Balatsky}}\ and\ \citenamefont
  {{Abrahams}}(1992)}]{balatsky1992}%
  \BibitemOpen
  \bibfield  {author} {\bibinfo {author} {\bibfnamefont {A.}~\bibnamefont
  {{Balatsky}}}\ and\ \bibinfo {author} {\bibfnamefont {E.}~\bibnamefont
  {{Abrahams}}},\ }\bibfield  {title} {\enquote {\bibinfo {title} {{New class
  of singlet superconductors which break the time reversal and parity}},}\
  }\href@noop {} {\bibfield  {journal} {\bibinfo  {journal} {Phys. Rev. B}\
  }\textbf {\bibinfo {volume} {45}},\ \bibinfo {pages} {13125--13128} (\bibinfo
  {year} {1992})}\BibitemShut {NoStop}%
\bibitem [{\citenamefont {Coleman}\ \emph {et~al.}(1993)\citenamefont
  {Coleman}, \citenamefont {Miranda},\ and\ \citenamefont
  {Tsvelik}}]{coleman_1993_prl}%
  \BibitemOpen
  \bibfield  {author} {\bibinfo {author} {\bibfnamefont {P.}~\bibnamefont
  {Coleman}}, \bibinfo {author} {\bibfnamefont {E.}~\bibnamefont {Miranda}}, \
  and\ \bibinfo {author} {\bibfnamefont {A.}~\bibnamefont {Tsvelik}},\
  }\bibfield  {title} {\enquote {\bibinfo {title} {Possible realization of
  odd-frequency pairing in heavy fermion compounds},}\ }\href {\doibase
  10.1103/PhysRevLett.70.2960} {\bibfield  {journal} {\bibinfo  {journal}
  {Phys. Rev. Lett.}\ }\textbf {\bibinfo {volume} {70}},\ \bibinfo {pages}
  {2960--2963} (\bibinfo {year} {1993})}\BibitemShut {NoStop}%
\bibitem [{\citenamefont {Coleman}\ \emph {et~al.}(1995)\citenamefont
  {Coleman}, \citenamefont {Miranda},\ and\ \citenamefont
  {Tsvelik}}]{coleman_1995_prl}%
  \BibitemOpen
  \bibfield  {author} {\bibinfo {author} {\bibfnamefont {P.}~\bibnamefont
  {Coleman}}, \bibinfo {author} {\bibfnamefont {E.}~\bibnamefont {Miranda}}, \
  and\ \bibinfo {author} {\bibfnamefont {A.}~\bibnamefont {Tsvelik}},\
  }\bibfield  {title} {\enquote {\bibinfo {title} {Three-body bound states and
  the development of odd-frequency pairing},}\ }\href {\doibase
  10.1103/PhysRevLett.74.1653} {\bibfield  {journal} {\bibinfo  {journal}
  {Phys. Rev. Lett.}\ }\textbf {\bibinfo {volume} {74}},\ \bibinfo {pages}
  {1653--1656} (\bibinfo {year} {1995})}\BibitemShut {NoStop}%
\bibitem [{\citenamefont {Belitz}\ and\ \citenamefont
  {Kirkpatrick}(1999)}]{belitz_1999_prb}%
  \BibitemOpen
  \bibfield  {author} {\bibinfo {author} {\bibfnamefont {D.}~\bibnamefont
  {Belitz}}\ and\ \bibinfo {author} {\bibfnamefont {T.~R.}\ \bibnamefont
  {Kirkpatrick}},\ }\bibfield  {title} {\enquote {\bibinfo {title} {Properties
  of spin-triplet, even-parity superconductors},}\ }\href {\doibase
  10.1103/PhysRevB.60.3485} {\bibfield  {journal} {\bibinfo  {journal} {Phys.
  Rev. B}\ }\textbf {\bibinfo {volume} {60}},\ \bibinfo {pages} {3485--3498}
  (\bibinfo {year} {1999})}\BibitemShut {NoStop}%
\bibitem [{\citenamefont {Abrahams}\ \emph {et~al.}(1993)\citenamefont
  {Abrahams}, \citenamefont {Balatsky}, \citenamefont {Schrieffer},\ and\
  \citenamefont {Allen}}]{abrahams1993interactions}%
  \BibitemOpen
  \bibfield  {author} {\bibinfo {author} {\bibfnamefont {Elihu}\ \bibnamefont
  {Abrahams}}, \bibinfo {author} {\bibfnamefont {Alexander}\ \bibnamefont
  {Balatsky}}, \bibinfo {author} {\bibfnamefont {JR}~\bibnamefont
  {Schrieffer}}, \ and\ \bibinfo {author} {\bibfnamefont {Philip~B}\
  \bibnamefont {Allen}},\ }\bibfield  {title} {\enquote {\bibinfo {title}
  {Interactions for odd-$\omega$-gap singlet superconductors},}\ }\href@noop {}
  {\bibfield  {journal} {\bibinfo  {journal} {Phys. Rev. B}\ }\textbf {\bibinfo
  {volume} {47}},\ \bibinfo {pages} {513} (\bibinfo {year} {1993})}\BibitemShut
  {NoStop}%
\bibitem [{\citenamefont {Coleman}\ \emph {et~al.}(1994)\citenamefont
  {Coleman}, \citenamefont {Miranda},\ and\ \citenamefont
  {Tsvelik}}]{coleman1994}%
  \BibitemOpen
  \bibfield  {author} {\bibinfo {author} {\bibfnamefont {P.}~\bibnamefont
  {Coleman}}, \bibinfo {author} {\bibfnamefont {E.}~\bibnamefont {Miranda}}, \
  and\ \bibinfo {author} {\bibfnamefont {A.}~\bibnamefont {Tsvelik}},\
  }\bibfield  {title} {\enquote {\bibinfo {title} {Odd-frequency pairing in the
  kondo-lattice},}\ }\href@noop {} {\bibfield  {journal} {\bibinfo  {journal}
  {Phys. Rev. B}\ }\textbf {\bibinfo {volume} {49}},\ \bibinfo {pages}
  {8955--8982} (\bibinfo {year} {1994})}\BibitemShut {NoStop}%
\bibitem [{\citenamefont {Dolgov}\ and\ \citenamefont
  {Losyakov}(1994)}]{dolgov1994renormalization}%
  \BibitemOpen
  \bibfield  {author} {\bibinfo {author} {\bibfnamefont {OV}~\bibnamefont
  {Dolgov}}\ and\ \bibinfo {author} {\bibfnamefont {VV}~\bibnamefont
  {Losyakov}},\ }\bibfield  {title} {\enquote {\bibinfo {title}
  {Renormalization factor and odd-$\omega$ gap singlet superconductivity},}\
  }\href@noop {} {\bibfield  {journal} {\bibinfo  {journal} {Phys. Lett. A}\
  }\textbf {\bibinfo {volume} {190}},\ \bibinfo {pages} {189--190} (\bibinfo
  {year} {1994})}\BibitemShut {NoStop}%
\bibitem [{\citenamefont {Heid}(1995)}]{heid1995}%
  \BibitemOpen
  \bibfield  {author} {\bibinfo {author} {\bibfnamefont {R.}~\bibnamefont
  {Heid}},\ }\bibfield  {title} {\enquote {\bibinfo {title} {On the
  thermodynamic stability of odd-frequency superconductors},}\ }\href@noop {}
  {\bibfield  {journal} {\bibinfo  {journal} {Z. Fur Physik B-condensed
  Matter}\ }\textbf {\bibinfo {volume} {99}},\ \bibinfo {pages} {15--18}
  (\bibinfo {year} {1995})}\BibitemShut {NoStop}%
\bibitem [{\citenamefont {Bergeret}\ \emph {et~al.}(2001)\citenamefont
  {Bergeret}, \citenamefont {Volkov},\ and\ \citenamefont
  {Efetov}}]{BergeretPRL2001}%
  \BibitemOpen
  \bibfield  {author} {\bibinfo {author} {\bibfnamefont {F.S.}\ \bibnamefont
  {Bergeret}}, \bibinfo {author} {\bibfnamefont {A.F.}\ \bibnamefont {Volkov}},
  \ and\ \bibinfo {author} {\bibfnamefont {K.B.}\ \bibnamefont {Efetov}},\
  }\bibfield  {title} {\enquote {\bibinfo {title} {Long-range proximity effects
  in superconductor-ferromagnet structures},}\ }\href@noop {} {\bibfield
  {journal} {\bibinfo  {journal} {Phys. Rev. Lett.}\ }\textbf {\bibinfo
  {volume} {86}},\ \bibinfo {pages} {4096} (\bibinfo {year}
  {2001})}\BibitemShut {NoStop}%
\bibitem [{\citenamefont {Halterman}\ \emph {et~al.}(2007)\citenamefont
  {Halterman}, \citenamefont {Barsic},\ and\ \citenamefont
  {Valls}}]{halterman2007odd}%
  \BibitemOpen
  \bibfield  {author} {\bibinfo {author} {\bibfnamefont {Klaus}\ \bibnamefont
  {Halterman}}, \bibinfo {author} {\bibfnamefont {Paul~H}\ \bibnamefont
  {Barsic}}, \ and\ \bibinfo {author} {\bibfnamefont {Oriol~T}\ \bibnamefont
  {Valls}},\ }\bibfield  {title} {\enquote {\bibinfo {title} {Odd triplet
  pairing in clean superconductor/ferromagnet heterostructures},}\ }\href@noop
  {} {\bibfield  {journal} {\bibinfo  {journal} {Phys. Rev. Lett.}\ }\textbf
  {\bibinfo {volume} {99}},\ \bibinfo {pages} {127002} (\bibinfo {year}
  {2007})}\BibitemShut {NoStop}%
\bibitem [{\citenamefont {Yokoyama}\ \emph {et~al.}(2007)\citenamefont
  {Yokoyama}, \citenamefont {Tanaka},\ and\ \citenamefont
  {Golubov}}]{yokoyama2007manifestation}%
  \BibitemOpen
  \bibfield  {author} {\bibinfo {author} {\bibfnamefont {T.}~\bibnamefont
  {Yokoyama}}, \bibinfo {author} {\bibfnamefont {Y.}~\bibnamefont {Tanaka}}, \
  and\ \bibinfo {author} {\bibfnamefont {A.~A.}\ \bibnamefont {Golubov}},\
  }\bibfield  {title} {\enquote {\bibinfo {title} {Manifestation of the
  odd-frequency spin-triplet pairing state in diffusive
  ferromagnet/superconductor junctions},}\ }\href {\doibase
  10.1103/PhysRevB.75.134510} {\bibfield  {journal} {\bibinfo  {journal} {Phys.
  Rev. B}\ }\textbf {\bibinfo {volume} {75}},\ \bibinfo {pages} {134510}
  (\bibinfo {year} {2007})}\BibitemShut {NoStop}%
\bibitem [{\citenamefont {Houzet}(2008)}]{houzet2008ferromagnetic}%
  \BibitemOpen
  \bibfield  {author} {\bibinfo {author} {\bibfnamefont {Manuel}\ \bibnamefont
  {Houzet}},\ }\bibfield  {title} {\enquote {\bibinfo {title} {Ferromagnetic
  josephson junction with precessing magnetization},}\ }\href {\doibase
  10.1103/PhysRevLett.101.057009} {\bibfield  {journal} {\bibinfo  {journal}
  {Phys. Rev. Lett.}\ }\textbf {\bibinfo {volume} {101}},\ \bibinfo {pages}
  {057009} (\bibinfo {year} {2008})}\BibitemShut {NoStop}%
\bibitem [{\citenamefont {Eschrig}\ and\ \citenamefont
  {L{\"o}fwander}(2008)}]{EschrigNat2008}%
  \BibitemOpen
  \bibfield  {author} {\bibinfo {author} {\bibfnamefont {M.}~\bibnamefont
  {Eschrig}}\ and\ \bibinfo {author} {\bibfnamefont {T.}~\bibnamefont
  {L{\"o}fwander}},\ }\bibfield  {title} {\enquote {\bibinfo {title} {Triplet
  supercurrents in clean and disordered half-metallic ferromagnets},}\
  }\href@noop {} {\bibfield  {journal} {\bibinfo  {journal} {Nature Physics}\
  }\textbf {\bibinfo {volume} {4}},\ \bibinfo {pages} {138--143} (\bibinfo
  {year} {2008})}\BibitemShut {NoStop}%
\bibitem [{\citenamefont {Linder}\ \emph {et~al.}(2008)\citenamefont {Linder},
  \citenamefont {Yokoyama},\ and\ \citenamefont {Sudb{\o}}}]{LinderPRB2008}%
  \BibitemOpen
  \bibfield  {author} {\bibinfo {author} {\bibfnamefont {J.}~\bibnamefont
  {Linder}}, \bibinfo {author} {\bibfnamefont {T.}~\bibnamefont {Yokoyama}}, \
  and\ \bibinfo {author} {\bibfnamefont {A.}~\bibnamefont {Sudb{\o}}},\
  }\bibfield  {title} {\enquote {\bibinfo {title} {Role of interface
  transparency and spin-dependent scattering in diffusive
  ferromagnet/superconductor heterostructures},}\ }\href@noop {} {\bibfield
  {journal} {\bibinfo  {journal} {Phys. Rev. B}\ }\textbf {\bibinfo {volume}
  {77}},\ \bibinfo {pages} {174514} (\bibinfo {year} {2008})}\BibitemShut
  {NoStop}%
\bibitem [{\citenamefont {Triola}\ \emph {et~al.}(2014)\citenamefont {Triola},
  \citenamefont {Rossi},\ and\ \citenamefont {Balatsky}}]{TriolaPRB2014}%
  \BibitemOpen
  \bibfield  {author} {\bibinfo {author} {\bibfnamefont {Christopher}\
  \bibnamefont {Triola}}, \bibinfo {author} {\bibfnamefont {E}~\bibnamefont
  {Rossi}}, \ and\ \bibinfo {author} {\bibfnamefont {Alexander~V}\ \bibnamefont
  {Balatsky}},\ }\bibfield  {title} {\enquote {\bibinfo {title} {{Effect of a
  spin-active interface on proximity-induced superconductivity in topological
  insulators}},}\ }\href {\doibase 10.1103/PhysRevB.89.165309} {\bibfield
  {journal} {\bibinfo  {journal} {Physical Review B}\ }\textbf {\bibinfo
  {volume} {89}},\ \bibinfo {pages} {165309} (\bibinfo {year}
  {2014})}\BibitemShut {NoStop}%
\bibitem [{\citenamefont {Cr\'epin}\ \emph {et~al.}(2015)\citenamefont
  {Cr\'epin}, \citenamefont {Burset},\ and\ \citenamefont
  {Trauzettel}}]{crepin2015odd}%
  \BibitemOpen
  \bibfield  {author} {\bibinfo {author} {\bibfnamefont {F}~\bibnamefont
  {Cr\'epin}}, \bibinfo {author} {\bibfnamefont {Pablo}\ \bibnamefont
  {Burset}}, \ and\ \bibinfo {author} {\bibfnamefont {Bj\"orn}\ \bibnamefont
  {Trauzettel}},\ }\bibfield  {title} {\enquote {\bibinfo {title}
  {Odd-frequency triplet superconductivity at the helical edge of a topological
  insulator},}\ }\href {\doibase 10.1103/PhysRevB.92.100507} {\bibfield
  {journal} {\bibinfo  {journal} {Phys. Rev. B}\ }\textbf {\bibinfo {volume}
  {92}},\ \bibinfo {pages} {100507} (\bibinfo {year} {2015})}\BibitemShut
  {NoStop}%
\bibitem [{\citenamefont {Petrashov}\ \emph {et~al.}(1994)\citenamefont
  {Petrashov}, \citenamefont {Antonov}, \citenamefont {Maksimov},\ and\
  \citenamefont {Shaikhaidarov}}]{petrashov1994conductivity}%
  \BibitemOpen
  \bibfield  {author} {\bibinfo {author} {\bibfnamefont {VT}~\bibnamefont
  {Petrashov}}, \bibinfo {author} {\bibfnamefont {VN}~\bibnamefont {Antonov}},
  \bibinfo {author} {\bibfnamefont {SV}~\bibnamefont {Maksimov}}, \ and\
  \bibinfo {author} {\bibfnamefont {R~Sh}\ \bibnamefont {Shaikhaidarov}},\
  }\bibfield  {title} {\enquote {\bibinfo {title} {Conductivity of mesoscopic
  structures with ferromagnetic and superconducting regions},}\ }\href@noop {}
  {\bibfield  {journal} {\bibinfo  {journal} {JETP Lett.}\ }\textbf {\bibinfo
  {volume} {59}},\ \bibinfo {pages} {551--555} (\bibinfo {year}
  {1994})}\BibitemShut {NoStop}%
\bibitem [{\citenamefont {Giroud}\ \emph {et~al.}(1998)\citenamefont {Giroud},
  \citenamefont {Courtois}, \citenamefont {Hasselbach}, \citenamefont
  {Mailly},\ and\ \citenamefont {Pannetier}}]{giroud1998superconducting}%
  \BibitemOpen
  \bibfield  {author} {\bibinfo {author} {\bibfnamefont {M}~\bibnamefont
  {Giroud}}, \bibinfo {author} {\bibfnamefont {H}~\bibnamefont {Courtois}},
  \bibinfo {author} {\bibfnamefont {K}~\bibnamefont {Hasselbach}}, \bibinfo
  {author} {\bibfnamefont {D}~\bibnamefont {Mailly}}, \ and\ \bibinfo {author}
  {\bibfnamefont {B}~\bibnamefont {Pannetier}},\ }\bibfield  {title} {\enquote
  {\bibinfo {title} {Superconducting proximity effect in a mesoscopic
  ferromagnetic wire},}\ }\href@noop {} {\bibfield  {journal} {\bibinfo
  {journal} {Phys. Rev. B}\ }\textbf {\bibinfo {volume} {58}},\ \bibinfo
  {pages} {R11872} (\bibinfo {year} {1998})}\BibitemShut {NoStop}%
\bibitem [{\citenamefont {Petrashov}\ \emph {et~al.}(1999)\citenamefont
  {Petrashov}, \citenamefont {Sosnin}, \citenamefont {Cox}, \citenamefont
  {Parsons},\ and\ \citenamefont {Troadec}}]{petrashov1999giant}%
  \BibitemOpen
  \bibfield  {author} {\bibinfo {author} {\bibfnamefont {VT}~\bibnamefont
  {Petrashov}}, \bibinfo {author} {\bibfnamefont {IA}~\bibnamefont {Sosnin}},
  \bibinfo {author} {\bibfnamefont {I}~\bibnamefont {Cox}}, \bibinfo {author}
  {\bibfnamefont {A}~\bibnamefont {Parsons}}, \ and\ \bibinfo {author}
  {\bibfnamefont {C}~\bibnamefont {Troadec}},\ }\bibfield  {title} {\enquote
  {\bibinfo {title} {Giant mutual proximity effects in
  ferromagnetic/superconducting nanostructures},}\ }\href@noop {} {\bibfield
  {journal} {\bibinfo  {journal} {Phys. Rev. Lett.}\ }\textbf {\bibinfo
  {volume} {83}},\ \bibinfo {pages} {3281} (\bibinfo {year}
  {1999})}\BibitemShut {NoStop}%
\bibitem [{\citenamefont {Aumentado}\ and\ \citenamefont
  {Chandrasekhar}(2001)}]{aumentado2001mesoscopic}%
  \BibitemOpen
  \bibfield  {author} {\bibinfo {author} {\bibfnamefont {Jose}\ \bibnamefont
  {Aumentado}}\ and\ \bibinfo {author} {\bibfnamefont {Venkat}\ \bibnamefont
  {Chandrasekhar}},\ }\bibfield  {title} {\enquote {\bibinfo {title}
  {Mesoscopic ferromagnet-superconductor junctions and the proximity effect},}\
  }\href@noop {} {\bibfield  {journal} {\bibinfo  {journal} {Phys. Rev. B}\
  }\textbf {\bibinfo {volume} {64}},\ \bibinfo {pages} {054505} (\bibinfo
  {year} {2001})}\BibitemShut {NoStop}%
\bibitem [{\citenamefont {Zhu}\ \emph {et~al.}(2010)\citenamefont {Zhu},
  \citenamefont {Krivorotov}, \citenamefont {Halterman},\ and\ \citenamefont
  {Valls}}]{zhu2010angular}%
  \BibitemOpen
  \bibfield  {author} {\bibinfo {author} {\bibfnamefont {Jian}\ \bibnamefont
  {Zhu}}, \bibinfo {author} {\bibfnamefont {Ilya~N}\ \bibnamefont
  {Krivorotov}}, \bibinfo {author} {\bibfnamefont {Klaus}\ \bibnamefont
  {Halterman}}, \ and\ \bibinfo {author} {\bibfnamefont {Oriol~T}\ \bibnamefont
  {Valls}},\ }\bibfield  {title} {\enquote {\bibinfo {title} {Angular
  dependence of the superconducting transition temperature in
  ferromagnet-superconductor-ferromagnet trilayers},}\ }\href@noop {}
  {\bibfield  {journal} {\bibinfo  {journal} {Phys. Rev. Lett.}\ }\textbf
  {\bibinfo {volume} {105}},\ \bibinfo {pages} {207002} (\bibinfo {year}
  {2010})}\BibitemShut {NoStop}%
\bibitem [{\citenamefont {Di~Bernardo}\ \emph
  {et~al.}(2015{\natexlab{a}})\citenamefont {Di~Bernardo}, \citenamefont
  {Diesch}, \citenamefont {Gu}, \citenamefont {Linder}, \citenamefont
  {Divitini}, \citenamefont {Ducati}, \citenamefont {Scheer}, \citenamefont
  {Blamire},\ and\ \citenamefont {Robinson}}]{di2015signature}%
  \BibitemOpen
  \bibfield  {author} {\bibinfo {author} {\bibfnamefont {Angelo}\ \bibnamefont
  {Di~Bernardo}}, \bibinfo {author} {\bibfnamefont {Simon}\ \bibnamefont
  {Diesch}}, \bibinfo {author} {\bibfnamefont {Y}~\bibnamefont {Gu}}, \bibinfo
  {author} {\bibfnamefont {Jacob}\ \bibnamefont {Linder}}, \bibinfo {author}
  {\bibfnamefont {Giorgio}\ \bibnamefont {Divitini}}, \bibinfo {author}
  {\bibfnamefont {Caterina}\ \bibnamefont {Ducati}}, \bibinfo {author}
  {\bibfnamefont {Elke}\ \bibnamefont {Scheer}}, \bibinfo {author}
  {\bibfnamefont {Mark~G}\ \bibnamefont {Blamire}}, \ and\ \bibinfo {author}
  {\bibfnamefont {Jason~WA}\ \bibnamefont {Robinson}},\ }\bibfield  {title}
  {\enquote {\bibinfo {title} {Signature of magnetic-dependent gapless odd
  frequency states at superconductor/ferromagnet interfaces},}\ }\href
  {\doibase 10.1038/ncomms9053} {\bibfield  {journal} {\bibinfo  {journal}
  {Nat. Commun.}\ }\textbf {\bibinfo {volume} {6}},\ \bibinfo {pages} {8053}
  (\bibinfo {year} {2015}{\natexlab{a}})}\BibitemShut {NoStop}%
\bibitem [{\citenamefont {Di~Bernardo}\ \emph
  {et~al.}(2015{\natexlab{b}})\citenamefont {Di~Bernardo}, \citenamefont
  {Salman}, \citenamefont {Wang}, \citenamefont {Amado}, \citenamefont
  {Egilmez}, \citenamefont {Flokstra}, \citenamefont {Suter}, \citenamefont
  {Lee}, \citenamefont {Zhao}, \citenamefont {Prokscha}, \citenamefont
  {Morenzoni}, \citenamefont {Blamire}, \citenamefont {Linder},\ and\
  \citenamefont {Robinson}}]{di2015intrinsic}%
  \BibitemOpen
  \bibfield  {author} {\bibinfo {author} {\bibfnamefont {A.}~\bibnamefont
  {Di~Bernardo}}, \bibinfo {author} {\bibfnamefont {Z.}~\bibnamefont {Salman}},
  \bibinfo {author} {\bibfnamefont {X.~L.}\ \bibnamefont {Wang}}, \bibinfo
  {author} {\bibfnamefont {M.}~\bibnamefont {Amado}}, \bibinfo {author}
  {\bibfnamefont {M.}~\bibnamefont {Egilmez}}, \bibinfo {author} {\bibfnamefont
  {M.~G.}\ \bibnamefont {Flokstra}}, \bibinfo {author} {\bibfnamefont
  {A.}~\bibnamefont {Suter}}, \bibinfo {author} {\bibfnamefont {S.~L.}\
  \bibnamefont {Lee}}, \bibinfo {author} {\bibfnamefont {J.~H.}\ \bibnamefont
  {Zhao}}, \bibinfo {author} {\bibfnamefont {T.}~\bibnamefont {Prokscha}},
  \bibinfo {author} {\bibfnamefont {E.}~\bibnamefont {Morenzoni}}, \bibinfo
  {author} {\bibfnamefont {M.~G.}\ \bibnamefont {Blamire}}, \bibinfo {author}
  {\bibfnamefont {J.}~\bibnamefont {Linder}}, \ and\ \bibinfo {author}
  {\bibfnamefont {J.~W.~A.}\ \bibnamefont {Robinson}},\ }\bibfield  {title}
  {\enquote {\bibinfo {title} {Intrinsic paramagnetic meissner effect due to
  $s$-wave odd-frequency superconductivity},}\ }\href {\doibase
  10.1103/PhysRevX.5.041021} {\bibfield  {journal} {\bibinfo  {journal} {Phys.
  Rev. X}\ }\textbf {\bibinfo {volume} {5}},\ \bibinfo {pages} {041021}
  (\bibinfo {year} {2015}{\natexlab{b}})}\BibitemShut {NoStop}%
\bibitem [{\citenamefont {Tanaka}\ and\ \citenamefont
  {Golubov}(2007)}]{tanaka2007theory}%
  \BibitemOpen
  \bibfield  {author} {\bibinfo {author} {\bibfnamefont {Y.}~\bibnamefont
  {Tanaka}}\ and\ \bibinfo {author} {\bibfnamefont {A.~A.}\ \bibnamefont
  {Golubov}},\ }\bibfield  {title} {\enquote {\bibinfo {title} {Theory of the
  proximity effect in junctions with unconventional superconductors},}\ }\href
  {\doibase 10.1103/PhysRevLett.98.037003} {\bibfield  {journal} {\bibinfo
  {journal} {Phys. Rev. Lett.}\ }\textbf {\bibinfo {volume} {98}},\ \bibinfo
  {pages} {037003} (\bibinfo {year} {2007})}\BibitemShut {NoStop}%
\bibitem [{\citenamefont {Tanaka}\ \emph {et~al.}(2007)\citenamefont {Tanaka},
  \citenamefont {Tanuma},\ and\ \citenamefont {Golubov}}]{TanakaPRB2007}%
  \BibitemOpen
  \bibfield  {author} {\bibinfo {author} {\bibfnamefont {Y.}~\bibnamefont
  {Tanaka}}, \bibinfo {author} {\bibfnamefont {Y.}~\bibnamefont {Tanuma}}, \
  and\ \bibinfo {author} {\bibfnamefont {A.A.}\ \bibnamefont {Golubov}},\
  }\bibfield  {title} {\enquote {\bibinfo {title} {Odd-frequency pairing in
  normal-metal/superconductor junctions},}\ }\href@noop {} {\bibfield
  {journal} {\bibinfo  {journal} {Phys. Rev. B}\ }\textbf {\bibinfo {volume}
  {76}},\ \bibinfo {pages} {054522} (\bibinfo {year} {2007})}\BibitemShut
  {NoStop}%
\bibitem [{\citenamefont {Rowell}\ and\ \citenamefont
  {McMillan}(1966)}]{rowell1966electron}%
  \BibitemOpen
  \bibfield  {author} {\bibinfo {author} {\bibfnamefont {JM}~\bibnamefont
  {Rowell}}\ and\ \bibinfo {author} {\bibfnamefont {WL}~\bibnamefont
  {McMillan}},\ }\bibfield  {title} {\enquote {\bibinfo {title} {Electron
  interference in a normal metal induced by superconducting contracts},}\
  }\href@noop {} {\bibfield  {journal} {\bibinfo  {journal} {Phys. Rev. Lett.}\
  }\textbf {\bibinfo {volume} {16}},\ \bibinfo {pages} {453} (\bibinfo {year}
  {1966})}\BibitemShut {NoStop}%
\bibitem [{\citenamefont {Rowell}(1973)}]{rowell1973tunneling}%
  \BibitemOpen
  \bibfield  {author} {\bibinfo {author} {\bibfnamefont {JM}~\bibnamefont
  {Rowell}},\ }\bibfield  {title} {\enquote {\bibinfo {title} {Tunneling
  observation of bound states in a normal metal—superconductor sandwich},}\
  }\href@noop {} {\bibfield  {journal} {\bibinfo  {journal} {Phys. Rev. Lett.}\
  }\textbf {\bibinfo {volume} {30}},\ \bibinfo {pages} {167} (\bibinfo {year}
  {1973})}\BibitemShut {NoStop}%
\bibitem [{\citenamefont {Alff}\ \emph {et~al.}(1997)\citenamefont {Alff},
  \citenamefont {Takashima}, \citenamefont {Kashiwaya}, \citenamefont {Terada},
  \citenamefont {Ihara}, \citenamefont {Tanaka}, \citenamefont {Koyanagi},\
  and\ \citenamefont {Kajimura}}]{alff1997spatially}%
  \BibitemOpen
  \bibfield  {author} {\bibinfo {author} {\bibfnamefont {L}~\bibnamefont
  {Alff}}, \bibinfo {author} {\bibfnamefont {H}~\bibnamefont {Takashima}},
  \bibinfo {author} {\bibfnamefont {S}~\bibnamefont {Kashiwaya}}, \bibinfo
  {author} {\bibfnamefont {N}~\bibnamefont {Terada}}, \bibinfo {author}
  {\bibfnamefont {H}~\bibnamefont {Ihara}}, \bibinfo {author} {\bibfnamefont
  {Y}~\bibnamefont {Tanaka}}, \bibinfo {author} {\bibfnamefont {M}~\bibnamefont
  {Koyanagi}}, \ and\ \bibinfo {author} {\bibfnamefont {K}~\bibnamefont
  {Kajimura}},\ }\bibfield  {title} {\enquote {\bibinfo {title} {Spatially
  continuous zero-bias conductance peak on (110) yba 2 cu 3 o 7- $\delta$
  surfaces},}\ }\href@noop {} {\bibfield  {journal} {\bibinfo  {journal} {Phys.
  Rev. B}\ }\textbf {\bibinfo {volume} {55}},\ \bibinfo {pages} {R14757}
  (\bibinfo {year} {1997})}\BibitemShut {NoStop}%
\bibitem [{\citenamefont {Covington}\ \emph {et~al.}(1997)\citenamefont
  {Covington}, \citenamefont {Aprili}, \citenamefont {Paraoanu}, \citenamefont
  {Greene}, \citenamefont {Xu}, \citenamefont {Zhu},\ and\ \citenamefont
  {Mirkin}}]{covington1997observation}%
  \BibitemOpen
  \bibfield  {author} {\bibinfo {author} {\bibfnamefont {M}~\bibnamefont
  {Covington}}, \bibinfo {author} {\bibfnamefont {M}~\bibnamefont {Aprili}},
  \bibinfo {author} {\bibfnamefont {E}~\bibnamefont {Paraoanu}}, \bibinfo
  {author} {\bibfnamefont {LH}~\bibnamefont {Greene}}, \bibinfo {author}
  {\bibfnamefont {F}~\bibnamefont {Xu}}, \bibinfo {author} {\bibfnamefont
  {J}~\bibnamefont {Zhu}}, \ and\ \bibinfo {author} {\bibfnamefont {Chad~A}\
  \bibnamefont {Mirkin}},\ }\bibfield  {title} {\enquote {\bibinfo {title}
  {Observation of surface-induced broken time-reversal symmetry in yba 2 cu 3 o
  7 tunnel junctions},}\ }\href@noop {} {\bibfield  {journal} {\bibinfo
  {journal} {Phys. Rev. Lett.}\ }\textbf {\bibinfo {volume} {79}},\ \bibinfo
  {pages} {277} (\bibinfo {year} {1997})}\BibitemShut {NoStop}%
\bibitem [{\citenamefont {Wei}\ \emph {et~al.}(1998)\citenamefont {Wei},
  \citenamefont {Yeh}, \citenamefont {Garrigus},\ and\ \citenamefont
  {Strasik}}]{wei1998directional}%
  \BibitemOpen
  \bibfield  {author} {\bibinfo {author} {\bibfnamefont {JYT}\ \bibnamefont
  {Wei}}, \bibinfo {author} {\bibfnamefont {N-C}\ \bibnamefont {Yeh}}, \bibinfo
  {author} {\bibfnamefont {DF}~\bibnamefont {Garrigus}}, \ and\ \bibinfo
  {author} {\bibfnamefont {M}~\bibnamefont {Strasik}},\ }\bibfield  {title}
  {\enquote {\bibinfo {title} {Directional tunneling and andreev reflection on
  yba 2 cu 3 o 7- $\delta$ single crystals: predominance of d-wave pairing
  symmetry verified with the generalized blonder, tinkham, and klapwijk
  theory},}\ }\href@noop {} {\bibfield  {journal} {\bibinfo  {journal} {Phys.
  Rev. Lett.}\ }\textbf {\bibinfo {volume} {81}},\ \bibinfo {pages} {2542}
  (\bibinfo {year} {1998})}\BibitemShut {NoStop}%
\bibitem [{\citenamefont {Linder}\ \emph
  {et~al.}(2010{\natexlab{a}})\citenamefont {Linder}, \citenamefont {Tanaka},
  \citenamefont {Yokoyama}, \citenamefont {Sudbo},\ and\ \citenamefont
  {Nagaosa}}]{linder2010}%
  \BibitemOpen
  \bibfield  {author} {\bibinfo {author} {\bibfnamefont {J.}~\bibnamefont
  {Linder}}, \bibinfo {author} {\bibfnamefont {Y.}~\bibnamefont {Tanaka}},
  \bibinfo {author} {\bibfnamefont {T.}~\bibnamefont {Yokoyama}}, \bibinfo
  {author} {\bibfnamefont {A.}~\bibnamefont {Sudbo}}, \ and\ \bibinfo {author}
  {\bibfnamefont {N.}~\bibnamefont {Nagaosa}},\ }\bibfield  {title} {\enquote
  {\bibinfo {title} {Unconventional {Superconductivity} on a {Topological}
  insulator},}\ }\href@noop {} {\bibfield  {journal} {\bibinfo  {journal}
  {Phys. Rev. Lett.}\ }\textbf {\bibinfo {volume} {104}},\ \bibinfo {pages}
  {067001} (\bibinfo {year} {2010}{\natexlab{a}})}\BibitemShut {NoStop}%
\bibitem [{\citenamefont {Linder}\ \emph
  {et~al.}(2010{\natexlab{b}})\citenamefont {Linder}, \citenamefont
  {Black-Schaffer},\ and\ \citenamefont {Sudbo}}]{linder2010_prb_graphene}%
  \BibitemOpen
  \bibfield  {author} {\bibinfo {author} {\bibfnamefont {J.}~\bibnamefont
  {Linder}}, \bibinfo {author} {\bibfnamefont {A.~M.}\ \bibnamefont
  {Black-Schaffer}}, \ and\ \bibinfo {author} {\bibfnamefont {A.}~\bibnamefont
  {Sudbo}},\ }\bibfield  {title} {\enquote {\bibinfo {title} {Triplet proximity
  effect and odd-frequency pairing in graphene},}\ }\href@noop {} {\bibfield
  {journal} {\bibinfo  {journal} {Phys. Rev. B}\ }\textbf {\bibinfo {volume}
  {82}},\ \bibinfo {pages} {041409} (\bibinfo {year}
  {2010}{\natexlab{b}})}\BibitemShut {NoStop}%
\bibitem [{\citenamefont {Black-Schaffer}\ and\ \citenamefont
  {Balatsky}(2012)}]{Black-SchafferPRB2012}%
  \BibitemOpen
  \bibfield  {author} {\bibinfo {author} {\bibfnamefont {A.M.}\ \bibnamefont
  {Black-Schaffer}}\ and\ \bibinfo {author} {\bibfnamefont {A.V.}\ \bibnamefont
  {Balatsky}},\ }\bibfield  {title} {\enquote {\bibinfo {title} {Odd-frequency
  superconducting pairing in topological insulators},}\ }\href@noop {}
  {\bibfield  {journal} {\bibinfo  {journal} {Phys. Rev. B}\ }\textbf {\bibinfo
  {volume} {86}},\ \bibinfo {pages} {144506} (\bibinfo {year}
  {2012})}\BibitemShut {NoStop}%
\bibitem [{\citenamefont {Black-Schaffer}\ and\ \citenamefont
  {Balatsky}(2013{\natexlab{a}})}]{Black-SchafferPRB2013}%
  \BibitemOpen
  \bibfield  {author} {\bibinfo {author} {\bibfnamefont {A.M.}\ \bibnamefont
  {Black-Schaffer}}\ and\ \bibinfo {author} {\bibfnamefont {A.V.}\ \bibnamefont
  {Balatsky}},\ }\bibfield  {title} {\enquote {\bibinfo {title}
  {Proximity-induced unconventional superconductivity in topological
  insulators},}\ }\href@noop {} {\bibfield  {journal} {\bibinfo  {journal}
  {Phys. Rev. B}\ }\textbf {\bibinfo {volume} {87}},\ \bibinfo {pages}
  {220506(R)} (\bibinfo {year} {2013}{\natexlab{a}})}\BibitemShut {NoStop}%
\bibitem [{\citenamefont {Parhizgar}\ and\ \citenamefont
  {Black-Schaffer}(2014{\natexlab{a}})}]{parhizgar2014}%
  \BibitemOpen
  \bibfield  {author} {\bibinfo {author} {\bibfnamefont {F.}~\bibnamefont
  {Parhizgar}}\ and\ \bibinfo {author} {\bibfnamefont {A.~M.}\ \bibnamefont
  {Black-Schaffer}},\ }\bibfield  {title} {\enquote {\bibinfo {title}
  {Unconventional proximity-induced superconductivity in bilayer systems},}\
  }\href@noop {} {\bibfield  {journal} {\bibinfo  {journal} {Phys. Rev. B}\
  }\textbf {\bibinfo {volume} {90}},\ \bibinfo {pages} {184517} (\bibinfo
  {year} {2014}{\natexlab{a}})}\BibitemShut {NoStop}%
\bibitem [{\citenamefont {Kuzmanovski}\ and\ \citenamefont
  {Black-Schaffer}(2017)}]{kuzmanovski2017multiple}%
  \BibitemOpen
  \bibfield  {author} {\bibinfo {author} {\bibfnamefont {Dushko}\ \bibnamefont
  {Kuzmanovski}}\ and\ \bibinfo {author} {\bibfnamefont {Annica~M}\
  \bibnamefont {Black-Schaffer}},\ }\bibfield  {title} {\enquote {\bibinfo
  {title} {Multiple odd-frequency superconducting states in buckled quantum
  spin hall insulators with time-reversal symmetry},}\ }\href@noop {}
  {\bibfield  {journal} {\bibinfo  {journal} {Phys. Rev. B}\ }\textbf {\bibinfo
  {volume} {96}},\ \bibinfo {pages} {174509} (\bibinfo {year}
  {2017})}\BibitemShut {NoStop}%
\bibitem [{\citenamefont {Rahimi}\ \emph {et~al.}(2017)\citenamefont {Rahimi},
  \citenamefont {Moghaddam}, \citenamefont {Dykstra}, \citenamefont
  {Governale},\ and\ \citenamefont {Z{\"u}licke}}]{rahimi2017unconventional}%
  \BibitemOpen
  \bibfield  {author} {\bibinfo {author} {\bibfnamefont {MA}~\bibnamefont
  {Rahimi}}, \bibinfo {author} {\bibfnamefont {AG}~\bibnamefont {Moghaddam}},
  \bibinfo {author} {\bibfnamefont {C}~\bibnamefont {Dykstra}}, \bibinfo
  {author} {\bibfnamefont {M}~\bibnamefont {Governale}}, \ and\ \bibinfo
  {author} {\bibfnamefont {U}~\bibnamefont {Z{\"u}licke}},\ }\bibfield  {title}
  {\enquote {\bibinfo {title} {Unconventional superconductivity from magnetism
  in transition-metal dichalcogenides},}\ }\href@noop {} {\bibfield  {journal}
  {\bibinfo  {journal} {Phys. Rev. B}\ }\textbf {\bibinfo {volume} {95}},\
  \bibinfo {pages} {104515} (\bibinfo {year} {2017})}\BibitemShut {NoStop}%
\bibitem [{\citenamefont {Aliabad}\ and\ \citenamefont
  {Zare}(2018)}]{aliabad2018proximity}%
  \BibitemOpen
  \bibfield  {author} {\bibinfo {author} {\bibfnamefont {Mojtaba~Rahimi}\
  \bibnamefont {Aliabad}}\ and\ \bibinfo {author} {\bibfnamefont
  {Mohammad-Hossein}\ \bibnamefont {Zare}},\ }\bibfield  {title} {\enquote
  {\bibinfo {title} {Proximity-induced mixed odd-and even-frequency pairing in
  monolayer {NbSe$_2$}},}\ }\href@noop {} {\bibfield  {journal} {\bibinfo
  {journal} {Phys. Rev. B}\ }\textbf {\bibinfo {volume} {97}},\ \bibinfo
  {pages} {224503} (\bibinfo {year} {2018})}\BibitemShut {NoStop}%
\bibitem [{\citenamefont {Linder}\ \emph
  {et~al.}(2010{\natexlab{c}})\citenamefont {Linder}, \citenamefont {Sudbo},
  \citenamefont {Yokoyama}, \citenamefont {Grein},\ and\ \citenamefont
  {Eschrig}}]{linder2010_magnetic}%
  \BibitemOpen
  \bibfield  {author} {\bibinfo {author} {\bibfnamefont {J.}~\bibnamefont
  {Linder}}, \bibinfo {author} {\bibfnamefont {A.}~\bibnamefont {Sudbo}},
  \bibinfo {author} {\bibfnamefont {T.}~\bibnamefont {Yokoyama}}, \bibinfo
  {author} {\bibfnamefont {R.}~\bibnamefont {Grein}}, \ and\ \bibinfo {author}
  {\bibfnamefont {M.}~\bibnamefont {Eschrig}},\ }\bibfield  {title} {\enquote
  {\bibinfo {title} {Signature of odd-frequency pairing correlations induced by
  a magnetic interface},}\ }\href@noop {} {\bibfield  {journal} {\bibinfo
  {journal} {Phys. Rev. B}\ }\textbf {\bibinfo {volume} {81}},\ \bibinfo
  {pages} {214504} (\bibinfo {year} {2010}{\natexlab{c}})}\BibitemShut
  {NoStop}%
\bibitem [{\citenamefont {Gor'kov}\ and\ \citenamefont
  {Rashba}(2001)}]{Gorkov2001}%
  \BibitemOpen
  \bibfield  {author} {\bibinfo {author} {\bibfnamefont {Lev~P.}\ \bibnamefont
  {Gor'kov}}\ and\ \bibinfo {author} {\bibfnamefont {Emmanuel~I.}\ \bibnamefont
  {Rashba}},\ }\bibfield  {title} {\enquote {\bibinfo {title} {Superconducting
  2d system with lifted spin degeneracy: Mixed singlet-triplet state},}\ }\href
  {http://link.aps.org/doi/10.1103/PhysRevLett.87.037004} {\bibfield  {journal}
  {\bibinfo  {journal} {Phys. Rev. Lett.}\ }\textbf {\bibinfo {volume} {87}},\
  \bibinfo {pages} {037004} (\bibinfo {year} {2001})}\BibitemShut {NoStop}%
\bibitem [{\citenamefont {Linder}\ \emph {et~al.}(2009)\citenamefont {Linder},
  \citenamefont {Yokoyama}, \citenamefont {Sudb{\o}},\ and\ \citenamefont
  {Eschrig}}]{LinderPRL2009}%
  \BibitemOpen
  \bibfield  {author} {\bibinfo {author} {\bibfnamefont {J.}~\bibnamefont
  {Linder}}, \bibinfo {author} {\bibfnamefont {T.}~\bibnamefont {Yokoyama}},
  \bibinfo {author} {\bibfnamefont {A.}~\bibnamefont {Sudb{\o}}}, \ and\
  \bibinfo {author} {\bibfnamefont {M.}~\bibnamefont {Eschrig}},\ }\bibfield
  {title} {\enquote {\bibinfo {title} {Pairing symmetry conversion by
  spin-active interfaces in magnetic normal-metal-superconductor junctions},}\
  }\href@noop {} {\bibfield  {journal} {\bibinfo  {journal} {Phys. Rev. Lett.}\
  }\textbf {\bibinfo {volume} {102}},\ \bibinfo {pages} {107008} (\bibinfo
  {year} {2009})}\BibitemShut {NoStop}%
\bibitem [{\citenamefont {Linder}\ \emph
  {et~al.}(2010{\natexlab{d}})\citenamefont {Linder}, \citenamefont {Sudb{\o}},
  \citenamefont {Yokoyama}, \citenamefont {Grein},\ and\ \citenamefont
  {Eschrig}}]{LinderPRB2010_2}%
  \BibitemOpen
  \bibfield  {author} {\bibinfo {author} {\bibfnamefont {J.}~\bibnamefont
  {Linder}}, \bibinfo {author} {\bibfnamefont {A.}~\bibnamefont {Sudb{\o}}},
  \bibinfo {author} {\bibfnamefont {T.}~\bibnamefont {Yokoyama}}, \bibinfo
  {author} {\bibfnamefont {R.}~\bibnamefont {Grein}}, \ and\ \bibinfo {author}
  {\bibfnamefont {M.}~\bibnamefont {Eschrig}},\ }\bibfield  {title} {\enquote
  {\bibinfo {title} {Signature of odd-frequency pairing correlations induced by
  a magnetic interface},}\ }\href@noop {} {\bibfield  {journal} {\bibinfo
  {journal} {Phys. Rev. B}\ }\textbf {\bibinfo {volume} {81}},\ \bibinfo
  {pages} {214504} (\bibinfo {year} {2010}{\natexlab{d}})}\BibitemShut
  {NoStop}%
\bibitem [{\citenamefont {Haldane}(1988)}]{haldane1988model}%
  \BibitemOpen
  \bibfield  {author} {\bibinfo {author} {\bibfnamefont {F~Duncan~M}\
  \bibnamefont {Haldane}},\ }\bibfield  {title} {\enquote {\bibinfo {title}
  {Model for a quantum hall effect without landau levels: Condensed-matter
  realization of the" parity anomaly"},}\ }\href@noop {} {\bibfield  {journal}
  {\bibinfo  {journal} {Phys. Rev. Lett.}\ }\textbf {\bibinfo {volume} {61}},\
  \bibinfo {pages} {2015} (\bibinfo {year} {1988})}\BibitemShut {NoStop}%
\bibitem [{\citenamefont {Kane}\ and\ \citenamefont
  {Mele}(2005)}]{kane2005qshi}%
  \BibitemOpen
  \bibfield  {author} {\bibinfo {author} {\bibfnamefont {C.~L.}\ \bibnamefont
  {Kane}}\ and\ \bibinfo {author} {\bibfnamefont {E.~J.}\ \bibnamefont
  {Mele}},\ }\bibfield  {title} {\enquote {\bibinfo {title} {Quantum spin hall
  effect in graphene},}\ }\href@noop {} {\bibfield  {journal} {\bibinfo
  {journal} {Phys. Rev. Lett.}\ }\textbf {\bibinfo {volume} {95}},\ \bibinfo
  {pages} {226801} (\bibinfo {year} {2005})}\BibitemShut {NoStop}%
\bibitem [{\citenamefont {Black-Schaffer}\ and\ \citenamefont
  {Balatsky}(2013{\natexlab{b}})}]{black2013odd}%
  \BibitemOpen
  \bibfield  {author} {\bibinfo {author} {\bibfnamefont {Annica~M.}\
  \bibnamefont {Black-Schaffer}}\ and\ \bibinfo {author} {\bibfnamefont
  {Alexander~V.}\ \bibnamefont {Balatsky}},\ }\bibfield  {title} {\enquote
  {\bibinfo {title} {Odd-frequency superconducting pairing in multiband
  superconductors},}\ }\href {\doibase 10.1103/PhysRevB.88.104514} {\bibfield
  {journal} {\bibinfo  {journal} {Phys. Rev. B}\ }\textbf {\bibinfo {volume}
  {88}},\ \bibinfo {pages} {104514} (\bibinfo {year}
  {2013}{\natexlab{b}})}\BibitemShut {NoStop}%
\bibitem [{\citenamefont {Asano}\ and\ \citenamefont
  {Sasaki}(2015)}]{asano2015odd}%
  \BibitemOpen
  \bibfield  {author} {\bibinfo {author} {\bibfnamefont {Yasuhiro}\
  \bibnamefont {Asano}}\ and\ \bibinfo {author} {\bibfnamefont {Akihiro}\
  \bibnamefont {Sasaki}},\ }\bibfield  {title} {\enquote {\bibinfo {title}
  {Odd-frequency cooper pairs in two-band superconductors and their magnetic
  response},}\ }\href@noop {} {\bibfield  {journal} {\bibinfo  {journal} {Phys.
  Rev. B}\ }\textbf {\bibinfo {volume} {92}},\ \bibinfo {pages} {224508}
  (\bibinfo {year} {2015})}\BibitemShut {NoStop}%
\bibitem [{\citenamefont {Komendov\'a}\ \emph {et~al.}(2015)\citenamefont
  {Komendov\'a}, \citenamefont {Balatsky},\ and\ \citenamefont
  {Black-Schaffer}}]{komendova2015experimentally}%
  \BibitemOpen
  \bibfield  {author} {\bibinfo {author} {\bibfnamefont {L.}~\bibnamefont
  {Komendov\'a}}, \bibinfo {author} {\bibfnamefont {A.~V.}\ \bibnamefont
  {Balatsky}}, \ and\ \bibinfo {author} {\bibfnamefont {A.~M.}\ \bibnamefont
  {Black-Schaffer}},\ }\bibfield  {title} {\enquote {\bibinfo {title}
  {Experimentally observable signatures of odd-frequency pairing in multiband
  superconductors},}\ }\href {\doibase 10.1103/PhysRevB.92.094517} {\bibfield
  {journal} {\bibinfo  {journal} {Phys. Rev. B}\ }\textbf {\bibinfo {volume}
  {92}},\ \bibinfo {pages} {094517} (\bibinfo {year} {2015})}\BibitemShut
  {NoStop}%
\bibitem [{\citenamefont {Komendov\'a}\ and\ \citenamefont
  {Black-Schaffer}(2017)}]{komendova2017odd}%
  \BibitemOpen
  \bibfield  {author} {\bibinfo {author} {\bibfnamefont {L.}~\bibnamefont
  {Komendov\'a}}\ and\ \bibinfo {author} {\bibfnamefont {A.~M.}\ \bibnamefont
  {Black-Schaffer}},\ }\bibfield  {title} {\enquote {\bibinfo {title}
  {Odd-frequency superconductivity in ${\mathrm{sr}}_{2}{\mathrm{ruo}}_{4}$
  measured by kerr rotation},}\ }\href {\doibase
  10.1103/PhysRevLett.119.087001} {\bibfield  {journal} {\bibinfo  {journal}
  {Phys. Rev. Lett.}\ }\textbf {\bibinfo {volume} {119}},\ \bibinfo {pages}
  {087001} (\bibinfo {year} {2017})}\BibitemShut {NoStop}%
\bibitem [{\citenamefont {Asano}\ and\ \citenamefont
  {Golubov}(2018)}]{asano2018green}%
  \BibitemOpen
  \bibfield  {author} {\bibinfo {author} {\bibfnamefont {Yasuhiro}\
  \bibnamefont {Asano}}\ and\ \bibinfo {author} {\bibfnamefont {Alexander~A}\
  \bibnamefont {Golubov}},\ }\bibfield  {title} {\enquote {\bibinfo {title}
  {Green's-function theory of dirty two-band superconductivity},}\ }\href@noop
  {} {\bibfield  {journal} {\bibinfo  {journal} {Phys. Rev. B}\ }\textbf
  {\bibinfo {volume} {97}},\ \bibinfo {pages} {214508} (\bibinfo {year}
  {2018})}\BibitemShut {NoStop}%
\bibitem [{\citenamefont {Triola}\ and\ \citenamefont
  {Black-Schaffer}(2018)}]{triola2018odd}%
  \BibitemOpen
  \bibfield  {author} {\bibinfo {author} {\bibfnamefont {Christopher}\
  \bibnamefont {Triola}}\ and\ \bibinfo {author} {\bibfnamefont {Annica~M}\
  \bibnamefont {Black-Schaffer}},\ }\bibfield  {title} {\enquote {\bibinfo
  {title} {Odd-frequency pairing and kerr effect in the heavy-fermion
  superconductor upt 3},}\ }\href@noop {} {\bibfield  {journal} {\bibinfo
  {journal} {Phys. Rev. B}\ }\textbf {\bibinfo {volume} {97}},\ \bibinfo
  {pages} {064505} (\bibinfo {year} {2018})}\BibitemShut {NoStop}%
\bibitem [{\citenamefont {Parhizgar}\ and\ \citenamefont
  {Black-Schaffer}(2014{\natexlab{b}})}]{parhizgar_2014_prb}%
  \BibitemOpen
  \bibfield  {author} {\bibinfo {author} {\bibfnamefont {Fariborz}\
  \bibnamefont {Parhizgar}}\ and\ \bibinfo {author} {\bibfnamefont {Annica~M}\
  \bibnamefont {Black-Schaffer}},\ }\bibfield  {title} {\enquote {\bibinfo
  {title} {Unconventional proximity-induced superconductivity in bilayer
  systems},}\ }\href@noop {} {\bibfield  {journal} {\bibinfo  {journal}
  {Physical Review B}\ }\textbf {\bibinfo {volume} {90}},\ \bibinfo {pages}
  {184517} (\bibinfo {year} {2014}{\natexlab{b}})}\BibitemShut {NoStop}%
\bibitem [{\citenamefont {Sothmann}\ \emph {et~al.}(2014)\citenamefont
  {Sothmann}, \citenamefont {Weiss}, \citenamefont {Governale},\ and\
  \citenamefont {K{\"o}nig}}]{sothmann2014unconventional}%
  \BibitemOpen
  \bibfield  {author} {\bibinfo {author} {\bibfnamefont {Bj{\"o}rn}\
  \bibnamefont {Sothmann}}, \bibinfo {author} {\bibfnamefont {Stephan}\
  \bibnamefont {Weiss}}, \bibinfo {author} {\bibfnamefont {Michele}\
  \bibnamefont {Governale}}, \ and\ \bibinfo {author} {\bibfnamefont
  {J{\"u}rgen}\ \bibnamefont {K{\"o}nig}},\ }\bibfield  {title} {\enquote
  {\bibinfo {title} {Unconventional superconductivity in double quantum
  dots},}\ }\href@noop {} {\bibfield  {journal} {\bibinfo  {journal} {Phys.
  Rev. B}\ }\textbf {\bibinfo {volume} {90}},\ \bibinfo {pages} {220501}
  (\bibinfo {year} {2014})}\BibitemShut {NoStop}%
\bibitem [{\citenamefont {Burset}\ \emph {et~al.}(2016)\citenamefont {Burset},
  \citenamefont {Lu}, \citenamefont {Ebisu}, \citenamefont {Asano},\ and\
  \citenamefont {Tanaka}}]{burset2016all}%
  \BibitemOpen
  \bibfield  {author} {\bibinfo {author} {\bibfnamefont {Pablo}\ \bibnamefont
  {Burset}}, \bibinfo {author} {\bibfnamefont {Bo}~\bibnamefont {Lu}}, \bibinfo
  {author} {\bibfnamefont {Hiromi}\ \bibnamefont {Ebisu}}, \bibinfo {author}
  {\bibfnamefont {Yasuhiro}\ \bibnamefont {Asano}}, \ and\ \bibinfo {author}
  {\bibfnamefont {Yukio}\ \bibnamefont {Tanaka}},\ }\bibfield  {title}
  {\enquote {\bibinfo {title} {All-electrical generation and control of
  odd-frequency s-wave cooper pairs in double quantum dots},}\ }\href@noop {}
  {\bibfield  {journal} {\bibinfo  {journal} {Phys. Rev. B}\ }\textbf {\bibinfo
  {volume} {93}},\ \bibinfo {pages} {201402} (\bibinfo {year}
  {2016})}\BibitemShut {NoStop}%
\bibitem [{\citenamefont {Ebisu}\ \emph {et~al.}(2016)\citenamefont {Ebisu},
  \citenamefont {Lu}, \citenamefont {Klinovaja},\ and\ \citenamefont
  {Tanaka}}]{ebisu2016theory}%
  \BibitemOpen
  \bibfield  {author} {\bibinfo {author} {\bibfnamefont {Hiromi}\ \bibnamefont
  {Ebisu}}, \bibinfo {author} {\bibfnamefont {Bo}~\bibnamefont {Lu}}, \bibinfo
  {author} {\bibfnamefont {Jelena}\ \bibnamefont {Klinovaja}}, \ and\ \bibinfo
  {author} {\bibfnamefont {Yukio}\ \bibnamefont {Tanaka}},\ }\bibfield  {title}
  {\enquote {\bibinfo {title} {{Theory of time-reversal topological
  superconductivity in double Rashba wires: symmetries of Cooper pairs and
  Andreev bound states}},}\ }\href {\doibase 10.1093/ptep/ptw094} {\bibfield
  {journal} {\bibinfo  {journal} {Progress of Theoretical and Experimental
  Physics}\ }\textbf {\bibinfo {volume} {2016}},\ \bibinfo {pages} {083I01}
  (\bibinfo {year} {2016})}\BibitemShut {NoStop}%
\bibitem [{\citenamefont {Triola}\ and\ \citenamefont
  {Black-Schaffer}(2019)}]{triola2019oddnw}%
  \BibitemOpen
  \bibfield  {author} {\bibinfo {author} {\bibfnamefont {Christopher}\
  \bibnamefont {Triola}}\ and\ \bibinfo {author} {\bibfnamefont {Annica~M}\
  \bibnamefont {Black-Schaffer}},\ }\bibfield  {title} {\enquote {\bibinfo
  {title} {Odd-frequency pairing in a superconductor coupled to two parallel
  nanowires},}\ }\href@noop {} {\bibfield  {journal} {\bibinfo  {journal}
  {Phys. Rev. B}\ }\textbf {\bibinfo {volume} {100}},\ \bibinfo {pages}
  {024512} (\bibinfo {year} {2019})}\BibitemShut {NoStop}%
\bibitem [{\citenamefont {Ugeda}\ \emph {et~al.}(2015)\citenamefont {Ugeda},
  \citenamefont {Bradley}, \citenamefont {Zhang}, \citenamefont {Onishi},
  \citenamefont {Chen}, \citenamefont {Ruan}, \citenamefont
  {Ojeda-Aristizabal}, \citenamefont {Ryu}, \citenamefont {Edmonds},\ and\
  \citenamefont {Tsai}}]{ugeda2015}%
  \BibitemOpen
  \bibfield  {author} {\bibinfo {author} {\bibfnamefont {Miguel~M.}\
  \bibnamefont {Ugeda}}, \bibinfo {author} {\bibfnamefont {Aaron~J.}\
  \bibnamefont {Bradley}}, \bibinfo {author} {\bibfnamefont {Yi}~\bibnamefont
  {Zhang}}, \bibinfo {author} {\bibfnamefont {Seita}\ \bibnamefont {Onishi}},
  \bibinfo {author} {\bibfnamefont {Yi}~\bibnamefont {Chen}}, \bibinfo {author}
  {\bibfnamefont {Wei}\ \bibnamefont {Ruan}}, \bibinfo {author} {\bibfnamefont
  {Claudia}\ \bibnamefont {Ojeda-Aristizabal}}, \bibinfo {author}
  {\bibfnamefont {Hyejin}\ \bibnamefont {Ryu}}, \bibinfo {author}
  {\bibfnamefont {Mark~T.}\ \bibnamefont {Edmonds}}, \ and\ \bibinfo {author}
  {\bibfnamefont {Hsin-Zon et~al.}\ \bibnamefont {Tsai}},\ }\bibfield  {title}
  {\enquote {\bibinfo {title} {Characterization of collective ground states in
  single-layer {NbSe$_2$}},}\ }\href {\doibase 10.1038/nphys3527} {\bibfield
  {journal} {\bibinfo  {journal} {Nature Physics}\ }\textbf {\bibinfo {volume}
  {12}},\ \bibinfo {pages} {92--97} (\bibinfo {year} {2015})}\BibitemShut
  {NoStop}%
\bibitem [{\citenamefont {Xi}\ \emph {et~al.}(2015)\citenamefont {Xi},
  \citenamefont {Wang}, \citenamefont {Zhao}, \citenamefont {Park},
  \citenamefont {Law}, \citenamefont {Berger}, \citenamefont {Forró},
  \citenamefont {Shan},\ and\ \citenamefont {Mak}}]{xi2016}%
  \BibitemOpen
  \bibfield  {author} {\bibinfo {author} {\bibfnamefont {Xiaoxiang}\
  \bibnamefont {Xi}}, \bibinfo {author} {\bibfnamefont {Zefang}\ \bibnamefont
  {Wang}}, \bibinfo {author} {\bibfnamefont {Weiwei}\ \bibnamefont {Zhao}},
  \bibinfo {author} {\bibfnamefont {Ju-Hyun}\ \bibnamefont {Park}}, \bibinfo
  {author} {\bibfnamefont {Kam~Tuen}\ \bibnamefont {Law}}, \bibinfo {author}
  {\bibfnamefont {Helmuth}\ \bibnamefont {Berger}}, \bibinfo {author}
  {\bibfnamefont {László}\ \bibnamefont {Forró}}, \bibinfo {author}
  {\bibfnamefont {Jie}\ \bibnamefont {Shan}}, \ and\ \bibinfo {author}
  {\bibfnamefont {Kin~Fai}\ \bibnamefont {Mak}},\ }\bibfield  {title} {\enquote
  {\bibinfo {title} {Ising pairing in superconducting {NbSe$_2$}
  atomic layers},}\ }\href {\doibase 10.1038/nphys3538} {\bibfield  {journal}
  {\bibinfo  {journal} {Nature Physics}\ }\textbf {\bibinfo {volume} {12}},\
  \bibinfo {pages} {139--143} (\bibinfo {year} {2015})}\BibitemShut {NoStop}%
\bibitem [{\citenamefont {Dvir}\ \emph {et~al.}(2018)\citenamefont {Dvir},
  \citenamefont {Aprili}, \citenamefont {Quay},\ and\ \citenamefont
  {Steinberg}}]{dvir2018}%
  \BibitemOpen
  \bibfield  {author} {\bibinfo {author} {\bibfnamefont {Tom}\ \bibnamefont
  {Dvir}}, \bibinfo {author} {\bibfnamefont {Marco}\ \bibnamefont {Aprili}},
  \bibinfo {author} {\bibfnamefont {Charis H.~L.}\ \bibnamefont {Quay}}, \ and\
  \bibinfo {author} {\bibfnamefont {Hadar}\ \bibnamefont {Steinberg}},\
  }\bibfield  {title} {\enquote {\bibinfo {title} {Tunneling into the vortex
  state of {NbSe$_2$} with van der {Waals} junctions},}\ }\href {\doibase
  10.1021/acs.nanolett.8b03605} {\bibfield  {journal} {\bibinfo  {journal}
  {Nano Letters}\ }\textbf {\bibinfo {volume} {18}},\ \bibinfo {pages}
  {7845--7850} (\bibinfo {year} {2018})}\BibitemShut {NoStop}%
\end{thebibliography}%


\end{document}